\documentclass[11pt,onecolumn,english,onecolumn]{IEEEtran}
\usepackage[T1]{fontenc}
\usepackage[latin9]{inputenc}
\usepackage{xcolor}
\usepackage{units}
\usepackage{amsmath}
\usepackage{amsthm}
\usepackage{amssymb}
\usepackage{graphicx}
\usepackage{geometry}
\usepackage{setspace}
\PassOptionsToPackage{normalem}{ulem}
\usepackage{ulem}
\makeatletter

\providecolor{lyxadded}{rgb}{0,0,1}
\providecolor{lyxdeleted}{rgb}{1,0,0}
\DeclareRobustCommand{\mklyxadded}[1]{\textcolor{lyxadded}\bgroup#1\egroup}
\DeclareRobustCommand{\mklyxdeleted}[1]{\textcolor{lyxdeleted}\bgroup\mklyxsout{#1}\egroup}
\DeclareRobustCommand{\mklyxsout}[1]{\ifx\\#1\else\sout{#1}\fi}
\theoremstyle{plain}
\newtheorem{thm}{\protect\theoremname}
\theoremstyle{definition}
\newtheorem{defn}[thm]{\protect\definitionname}
\theoremstyle{plain}
\newtheorem{lem}[thm]{\protect\lemmaname}
\theoremstyle{plain}
\newtheorem{prop}[thm]{\protect\propositionname}
\theoremstyle{definition}
\newtheorem{example}[thm]{\protect\examplename}

\usepackage[pagebackref]{hyperref}
\usepackage{enumitem}
\usepackage{breqn}
\usepackage{bbm} 
\usepackage{cite}

\renewcommand\[{\begin{equation}}
\renewcommand\]{\end{equation}}

\DeclareMathOperator*{\argmax}{arg\,max} 
\DeclareMathOperator{\supp}{supp}

\allowdisplaybreaks

\global\long\def\eqd{\stackrel{d}{=}}
\global\long\def\P{\mathbb{P}}
\global\long\def\E{\mathbb{E}}
\global\long\def\V{\mathbb{V}}

\global\long\def\I{\mathbbm{1}}
\global\long\def\d{\mathrm{d}}

\global\long\def\trre[#1,#2]{\overset{{\scriptstyle (#2)}}{#1}} % transition explained with reason

\author{
\IEEEauthorblockN{Nir Luria and Nir Weinberger}

\IEEEauthorblockA{The Viterbi Faculty of Electrical and Computer Engineering\\
  	    Technion - Israel Institute of Technology\\
Technion City, Haifa 3200004, Israel
} \\
\IEEEauthorblockA{nir.luria@campus.technion.ac.il, nirwein@technion.ac.il.}\\
}

\makeatother

\usepackage{babel}
\providecommand{\definitionname}{Definition}
\providecommand{\examplename}{Example}
\providecommand{\lemmaname}{Lemma}
\providecommand{\propositionname}{Proposition}
\providecommand{\theoremname}{Theorem}

\begin{document}
\title{Optimal Overlap Detection of Shotgun Reads\thanks{Nir Luria and Nir Weinberger are with the Viterbi Faculty of Electrical
and Computer Engineering, Technion-Israel Institute of Technology,
Haifa 3200004, Israel (Emails: \texttt{nir.luria@campus.technion.ac.il},
\texttt{nirwein@technion.ac.il.}). Their work is supported by the
Israel Science Foundation (ISF), grant no. 1782/22. A short version
of this paper was submitted to the 2025 IEEE International Symposium
on Information Theory. This work has been submitted to the IEEE for
possible publication. Copyright may be transferred without notice,
after which this version may no longer be accessible.}}
\maketitle
\begin{abstract}
We consider the problem of detecting the overlap between a pair of
short fragments sampled in random locations from an exponentially
longer sequence, via their possibly noisy reads. We consider a noiseless
setting, in which the reads are noiseless, and the sequence is only
assumed to be stationary and ergodic. Under mild conditions on the
mixing property of the process generating the sequence, we characterize
exactly the asymptotic error probability of the optimal Bayesian detector.
Similarly, we consider a noisy setting, in which the reads are noisy
versions of the sampled fragments obtained via a memoryless channel.
We further assume that the sequence is stationary and memoryless,
and similarly characterize exactly the asymptotic error probability
of the optimal Bayesian detector for this case. 
\end{abstract}


\section{Introduction}

Consider an archaeologist who discovers two short torn fragments from
an ancient long text corpus. When it can be reliably determined that
these two fragments come from the same location in the text, and thus
should be stitched together to form a longer contiguous fragment?
Alternatively, when it can be determined that these two fragments
are parts of different locations in the text? A correct answer to
this question would affect the conclusions that can be made from this
discovery. Notwithstanding the above example, the most well-known
instance of this problem arises in \emph{DNA shotgun sequencing} \cite{reuter2015high},
in which the goal is to assembly a long genomic sequence based on
\emph{shotgun reads}. In this setup, short fragments are sampled from
random starting locations on the sequence, and then each of them is
independently sequenced to obtain a \emph{read} -- a possibly corrupted
version of this fragment. The set of noisy reads is then input to
an assembly algorithm, which outputs a reconstruction of the sequence.
From a more general perspective, alignment of short signals is a widely
used pre-processing technique. In image processing, one can crate
a panoramic view from smaller images, or connect localized medical
scans to form one whole medical image. In hardware security, many
power traces are recorded and should be aligned in order to understand
the power behavior of the design. In this paper, we address a distilled
version of this setting: Given a \emph{single pair} of reads from
a long sequence, what are the fundamental limits of the detection
of the overlap length, or the relative alignment, between these reads?
To address this question, we develop the optimal Bayesian detector
for this problem, and derive exact asymptotic expressions for its
optimal Bayesian error probability. 

\subsection{Motivation and Background: Fundamental Limits of DNA Shotgun Sequencing}

As said, a prominent example of the alignment problem is DNA sequencing,
and so we next focus on this setting. Our motivation for addressing
the alignment problem in isolation arises, because, as we shall next
justify, an analysis of the fundamental limits of the reconstruction
of the full sequence from multiple reads masks this basic question.
To this end, we briefly review the information-theoretic study of
fundamental limits of DNA shotgun sequencing, pioneered in \cite{motahari2013information}.
The genome $\mathbf{X}$ is modeled as a length $n$ sequence over
an alphabet ${\cal X}=\{\text{A,C,G,T}\}$. Each of the $m$ reads
has length $\ell$, and a starting point that is chosen uniformly
over the sequence,\footnote{That is, uniformly over $[n]:=\{1,2,\ldots,n\}$, with some adjustment
at the boundaries; \cite{motahari2013information} analyzed a cyclic
sequence model, and we will consider here a slightly different model
at the boundaries. Both choice are made mainly for simplicity, and
do not have any impact on the asymptotic analytical results. } independently of the starting points of all other reads. The interesting
analytical regime is when the reads are exponentially shorter than
the sequence, concretely $\ell=\beta\log n$, for some fixed scaling
constant $\beta>0$. Trivially, if $m\leq\frac{n}{\ell}$ then the
sequence is not even fully covered by the reads, and thus cannot be
fully reconstructed from the $m$ reads. In their classic paper \cite{lander1988genomic},
Lander and Waterman used a coupon-collection argument to show that
if the \emph{coverage condition} $m\gtrsim\frac{n}{\ell}\log(\frac{n}{\ell\epsilon})\sim\frac{n}{\beta}$
holds, then $m$ reads will fully cover the entire sequence with probability
$1-\epsilon$. As argued later on in \cite{arratia1996poisson}, this
does not suffice to assure a reliable reconstruction of the sequence,
due possible repetitions in the sequence. Indeed, it was shown in
\cite{ukkonen1992approximate} that if there is either a substring
of length $\ell-1$ that repeats itself \emph{three} times, or there
are two interleaved pairs of repeats of length $\ell-1$, then perfect
reconstruction is impossible, even given the $\ell$-spectrum --
the set of all reads of length $\ell$ in the sequence (see \cite{bresler2013optimal}
for refined conditions on the required coverage depth). It was proved
in \cite{arratia1996poisson} that for a sequence generated from a
memoryless process with a symbol probability mass function (p.m.f.)
$P_{X}$, such interleaved repeats will occur with high probability
whenever $\beta<\frac{2}{H_{2}(P_{X})}$, where $H_{2}(P_{X})$ is
the second-order R\'{e}nyi entropy $H_{2}(P_{X}):=-\log\sum_{x\in{\cal X}}P_{X}^{2}(x)$.
It was then proved in \cite{motahari2013information} that if the
sequence is a memoryless random process, and the \emph{repeat condition}
$\beta\geq\frac{2}{H_{2}(P_{X})}$ holds on top of the coverage condition,
it is possible to perfectly reconstruct the sequence from $m$ reads. 

Importantly, the requirement of the covering condition for reliable
reconstruction greatly influenced the reconstruction algorithm proposed
and analyzed in \cite{motahari2013information}. This algorithm is
greedy in its nature, and keeps track of a set of \emph{contigs} --
a contiguous fragment of overlapping reads. At first, the set of contigs
is initialized to the set of sequenced reads. Then, at each of $m-1$
steps, the algorithm finds a pair of contigs with maximal overlap
and merges them into a single contig. The analysis of the probability
that the algorithm fails to perfectly reconstruct the sequence leads
to the necessary repeat-condition $\beta\geq\frac{2}{H_{2}(P_{X})}$.
Intuitively speaking, it is based on the following arguments: (i)
Long overlaps between reads from distant locations are unlikely. Indeed,
consider overlaps of length $\alpha\ell$ for some $\alpha\in(0,1)$.
For a memoryless process over $|{\cal X}|$, the probability that
two reads from distant locations in the sequence will share the exact
same symbols is $|{\cal X}|^{-H_{2}(P_{X})\alpha\ell}$, and since
$\ell=\beta\log n$, this probability polynomially decays with $n$
(with degree $\alpha\beta H_{2}(P_{X})$). This probability decays
fast enough if $\alpha$ is large enough. (ii) It is highly unlikely
for a specific symbol in the sequence \emph{not} to be covered by
at least one pair of reads that has a \emph{long} overlap. Indeed,
given the minimum number of reads to satisfy the coverage condition,
on the average, each symbol is covered by $\frac{m\ell}{n}\sim\log n$
reads. Since the relative alignment between the reads is uniformly,
it is highly unlikely that neither pair from these reads does not
have a long overlap. Therefore, reliable reconstruction is essentially
obtained by almost exclusively merging reads with \emph{long} overlaps. 

So, since the algorithm is based on merging reads with sufficiently
long overlaps, the analysis of this algorithm masks the basic question
regarding the ability to detect an overlap, without any prior guarantee
that this overlap is long enough. In fact, detection of short overlaps
is more efficient, because they result longer contigs. Interestingly,
\cite{lander1988genomic}, dated to the early days of the genome sequencing
projects, considered the case in which the genome is not fully covered
(without a rigorous analysis). In the parlance of \cite{lander1988genomic},
an ideal alignment of the available reads in such cases will result
in a set of contigs, possibly separated by uncovered symbols. The
reconstruction algorithm results \emph{islands}, whose correct order
cannot be determined from the available reads. In this under-coverage
regime, $m\leq\frac{n}{\ell}$, it is possible for a read to have
a relatively short maximal overlap with all other reads. Such shorts
overlaps are not used in the reconstruction algorithm of \cite{motahari2013information}
(with high probability), as it is unclear when a detection of such
an overlap is actually reliable, and, as said, it is also necessary
to use such overlaps whenever the coverage condition holds. This question
remains a timely challenge, even with contemporary high-throughput
DNA sequencing technologies: In \emph{metagenomics analysis} such
as that of the Human Microbiome Project \cite{le2013richness}, multiple
sequences from different genomes are analyzed at once, and typically
the total number of reads does not suffice for a full coverage of
all the genomes \cite{kang2015metabat,greenberg2019metagenomic}.\footnote{In metagenomics, even the identifiability of the sequences from the
reads is a challenging question \cite{herring2022probabilistic}. }

The reconstruction algorithm and analysis were extended in \cite{motahari2013information}
in two directions. First, the assumption on the input process generating
the sequence was generalized from a memoryless process to a Markov
process (assumed for simplicity, to be a first-order Markov process),
in which the second-order R\'{e}nyi entropy rate\footnote{To be formally defined later on.}
${\cal H}_{2}(\mathbf{X})$ replaces $H_{2}(P_{X})$ used in the memoryless
case, and was shown to be optimal \cite[Th. 6]{motahari2013information}
(ii) Noisy sequencing case of a memoryless sequence, in which the
clean fragments are corrupted by a memoryless channel. In this case,
$H_{2}(P_{X})$ is replaced by a information measure denoted $I^{*}$
\cite[Th. 10]{motahari2013information}, without a claim for optimality.
Evidently, while the \emph{scaling} in all cases is the same, it is
the exact minimal value of $\beta$ required for perfect reconstruction
that is affected by the statistics of the process generating the sequence
or the reads ($H_{2}(P_{X})$, ${\cal H}_{2}(\mathbf{X})$ or $I^{*}$).
As we show, a similar property holds in our setting as well. 

In general, a central part of the operation of an assembly algorithm
of a sequence form reads is merging of pairs of fragments based on
their overlap (either reads, or shorter fragments called $k$-mers).
Various papers have shown that the ability to do so successfully fundamentally
depends on the statistics of repeats in the long sequence. For example,
in \cite{bresler2013optimal}, a multi-bridging algorithm based on
$k$-mers was proposed, and conditions on $\beta$ were derived based
on the appearance of repeats in the sequence. This idea was extended
to noisy reads in \cite{lam2014near}, and to an assembly algorithm
based on the reads themselves in \cite{shomorony2016information}.
Information-theoretic limits on assembly for the noisy case were developed
in \cite{motahari2013optimal} for independent and identically distributed
(i.i.d.) sequences and in \cite{shomorony2016fundamental} for arbitrary
sequences under an adversarial noise model, showing the effect of
noise on the assembly algorithm performance. In \cite{shomorony2021sketching},
the computational problem of overlap detection was addressed based
on sketches, which are compressed versions of the pair of reads. The
optimal number of bits required for reliable approximate overlap was
tightly bounded, where the approximation is under the mean squared
error loss function. In \cite{shomorony2021sketching}, it was \emph{assumed}
that the alignment parameter is larger than some $\theta_{0}\ell$
for some constant $\theta_{0}\in(0,1)$. Here, we show that sufficiently
short overlaps cannot be reliably detected. See also \cite{baharav2020spectral,kamath2020adaptive,shomorony2021sketching,sahlin2022strobealign}
and references therein for practical overlap detection algorithms. 

\subsection{Our Results}

We focus on the distilled problem of overlap detection between a \emph{single}
pair of reads, and analyze a detector that is required to decide on
the length of the overlap between them. We emphasize here the Bayesian
setting, because each of the possibilities have a substantially different
prior. Given a length $n$ sequence $X_{1}^{n}$ and two reads of
length $\ell=\beta\log n$ starting at random locations, the probability
that they overlap at a given length $T\in\{-[\ell-1]\}\cup[\ell]$
is $\frac{1}{n}$,\footnote{We distinguish between a positive and negative overlaps, to wit, between
the case that the first read appears before the second one and vice-versa. } whereas the probability that they do not overlap at all is $\frac{2\ell-1}{n}\approx\frac{2\beta\log n}{n}$,
an event which we denote by $T=0$. Therefore, for a detector $\hat{T}$,
we let the type-I error be the \emph{false-overlap-detection} error
event, whose probability is $p_{\text{type-I}}(\hat{T}):=\P[\hat{T}\neq0\mid T=0]$,
and the type-II error be the \emph{misdetection} \emph{or erroneous-overlap}
error event, whose probability is $p_{\text{type-II}}(\hat{T}):=\P[\hat{T}\neq T\mid T\neq0]$
. The total error probability is thus tightly upper bounded as 
\[
P_{\text{error}}(\hat{T})\lessapprox p_{\text{type-I}}(\hat{T})+\frac{2\beta\log n}{n}p_{\text{type-II}}(\hat{T}),
\]
which highlights the non-equal contribution of the two types of error
to the total error probability. 

Similarly to the reconstruction problem, we characterize the scaling
as well as the exact \emph{constant} of the error probability (see
\cite[Section I.C]{motahari2013information}). It is clear here that
the scaling of the optimal error probability is $P_{\text{error}}(\hat{T})=\Theta(\frac{\log n}{n})$,
because a trivial detector $\hat{T}_{0}\equiv0$, which always decides
``no-overlap'', has an error probability of $P_{\text{error}}(\hat{T}_{0})=\P[T\neq0]\sim2\beta\cdot\frac{\log n}{n}$.
Therefore, the main problem is to characterize the optimal constant
\[
\phi:=\min_{\hat{T}}\lim_{n\to\infty}\frac{P_{\text{error}}(\hat{T})}{\nicefrac{(\log n)}{n}},
\]
for different probabilistic models of the process generating the sequence.
That is, taking into account memory in the process generating the
sequence, which might lead to various fragment repeat properties,
as well as the noise in the reading mechanism of the fragments. For
tractability, we focus on two settings -- the \emph{noiseless setting}
and the \emph{noisy setting} -- for which the optimal Bayesian detector
can be easily derived and the error probability analysis is tractable. 

In the first setting, referred to for brevity as the \emph{noiseless
setting}, we assume that the reading process is noiseless, but the
sequence is generated by a stationary and ergodic process. We do not
assume that the process is memoryless or a Markov process of a finite
order. We show that under a mild condition on the strong mixing coefficient
of the process, as well as a mild condition on the recurrence probability
of the process (to be formally defined later on), it holds that 
\[
\phi=2\cdot\left[\beta\wedge\frac{1}{{\cal H}_{1}(\mathbf{X})}\right],
\]
where ${\cal H}_{1}(\mathbf{X})$ is the Shannon entropy rate of the
process. 

Interestingly, the constant is \emph{not} determined by the second-order
R\'{e}nyi entropy rate ${\cal H}_{2}(\mathbf{X})$, as might be hypothesized
from the analysis of sequence reconstruction discussed above. As we
show and discuss in length, the root of this is directly related to
the imbalance in the prior probability. Indeed, a simplified, yet
very useful, way of thinking on the alignment detection problem is
as a collection of $2\ell-1$ binary detection problems between $T=0$
and $T=t$, for some $t\in\{-[\ell-1]\}\cup[\ell]$, with a prior
probability on which detection problem is active. The testing problem
is more difficult for short alignments, and so we upper bound the
error probability by $1$ if the active detection problem is in one
of the most $2t^{*}$ difficult problems (those of shortest overlaps).
The error probability upper bound is thus at least the probability
that one of these difficult detection problems is active, which is
$\sim2t^{*}\cdot\frac{\log n}{n}$. The challenge in the proof is
to accurately identify that $t^{*}$ for which both the probability
of both false-overlap-detection, as well as the probability to err
from one detection problem to another (that is, a wrong alignment
value), is negligible compared to this probability. While this intuition
may appear to be just a bounding technique, it is actually also used
to obtain a matching lower bound. 

In a nutshell, the optimal Bayesian detector will detect an overlap
of some specific $T=t\neq0$ only if the likelihood for $T=t$ compared
to the likelihood for $T=0$ (no overlap) is larger than the ratio
between their inverse prior probabilities $\frac{\P[T=0]}{\P[T=t]}\sim\frac{1}{1/n}=n$.
However, for overlaps shorter than $t\sim\frac{\log n}{{\cal H}_{1}(\mathbf{X})}$,
it holds that the expectation of this log-likelihood-ratio is less
than $n$, and so, such overlaps are not likely to be detected by
the optimal detector. Consequently, whenever $1\leq|T|\leq\frac{\log n}{{\cal H}_{1}(\mathbf{X})}$
the conditional error probability is $1$. Evidently, when $\beta>\frac{1}{{\cal H}_{1}(\mathbf{X})}$
this is the dominating error event. The technical challenge in upper
bounding the error probability is to show that all other error events,
either type-I error event or other sub-events of the type-II error
event, have negligible $o(\frac{\log n}{n})$ probability compared
to the probability of this event, without making strong assumptions
on the process generating the sequence. 

An easy generalization of our upper bounds leads to a different scaling
of the contributions of the two type of errors. If we replace the
prior probability ratio $\frac{\P[T=0]}{\P[T=t]}\sim n$ by $n^{\mu}$
for $\mu>1$ then we decrease the type-I error probability to 
\[
p_{\text{type-I}}(\hat{T})=o\left(\frac{\log n}{n^{\mu}}\right),
\]
a faster decay rate, with power almost $\mu$, whereas overlaps $|T|\leq\frac{\mu\log n}{{\cal H}_{1}(\mathbf{X})}$
are not detected, and in turn, the contribution of the type-II error,
and so also the total error probability is increased by a factor of
$\mu$, it is 
\[
P_{\text{error}}(\hat{T})\sim2\mu\cdot\left[\beta\wedge\frac{1}{{\cal H}_{1}(\mathbf{X})}\right]\cdot\frac{\log n}{n}.
\]

In the second setting, referred to for brevity as the \emph{noisy
setting}, we assume that the process is generated by a stationary
memoryless process with p.m.f. $P_{X}$, and that the reads are obtained
by observing the fragments through a discrete memoryless channel $P_{Y\mid X}$
(independently for each read). Let $Y,\tilde{Y}$ be two independent
observations of the same $X\sim P_{X}$ through $P_{Y\mid X}$ so
that the Markov relation $Y-X-\tilde{Y}$ holds. Here we show that
\[
\phi=2\cdot\left[\beta\wedge\frac{1}{I(Y;\tilde{Y})}\right],
\]
where $I(Y;\tilde{Y})$ is the mutual information. Similarly to the
noiseless setting, this is related to the log-likelihood-ratio between
an overlap, characterized by the joint distribution $P_{Y\tilde{Y}}$
and no-overlap, characterized by the product distribution $P_{Y}\otimes P_{\tilde{Y}}$.
The expectation of this log-likelihood ratio under $P_{Y\tilde{Y}}$
is exactly $I(Y;\tilde{Y})$, which explains why it appears in the
bound. 

There are various technical challenges associated with the error probability
analysis. First, it is challenging to show that the type-I error probability
is upper bounded as $o(\frac{\log n}{n})$ (thus negligible). In the
noiseless setting, this requires a tight analysis of a term of the
form 
\[
\sum_{x_{1}^{t}\in{\cal X}^{\otimes t}}P_{X_{1}^{t}}^{2}\left(x_{1}^{t}\right)\I\left[\frac{1}{P_{X_{1}^{t}}(x_{1}^{t})}\geq n\right]
\]
which is a partial sum of the sum involved in the second-order R\'{e}nyi
entropy of $P_{X_{1}^{t}}$, the p.m.f. of $X_{1}^{t}:=(X_{1},\ldots,X_{t})$.
Interestingly, an analysis that is based on trivially upper bounding
the indicator function by $1$, results an sub-optimal upper bound
on the constant of total error probability of $\phi\leq\beta\wedge\frac{1}{{\cal H}_{2}(\mathbf{X})}$.
In the noisy setting, the type-I error probability pertains to the
probability that a sum of independent random variables (r.v.'s) crosses
a threshold. Here, an application of the Chernoff bound does not suffice,
despite the fact that it is exponentially tight. We thus turn to stronger
versions, similar to that of Bahadur--Rao \cite{Bahadur1960on} \cite[Th. 3.7.4.]{dembo2009large}
(see also \cite{tan2014asymptotic,merhav2024toolbox}), utilizing
their accurate pre-exponent. Second, there are challenge in establishing
that the probability of an erroneous overlap is negligible. This is
because conditioned on the event that $T=t>0$, the analysis of the
probability of the event that the detector decides $\hat{T}\neq t$
involves analyzing sum of possibly \emph{dependent} r.v.'s. The proofs
propose methods to circumvent the need to analyze these sums of dependent
r.v.'s.

\subsection{Paper Outline}

The outline of the rest of the paper is as follows. In Sec. \ref{sec:Problem-Formulation}
we state notation conventions and formulate the problem. In Sec. \ref{sec:Alignment-Detection-for noiseless}
we address noiseless reading kernels, and state a tight asymptotic
expression for the Bayesian error probability. In Sec. \ref{sec:Alignment-Detection-for noisy}
we similarly address noisy reading kernels, yet restrict to memoryless
processes. In Sec. \ref{sec:Conclusion} we conclude the paper. In
the appendices, we provide the proofs for the statements made throughout
the paper. 

\section{Problem Formulation \label{sec:Problem-Formulation}}

\subsection{Notation conventions}

We denote r.v.'s by capital letters, specific values they may take
by the corresponding lower case letters, and their alphabets by calligraphic
letters. We superscript random sequences and their realizations by
their dimension. For example, the random sequence $X_{1}^{n}=(X_{1},\ldots,X_{n})$,
where $n$ is a positive integer, may take a specific vector value
$x^{n}=(x_{1},\ldots,x_{n})\in{\cal X}^{\otimes n}$, the $n$th order
Cartesian power of ${\cal X}$, which is the alphabet of each component
of this vector. We denote a two-sided random process by $\mathbf{X}:=X_{-\infty}^{\infty}=\{X_{i}\}_{i\in\mathbb{Z}}$.
We denote the Cartesian product of ${\cal X}$ and ${\cal Y}$ (finite
alphabets) by ${\cal X}\otimes{\cal Y}$, the p.m.f. of $X$ by $P_{X}$,
the p.m.f. of $X_{i}^{j}$ by $P_{X_{i}^{j}}$, and by the joint p.m.f.
of $(X,Y)$ by $P_{XY}$, and by $P_{Y\mid X}$ the conditional p.m.f.
of $Y$ conditioned on $X$. We denote the complement of an event
${\cal A}$ (or a set) by ${\cal A}^{c}$. 

We denote the order-$\alpha$ R\'{e}nyi entropy of a p.m.f. $P_{X}$
by 
\begin{align}
H_{\alpha}(P_{X}) & :=\begin{cases}
\frac{1}{1-\alpha}\log\sum_{x\in{\cal X}}P_{X}^{\alpha}(x), & \alpha\neq1\\
-\sum_{x\in{\cal X}}P_{X}(x)\log P_{X}(x), & \alpha=1
\end{cases},\label{eq: Renyi entropy}
\end{align}
where we notice that $H_{1}(P_{X})$ is the Shannon entropy. We will
also use the minus-infinity-order R\'{e}nyi entropy, defined as 
\[
H_{-\infty}:=\lim_{\alpha\downarrow-\infty}H_{\alpha}(P_{X})=-\log\min_{x\in\supp(P_{X})}P_{X}(x),
\]
which is determined by the minimal probability in the support of $P_{X}$.
For a given random process $\mathbf{X}$, with marginal p.m.f.'s $X_{-t}^{t}\sim P_{X_{-t}^{t}}$,
we let 
\[
{\cal \overline{H}}_{\alpha}(\mathbf{X}):=\limsup_{t\to\infty}\frac{1}{2t+1}H_{\alpha}(P_{X_{-t}^{t}})
\]
and 
\[
{\cal \underline{H}}_{\alpha}(\mathbf{X}):=\liminf_{t\to\infty}\frac{1}{2t+1}H_{\alpha}(P_{X_{-t}^{t}})
\]
be the supremum (resp. infimum) order-$\alpha$ R\'{e}nyi entropy
rate. If the limit exists and ${\cal \underline{H}}_{\alpha}(\mathbf{X})={\cal \overline{H}}_{\alpha}(\mathbf{X})$
then the common limit ${\cal H}_{\alpha}(\mathbf{X})$ is the order
$\alpha$ R\'{e}nyi entropy rate. For the sake of simplicity, we
assume henceforth that all considered entropy rates exists. Otherwise,
one can replace the entropy rate ${\cal H}_{\alpha}(\mathbf{X})$
with either ${\cal \underline{H}}_{\alpha}(\mathbf{X})$ or ${\cal \overline{H}}_{\alpha}(\mathbf{X})$
(depending on the context).\footnote{An exception is Theorem \ref{thm: exponent lower bound noiseless},
in which the existence of ${\cal H}_{1}(\mathbf{X})$ is required
to be assumed.} 

We denote the R\'{e}nyi divergence pair of probability distributions
$P_{X}\ll Q_{X}$ over ${\cal X}$ by 
\begin{equation}
D_{\alpha}(P_{X}\mid\mid Q_{X}):=\begin{cases}
\frac{1}{\alpha-1}\log\left(\sum_{x\in{\cal X}}P_{X}^{\alpha}(x)Q_{X}^{1-\alpha}(x)\right), & \alpha\neq1\\
\sum_{x\in{\cal X}}P_{X}(x)\log\frac{P_{X}(x)}{Q_{X}(x)}, & \alpha=1
\end{cases}\label{eq: Renyi divergence}
\end{equation}
where we notice that $D_{1}(P_{X}\mid\mid Q_{X})$ is the Kullback--Leibler
divergence. We denote the total variation distance between $P_{X}$
and $Q_{X}$ by 
\[
\|P-Q\|_{\text{TV}}:=\frac{1}{2}\sum_{x\in{\cal X}}\left|P(x)-Q(x)\right|.
\]
For an integer $n\in\mathbb{N}_{+}$, we denote $[n]:=\{1,2,\ldots,n\}$
and for brevity, we also denote $-[n]:=\{-1,-2,\ldots,-n\}$ . We
denote $a\vee b:=\max\{a,b\}$ and $a\wedge b:=\min\{a,b\}$ for $a,b\in\mathbb{R}$.
Unless otherwise stated, we take logarithms to the base $|{\cal X}|$,
and we also denote $\exp_{a}(t):=a^{t}$. We will use the standard
Bachmann--Landau asymptotic notation, such as $O(\frac{\log n}{n})$,
$o_{n}(1)$ (where here the subscript indicates that this term is
vanishing when $n\to\infty$), and $\omega(\log n)$, and also use
$a_{n}\sim b_{n}$ to denote that $\frac{a_{n}}{b_{n}}\to1$ as $n\to\infty$.
We will use $\equiv$ to denote equivalence, mainly to simplify notation.
For example, we will denote an element from a sequence $\{\tau(n)\}_{n\in\mathbb{N}}$
by $\tau$, and indicate this by $\tau\equiv\tau(n)$. 

\subsection{Problem Setting}

We next model the process of sampling a pair of short substrings (sub-sequences
of consecutive symbols) called \emph{reads}, from a sequence, and
the possible overlap between them. Let ${\cal X}$ be a finite alphabet,
and let $\mathbf{X}:=\{X_{i}\}_{i\in\mathbb{Z}}$ be a random process
defined over ${\cal X}$. Let $\ell\equiv\ell(n)$ denote the length
of each read, where $n$ determines the length of the available sequence
$X_{1}^{n}$. We will assume that $\ell\ll n$, and concretely, we
will focus on the regime $\ell=\Theta(\log n)$. For a reason that
will be clarified next, we will slightly extend the available sequence
to $X_{2-\ell}^{n+2\ell-2}$, whose length is $n+3\ell-3=[1+o_{n}(1)]\cdot n$
in this regime, and thus is essentially $n$. 

\begin{figure}
\begin{centering}
\includegraphics[scale=1.25]{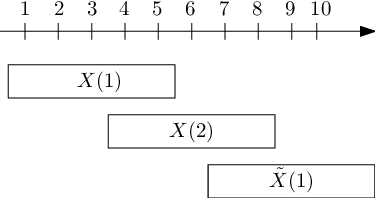}
\par\end{centering}
\caption{Reads of length $\ell=5$. $X(1)$ and $X(2)$ have a positive overlap
of $T=2$, whereas $\tilde{X}(1)$ and $X(2)$ have a negative overlap
$T=-2$. \label{fig: positive negative overlap}}
\end{figure}

Let $\{I(j)\}_{j\in[2]}$ be two starting indices, and let the \emph{noiseless
reads} be the substrings 
\[
X_{1}^{\ell}(j)=X_{I(j)}^{I(j)+\ell-1}
\]
for $j\in[2]$. The pair of reads \emph{overlap} if the indices of
the reads have at least a single common index, that is, if $|I(1)-I(2)|<\ell$.
We will be interested in the\emph{ overlap} between the reads. To
this end, we define the \emph{(signed) overlap} as
\[
T:=\begin{cases}
0\vee[\ell-(I(2)-I(1))], & I(2)\geq I(1)\\
-\left\{ 0\vee[\ell-(I(1)-I(2))]\right\} , & I(2)<I(1)
\end{cases}.
\]
Thus, $T=0$ if the reads do not overlap, the overlap is positive
$T\in[\ell]$ if the reads overlap and the last $T$ indices of $X_{1}^{\ell}(1)$
in the sequence are the same as the first $T$ indices of $X_{1}^{\ell}(2)$,
and the overlap is negative $T\in-[\ell-1]$ in the opposite case.\footnote{Note that when $T=\ell$, there is no distinction between the read
with smaller index and the read with larger index. We arbitrarily
attribute this case to a positive overlap of $\ell$. } See Fig. \ref{fig: positive negative overlap} for an illustration.

The direct way of modeling the random sampling process of the reads
is to independently choose two random starting indices $I(j)\sim\text{Uniform}[n]$.
As is, this read-sampling process results in an edge effect in the
distribution of the amount of overlap, occurring when $I(1)$ is sampled
close to the edges, i.e., $I(1)\in[\ell]$ or $I(1)\in[n]\backslash[n-\ell-1]$
(or, similarly, $I(2)$). Since we are interested in $n\gg\ell$,
these edge effects are negligible, and so we slightly modify the sampling
process in order to make later derivations less cumbersome. Concretely,
we assume that $I(1)\sim\text{Uniform}[n]$ (or even arbitrary chosen
in $[n]$), and that 
\begin{equation}
I(2)\mid I(1)\sim\begin{cases}
I(1)+\text{Uniform}\{-\ell+1,-\ell+2,\ldots,n-\ell\}, & 1\leq I(1)\leq\ell-1\\
\text{Uniform}[n], & \ell\leq I(1)\leq n-\ell+1\\
I(1)+\text{Uniform}\{\ell-n,\ell-n+1,\ldots,\ell-1\}, & n-\ell+2\le I(1)\leq n
\end{cases}\label{eq: I(2) conditioned on I(1)}
\end{equation}
So, the distribution of $I(2)$ is always uniform over a window of
length $n$ regardless of $I(1)$, the reads are always sampled from
the extended available sequence $X_{2-\ell}^{n+2\ell-2}$, and the
sampling window location of $I(2)$ only slightly changes at the edges,
whose total size is $2\ell\ll n$. With this slight simplification,
it holds that $T$ is an r.v. whose p.m.f. is 
\begin{align}
P_{T}(t) & =\begin{cases}
1-\frac{(2\ell-1)}{n}, & t=0\\
\frac{1}{n}, & t\in-[\ell-1]\cup[\ell]
\end{cases}\nonumber \\
 & =\begin{cases}
\frac{n_{\ell}}{n}, & t=0\\
\frac{1}{n}, & t\in-[\ell-1]\cup[\ell]
\end{cases}\label{eq: DNA shotgun sequencing overlap}
\end{align}
where, for brevity, we denoted $n_{\ell}:=n-(2\ell-1)$ and note that
$n_{\ell}\sim n$ whenever $\ell=o(n)$. 

Let ${\cal Y}$ be another finite alphabet, and let $P_{Y_{1}^{\ell}\mid X_{1}^{\ell}}$
be a Markov kernel from ${\cal X}^{\otimes\ell}$ to ${\cal Y}^{\otimes\ell}$.
For $j\in[2]$, let 
\[
Y_{1}^{\ell}(j)\sim P_{Y_{1}^{\ell}\mid X_{1}^{\ell}}\left(\cdot\mid x_{1}^{\ell}(j)\right)
\]
be the \emph{noisy read} of $x_{1}^{\ell}(j)$, where conditioned
on $\{X_{1}^{\ell}(j)=x_{1}^{\ell}(j)\}_{j\in[2]}$ it holds that
$\{Y_{1}^{\ell}(j)\}_{j\in[2]}$ are pairwise independent. The goal
is to detect the \emph{signed overlap} $T$ based on these noisy reads.
Letting ${\cal T}:=\{-[\ell-1]\}\cup\{0\}\cup[\ell]$, an overlap
detector based on a pair of noisy reads is a mapping $\hat{T}\colon{\cal Y}^{\otimes\ell}\times{\cal Y}^{\otimes\ell}\to{\cal T}$.
The Bayesian error probability of a detector $\hat{T}$ is given by
\begin{equation}
P_{\text{error}}(\hat{T}):=\P\left[\hat{T}\left(Y_{1}^{\ell}(1),Y_{1}^{\ell}(2)\right)\neq T\right].\label{eq: error probability}
\end{equation}

In many cases, there is a conceptual distinction between detection
of $\hat{T}=0$ -- no overlap, and $\hat{T}\neq0$ -- a discovery
of an overlap. Thus, with a slight deviation from the standard statistical
terminology, we define the type-I error of a detector $\hat{T}$ as
the event that $T=0$, but $\hat{T}$ is detected as a discovery overlap,
and denote its probability by 
\[
p_{\text{type-I}}(\hat{T}):=\P\left[\hat{T}\neq0\mid T=0\right].
\]
We then also define the type-II error as the event that $T=t$ for
some non-zero overlap $t\neq0$, and it is either misdetected or erroneously
detected, and denote its probability by 
\[
p_{\text{type-II}}(\hat{T};t):=\P\left[\hat{T}\neq t\mid T=t\right],
\]
where we also denote, again with a slight abuse of notation, 
\[
p_{\text{type-II}}(\hat{T})=\sum_{t\in\{-[\ell-1]\}\cup[\ell]}\P[T=t\mid T\neq0]\cdot p_{\text{type-II}}(\hat{T};t).
\]
The Bayesian error probability is then given by 
\begin{equation}
P_{\text{error}}(\hat{T})=P_{T}(0)\cdot p_{\text{type-I}}(\hat{T})+\P[T\neq0]\cdot\sum_{t\in\{-[\ell-1]\}\cup[\ell]}\P[T=t\mid T\neq0]\cdot p_{\text{type-II}}(\hat{T};t).\label{eq: error decomposition model}
\end{equation}
For the considered overlap p.m.f. $P_{T}$ in (\ref{eq: DNA shotgun sequencing overlap}),
the Bayesian error probability is (tightly) upper bounded as
\begin{equation}
P_{\text{error}}(\hat{T})\sim p_{\text{type-I}}(\hat{T})+\frac{2\ell}{n}\cdot p_{\text{type-II}}(\hat{T}).\label{eq: error decomposition model tight upper bound}
\end{equation}

We will be interested in the minimal error probability 
\[
P_{\text{error}}^{*}\equiv P_{\text{error}}^{*}\left(\beta,P_{\boldsymbol{X}},P_{Y_{1}^{\ell}\mid X_{1}^{\ell}}\right):=\min_{\hat{T}}P_{\text{error}}(\hat{T}).
\]
As the error probability (\ref{eq: error probability}) is averaged
over the prior distribution, the optimal detector is the maximum \emph{a
posteriori} (MAP) detector. For a given realization of a pair of noisy
reads $y_{1}^{\ell}(1),y_{1}^{\ell}(2)$, it is given by the Bayes
rule as 
\begin{align}
\hat{T}_{\text{MAP}}\left(y_{1}^{\ell}(1),y_{1}^{\ell}(2)\right) & =\argmax_{t\in{\cal T}}P_{T\mid Y_{1}^{\ell}(1)Y_{1}^{\ell}(2)}\left(t\mid y_{1}^{\ell}(1),y_{1}^{\ell}(2)\right)\\
 & =\argmax_{t\in{\cal T}}P_{T}(t)\cdot P_{Y_{1}^{\ell}(1)Y_{1}^{\ell}(2)\mid T}\left(y_{1}^{\ell}(1),y_{1}^{\ell}(2)\mid T=t\right),
\end{align}
where we denote the likelihood by 
\[
f\left(y_{1}^{\ell}(1),y_{1}^{\ell}(2);t\right):=P_{Y_{1}^{\ell}(1)Y_{1}^{\ell}(2)\mid T}\left(y_{1}^{\ell}(1),y_{1}^{\ell}(2)\mid T=t\right).
\]
For any given overlap $t\in[\ell]$, we may partition $Y_{1}^{\ell}(1)$
into two segments, given by $Y_{1}^{\ell-t}(1)$, which does not overlap
with $Y_{1}^{\ell}(2)$, and $Y_{\ell-t+1}^{\ell}(1)$, which overlaps
with $Y_{1}^{t}(2)$. Similarly, we also partition $Y_{1}^{\ell}(2)$
into $Y_{1}^{t}(2)$ which overlaps with $Y_{\ell-t+1}^{\ell}(1)$,
and $Y_{t+1}^{\ell}(2)$, which does not overlap with $Y_{1}^{\ell}(1)$.
The two reads may be analogously partitioned when $t\in-[\ell-1]$. 

We will assume throughout that the read length scales as $\ell=\beta\log n$,
where $\beta>0$ is a fixed parameter (in fact, the interesting regime
is in which $\beta>\log|{\cal X}|=1$). We will focus on two specific
settings:
\begin{defn}[Noiseless setting]
\label{def: Noiseless setting} $\mathbf{X}$ is a stationary and
ergodic process, and the reading kernel $P_{Y_{1}^{\ell}\mid X_{1}^{\ell}}$
is a noiseless channel, that is $Y_{1}^{\ell}=X_{1}^{\ell}$, with
probability $1$. 
\end{defn}
For simplicity, when analyzing this setting, we will use $\{X_{1}^{\ell}(j)\}_{j\in[2]}$
to denote the reads. 

\begin{defn}[Noisy setting]
\label{def: Noisy setting} $\mathbf{X}$ is stationary and memoryless,
that is, $P_{X^{n}}(x^{n})=\prod_{i=1}^{n}P_{X}(x_{i})$ for any $n\in\mathbb{N}_{+}$,
and the reading channel is also stationary and memoryless, that is,
$P_{Y_{1}^{\ell}\mid X_{1}^{\ell}}=\prod_{i=1}^{\ell}P_{Y\mid X}(y_{i}\mid x_{i})$
for all $\ell\in\mathbb{N}_{+}$. 
\end{defn}
Finally, we abbreviate $P_{\text{error}}^{*}\equiv P_{\text{error}}(\hat{T}_{\text{MAP}})$.
Our goal is to determine 
\[
\phi\left(\beta,P_{\boldsymbol{X}},P_{Y_{1}^{\ell}\mid X_{1}^{\ell}}\right):=\lim_{n\to\infty}\frac{n}{\log n}\cdot P_{\text{error}}^{*}
\]
 for both the noiseless and the noisy settings. 

\section{Overlap Detection for Noiseless Reading Kernels \label{sec:Alignment-Detection-for noiseless}}

In this section, we address the noiseless setting of Definition \ref{def: Noiseless setting}.

\subsection{The MAP Detection Rule \label{subsec:The-MAP-Detection noiseless}}

Our first step in deriving the MAP detection rule is to obtain an
expression for the likelihood function.
\begin{lem}
\label{lem: likelihood for noisless reads}Consider the noiseless
setting (Def. \ref{def: Noiseless setting}). Let 
\[
\tilde{\Gamma}_{+}(t;x_{1}^{\ell-t}):=\frac{1}{P_{X_{1}^{t}\mid X_{-(\ell-t-1)}^{0}}(x_{1}^{t}(2)\mid x_{1}^{\ell-t})}
\]
and let
\[
\tilde{\Gamma}_{-}(t;x_{1}^{\ell-t}):=\frac{1}{P_{X_{1}^{t}\mid X_{-(\ell-t-1)}^{0}}(x_{1}^{t}(1)\mid x_{1}^{\ell-t})}
\]
Then, the likelihood of overlap $t\in[\ell]$ given reads $x_{1}^{\ell}(1)$
and $x_{1}^{\ell}(2)$ is
\begin{equation}
f\left(x_{1}^{\ell}(1),x_{1}^{\ell}(2);t\right)\propto\I\left[x_{\ell-t+1}^{\ell}(1)=x_{1}^{t}(2)\right]\cdot P_{X_{1}^{\ell}\mid X_{-(\ell-t-1)}^{0}}\left(x_{1}^{\ell}(2)\mid x_{1}^{\ell-t}(1)\right)\cdot\tilde{\Gamma}_{+}(t;x_{1}^{\ell-t}(1)),\label{eq: likelihood for noisless reads positive}
\end{equation}
and for $t\in-[\ell-1]$ 
\begin{equation}
f\left(x_{1}^{\ell}(1),x_{1}^{\ell}(2);t\right)\propto\I\left[x_{\ell-t+1}^{\ell}(2)=x_{1}^{t}(1)\right]\cdot P_{X_{1}^{\ell}\mid X_{-(\ell-t-1)}^{0}}\left(x_{1}^{\ell}(1)\mid x_{1}^{\ell-t}(2)\right)\cdot\tilde{\Gamma}_{-}(t;x_{1}^{\ell-t}(2))\cdot,\label{eq: likelihood for noisless reads negative}
\end{equation}
and where we set $\Gamma_{\pm}(0)=1$. 
\end{lem}
\begin{IEEEproof}
It holds that 
\begin{align}
 & f(x_{1}^{\ell}(1),x_{1}^{\ell}(2);t)\nonumber \\
 & =P_{X_{1}^{\ell}}\left(x_{1}^{\ell}(1)\right)\cdot\I\left[x_{\ell-t+1}^{\ell}(1)=x_{1}^{t}(2)\right]P_{X_{t+1}^{\ell}\mid X_{-(\ell-t-1)}^{t}}\left(x_{t+1}^{\ell}(2)\mid(x_{1}^{t}(2),x_{1}^{\ell-t}(1))\right)\\
 & \trre[=,a]P_{X_{1}^{\ell}}\left(x_{1}^{\ell}(1)\right)\I\left[x_{\ell-t+1}^{\ell}(1)=x_{1}^{t}(2)\right]\cdot\frac{P_{X_{1}^{\ell}\mid X_{-(\ell-t-1)}^{0}}\left(x_{1}^{\ell}(2)\mid x_{1}^{\ell-t}(1)\right)}{P_{X_{1}^{t}\mid X_{-(\ell-t-1)}^{0}}(x_{1}^{t}(2)\mid x_{1}^{\ell-t}(1))}\\
 & \trre[=,b]\I\left[x_{\ell-t+1}^{\ell}(1)=x_{1}^{t}(2)\right]\cdot P_{X_{1}^{\ell}}\left(x_{1}^{\ell}(1)\right)\cdot P_{X_{1}^{\ell}\mid X_{-(\ell-t-1)}^{0}}\left(x_{1}^{\ell}(2)\mid x_{1}^{\ell-t}(1)\right)\cdot\tilde{\Gamma}_{+}(t;x_{1}^{\ell-t}(1)),
\end{align}
where $(a)$ follows from Bayes rule, and in $(b)$ we have used the
notation for $\tilde{\Gamma}_{+}(t;x_{1}^{\ell-t})$. This shows (\ref{eq: likelihood for noisless reads positive}),
and (\ref{eq: likelihood for noisless reads negative}) follows by
analogous arguments. 
\end{IEEEproof}
While the likelihood function appears to be rather cumbersome, and
leads to a complicated MAP rule, it is greatly simplified when $\mathbf{X}$
is a memoryless process. Moreover, it turns out that the MAP rule
in the memoryless case (in the form we state it) is a good approximation
in the case that the process $\mathbf{X}$ has memory, yet ergodic.
Specifically, in the memoryless case, the MAP detector takes the following
simple form:
\begin{prop}
\label{prop: Bayes estimator noiseless memoryless}Consider the noiseless
setting (Def. \ref{def: Noiseless setting}), and further assume that
$\boldsymbol{X}$ is a memoryless process. Let the maximum overlap
with strictly positive (resp., negative) posterior be
\begin{align}
t_{+} & :=\max_{t\in[\ell]}\I\left[x_{\ell-t+1}^{\ell}(1)=x_{1}^{t}(2)\right]\\
t_{-} & :=\max_{t\in[\ell-1]}\I\left[x_{\ell-t+1}^{\ell}(2)=x_{1}^{t}(1)\right]
\end{align}
and let
\begin{align}
\Gamma_{+}(t) & :=\frac{1}{P_{X_{1}^{t}}\left(x_{1}^{t}(2)\right)}\label{eq: Gamma plus memoryless noiseless}\\
\Gamma_{-}(t) & :=\frac{1}{P_{X_{1}^{t}}\left(x_{1}^{t}(1)\right)}.
\end{align}
Then,
\begin{equation}
\hat{T}_{\text{\emph{MAP}}}\left(x_{1}^{\ell}(1),x_{1}^{\ell}(2)\right)=\begin{cases}
t_{+}, & \Gamma_{+}(t_{+})\geq[\Gamma_{-}(t_{-})\vee n_{\ell}]\\
-t_{-}, & \Gamma_{-}(t_{-})\geq[\Gamma_{+}(t_{+})\vee n_{\ell}]\\
0, & \text{\emph{otherwise}}
\end{cases}.\label{eq: optimal detection rule noiseless memoryless}
\end{equation}
\end{prop}
\begin{IEEEproof}
The Bayesian detector compares the posterior probabilities among the
different possibilities $t\in-[\ell-1]$, $t=0$ and $t\in[\ell]$.
First, let us compare the posterior probabilities only among overlaps
in $t\in[\ell]$. Using (\ref{eq: likelihood for noisless reads positive})
from Lemma \ref{lem: likelihood for noisless reads}, it holds for
$t\in[\ell]$ under the assumption that $\boldsymbol{X}$ is a memoryless
process that 
\begin{align}
 & \argmax_{t\in[\ell]}f(x_{1}^{\ell}(1),x_{1}^{\ell}(2);t)\nonumber \\
 & =\argmax_{t\in[\ell]}\I\left[x_{\ell-t+1}^{\ell}(1)=x_{1}^{t}(2)\right]\cdot P_{X_{1}^{\ell}}\left(x_{1}^{\ell}(1)\right)\cdot P_{X_{1}^{\ell}}\left(x_{1}^{\ell}(2)\right)\cdot\Gamma_{+}(t),\label{eq: likelihood noiseless positive overlap}
\end{align}
where $\Gamma_{+}(t)$ is as defined in (\ref{eq: Gamma plus memoryless noiseless}).
Since the prior distribution $P_{T}(t)=\frac{1}{n}$ for all $t\in[\ell]$,
and as $\Gamma_{+}(t)$ is a monotonic non-decreasing function of
$t\in[\ell]$ we deduce that if $\hat{T}_{\text{MAP}}\in[\ell]$ then
$\hat{T}_{\text{MAP}}=t_{+}$. Analogously, if $\hat{T}_{\text{MAP}}\in-[\ell-1]$
then $\hat{T}_{\text{MAP}}=t_{-}$. The decision between $\hat{T}_{\text{MAP}}=0$,
$\hat{T}_{\text{MAP}}=t_{+}$ and $\hat{T}_{\text{MAP}}=t_{-}$ is
again according to the maximal posterior probability, which is proportional
to $P_{T}(t)\cdot f(x^{\ell}(1),x^{\ell}(2);t)$. As the likelihood
for $t=0$ is 
\begin{equation}
f\left(x_{1}^{\ell}(1),x_{1}^{\ell}(2);0\right)=P_{X_{1}^{\ell}}\left(x_{1}^{\ell}(1)\right)\cdot P_{X_{1}^{\ell}}\left(x_{1}^{\ell}(2)\right),\label{eq: likelihood noiseless zero overlap}
\end{equation}
and as the priors are $P_{T}(0)=\frac{n_{\ell}}{n}$ and $P_{T}(t)=\frac{1}{n}$
for $t\in-[\ell-1]\cup[\ell]$, utilizing (\ref{eq: likelihood noiseless positive overlap})
and (\ref{eq: likelihood noiseless zero overlap}) implies the stated
claim. 
\end{IEEEproof}
In words, the MAP detector in the memoryless case finds the maximal
positive overlap and the maximal negative overlap, and chooses between
them based on the maximal value of the corresponding inverse sequence
probability, computed at the potential overlapping segments. If this
value crosses the threshold $n_{\ell}$ then an overlap is detected.
Otherwise, the detector declares that there is no overlap. 

Comparing the likelihood in the memoryless case (\ref{eq: likelihood noiseless positive overlap})
and the case with memory (\ref{eq: likelihood for noisless reads positive}),
we see that $P_{X_{1}^{\ell}}(x_{1}^{\ell}(2))$ is replaced by $P_{X_{1}^{\ell}\mid X_{-(\ell-t-1)}^{0}}(x_{1}^{\ell}(2)\mid x_{1}^{\ell-t}(1))$,
and that $\Gamma_{+}(t)$ is replaced by $\tilde{\Gamma}_{+}(t;x_{1}^{\ell-t}(1))$.
To wit, both terms now have a dependency on past symbols $X_{-(\ell-t-1)}^{0}=x_{1}^{\ell-t}(1)$,
assuming $I(1)=1$ without loss of generality (w.l.o.g.). Moreover,
the time span of this past symbols depends on $t$, and so it cannot
be omitted from the likelihood, unlike the first term in the memoryless
case $P_{X_{1}^{\ell}}(x_{1}^{\ell}(2))$. An analogous expression
can be derived for $t\in-[\ell-1]$, which introduces an analogous
situation. Similarly, the expression for the likelihood condition
on $t=0$ is invalid, in general, since due to the memory in the process,
the reads $X_{1}^{\ell}(1)$ and $X_{1}^{l}(2)$ may be dependent
even if they do not overlap. Nonetheless, we argue that under mild
conditions, the MAP detector (\ref{eq: optimal detection rule noiseless memoryless})
derived in Prop. \ref{prop: Bayes estimator noiseless memoryless}
for the memoryless case is a viable option for the case with memory,
as it is a sufficiently good approximation to it. We will fully justify
this argument in what follows by showing that the detector  (\ref{eq: optimal detection rule noiseless memoryless})
of Prop. \ref{prop: Bayes estimator noiseless memoryless} achieves
asymptotically optimal error probability, as the exact MAP detector.
To make this notion precise, let us recall the following definition
mixing of a random process: 
\begin{defn}[Strong mixing coefficient]
\label{def: strong mixing coefficient}Let $\mathbf{X}:=\{X_{i}\}_{i\in\mathbb{Z}}$
be a stationary and ergodic random process. For a given delay $s\in\mathbb{N}_{+}$,
the \emph{strong mixing coefficient} is
\begin{equation}
d(s)\equiv d(s;\mathbf{X}):=\sup\left\{ \left|\P[{\cal F}\mid{\cal E}]-\P[{\cal F}]\right|\colon i\in\mathbb{Z},\;{\cal E}\in\sigma\left(X_{-\infty}^{i}\right),\;{\cal F}\in\sigma\left(X_{i+s}^{\infty}\right)\right\} ,\label{eq: strong mixing coefficient}
\end{equation}
where $\sigma(X_{i}^{j})$ is the sigma-algebra generated by $X_{i}^{j}$.
The process $\mathbf{X}$ is \emph{strongly mixing} if $d(s)\to0$
as $s\to\infty$. In this case, since we can always replace $d(s)$
with $\tilde{d}(s):=\sup_{s'\geq s}d(s)$, we may assume, w.l.o.g.,
that $d(s)$ is monotonically non-increasing. 
\end{defn}
Hence, for strongly mixing processes, we expect that the exact likelihood
for the case with memory is well approximated by the likelihood in
Prop. \ref{prop: Bayes estimator noiseless memoryless}. Indeed, for
$T=t\in[\ell]$ the conditioning on past symbols $X_{-(\ell-t-1)}^{0}=x_{1}^{\ell-t}(1)$
will only affect a relatively small number of symbols at the beginning
of the read, and so 
\[
f\left(x_{1}^{\ell}(1),x_{1}^{\ell}(2);t\right)\approx\I\left[x_{\ell-t+1}^{\ell}(1)=x_{1}^{t}(2)\right]\cdot\frac{P_{X_{1}^{\ell}}\left(x_{1}^{\ell}(2)\right)}{P_{X_{1}^{t}}(x_{1}^{t}(2))},
\]
as the likelihood in Prop. \ref{prop: Bayes estimator noiseless memoryless}.
Similarly, for $T=0$, since the reads length is $\ell=\Theta(\log n)$
and the length of the entire sequence is $n$, under mild assumption
on the mixing of process they will be approximately independent for
most instances of $I(1)$ and $I(2)$. Then, we may expect that 
\begin{equation}
f\left(x_{1}^{\ell}(1),x_{1}^{\ell}(2);0\right)\approx P_{X_{1}^{\ell}}\left(x_{1}^{\ell}(1)\right)\cdot P_{X_{1}^{\ell}}\left(x_{1}^{\ell}(2)\right),\label{eq: approximate likelihood for T=00003D0 when the process has memory}
\end{equation}
which is again, similar to the likelihood in Prop. \ref{prop: Bayes estimator noiseless memoryless}.
At this point, we do not get into precise conditions on $d(s)$ for
sufficient approximate independence to holds,\footnote{See Lemma \ref{lem: asymptotic independence condition on T=00003D0}
and its proof in what follows for the exact likelihood for $t=0$,
and a rigorous quantitative statement of the approximation (\ref{eq: approximate likelihood for T=00003D0 when the process has memory}).} but only say that following this line of thought, it makes sense
to use the detector (\ref{eq: optimal detection rule noiseless memoryless})
of Prop. \ref{prop: Bayes estimator noiseless memoryless} even when
the source has memory (note that $\Gamma_{\pm}(t)$ are well defined
for this case too). In this case, the error probability of this sub-optimal
detector is an upper bound on the Bayesian error probability. 

For the detector (\ref{eq: optimal detection rule noiseless memoryless})
of Prop. \ref{prop: Bayes estimator noiseless memoryless}, there
exists a minimal absolute overlap value that can be detected. Specifically,
let the \emph{minimal detectable overlap} be defined as
\begin{align}
t_{{\scriptscriptstyle \text{MDO}}}\equiv t_{{\scriptscriptstyle \text{MDO}}}(n) & :=\max\left\{ t\in[\ell]\colon\max_{x_{1}^{t}\in\supp(P_{X_{1}^{t}})}\frac{1}{P_{X_{1}^{t}}(x_{1}^{t})}\leq n_{\ell}\right\} \\
 & =\max\left\{ t\in[\ell]\colon\log\frac{1}{\min_{x_{1}^{t}\in\supp(P_{X_{1}^{t}})}P_{X_{1}^{t}}(x_{1}^{t})}\leq\log n_{\ell}\right\} \\
 & =\max\left\{ t\in[\ell]\colon H_{-\infty}(P_{X_{1}^{t}})\leq\log n_{\ell}\right\} .\label{eq: minmal detectable overlap}
\end{align}
Prop. \ref{prop: Bayes estimator noiseless memoryless} implies that
$|t|\leq t_{{\scriptscriptstyle \text{MDO}}}$ will never be detected
by the detector (\ref{eq: optimal detection rule noiseless memoryless}).
It is readily verified that $t_{{\scriptscriptstyle \text{MDO}}}(n)$
is a monotonic non-decreasing sequence, which also satisfies $\lim_{n\to\infty}t_{{\scriptscriptstyle \text{MDO}}}(n)=\infty$
(assuming ${\cal H}_{-\infty}(\mathbf{X})<\infty$). Furthermore,
if the minus-infinity-order R\'{e}nyi entropy rate exists, then 
\[
t_{{\scriptscriptstyle \text{MDO}}}(n)\sim\frac{\log n_{\ell}}{\frac{1}{t_{{\scriptscriptstyle \text{MDO}}}(n)}H_{-\infty}(P_{X_{1}^{t_{{\scriptscriptstyle {\scriptscriptstyle \text{MDO}}}}(n)}})}\sim\frac{\log n}{{\cal H_{-\infty}}(\mathbf{X})},
\]
which scales linearly with the read length whenever $\ell=\Theta(\log n)$.
Nonetheless, in the \emph{typical} case, even longer overlaps are
not expected to be reliably detected. The scaling of $t_{{\scriptscriptstyle \text{MDO}}}(n)=\Theta(\log n)$
shows that we need to consider the typical value of $\frac{1}{t}\log\frac{1}{P_{X_{1}^{t}}(x_{1}^{t})}$,
and indeed, if we take the \emph{expected} value of $\frac{1}{t}\log\frac{1}{P_{X_{1}^{t}}(x_{1}^{t})}$
instead of its maximal value in (\ref{eq: minmal detectable overlap}),
it will tend to ${\cal H}_{1}(\mathbf{X})$ (whenever this entropy
rate exists). Following this intuition, we may expect that 
\[
t^{*}\equiv t^{*}(n)\sim\frac{\log n}{{\cal H}_{1}(\mathbf{X})}
\]
is a \emph{minimal reliably detectable overlap}.\footnote{This is indeed a longer overlap, since the standard entropy rate ${\cal H}_{1}(\mathbf{X})$
(assuming it exists) is smaller or equal to ${\cal H_{-\infty}}(\mathbf{X})$. } As we shall see in the next section, this intuition is correct.

\subsection{Error Probability Analysis \label{subsec: noiseless Error-Probability-Analysis}}

We next separately state an upper bound on the Bayesian error probability,
and then a matching lower bound. We begin with the upper bound:
\begin{thm}
\label{thm: exponent upper bound noiseless}Consider the noiseless
setting (Def. \ref{def: Noiseless setting}), and assume that the
following conditions hold: 
\begin{enumerate}
\item Finite minus-infinity-order R\'{e}nyi entropy rate: ${\cal H}_{-\infty}(\mathbf{X})<\infty.$
\item The process is moderately-fast strongly mixing: $\sum_{s=0}^{\infty}d(s;\mathbf{X})<\infty.$
\item The process is aperiodic:
\[
{\cal R}(\mathbf{X}):=\sup_{s\in\mathbb{N}_{+}}\P\left[X_{1}=X_{1+s}\right]<1.
\]
\end{enumerate}
Then, 
\begin{equation}
P_{\text{\emph{error}}}^{*}\leq2\left[1+o_{n}(1)\right]\cdot\left\{ \beta\wedge\frac{1}{{\cal H}_{1}(\mathbf{X})}\right\} \cdot\frac{\log n}{n}.\label{eq: upper bound on the error probability noiseless}
\end{equation}

\end{thm}
The proof of Prop. \ref{prop: Bayes estimator noiseless memoryless},
is deferred to Appendix \ref{sec:Proofs-noiseless}, as all the other
proofs in this section. We next discuss the result of Theorem \ref{thm: exponent upper bound noiseless}. 

\paragraph*{The dependence of the error probability on $\beta$}

The root of this dependence is simple to explain: A trivial detector,
which always decides that there is no overlap, will err only if $T\neq0$.
The probability of this event is $\frac{2\ell}{n}\sim\frac{2\beta\log n}{n}$,
and thus the error probability of the optimal MAP detector is less
than $\frac{2\beta\log n}{n}$ (asymptotically). This explains the
upper bound in (\ref{eq: upper bound on the error probability noiseless})
when $\beta\leq\frac{1}{{\cal H}_{1}(\mathbf{X})}$, and so we assume
in the rest of the discussion that $\beta>\frac{1}{{\cal H}_{1}(\mathbf{X})}$. 

\paragraph*{The dependence of the error on the entropy rate ${\cal H}_{1}(\mathbf{X})$}

This dependence is much more subtle. Let us assume for the sake of
this discussion that the source is memoryless. In this case, the MAP
detector is $\hat{T}$ given in (\ref{eq: optimal detection rule noiseless memoryless})
(Prop. \ref{prop: Bayes estimator noiseless memoryless}). On the
face of it, one might expect that the error probability is determined
by the second-order R\'{e}nyi entropy, also known as the \emph{collision
entropy}, which is given by 
\[
H_{2}(P_{X})=-\log\sum_{x\in{\cal X}}P_{X}^{2}(x).
\]
This quantity indeed appears in various previous works on reads alignment,
e.g., \cite{motahari2013information}. This is because whenever $T=0$,
the two reads do not overlap, and the probability that two independent
substrings of length $t$ exactly match and a false overlap is detected
is $|{\cal X}|^{-tH_{2}(P_{X_{1}})}$. Following this line of thought,
one might expect, that the Bayesian error probability is upper bounded
as
\[
P_{\text{error}}^{*}\leq\left[1+o_{n}(1)\right]\cdot\frac{2}{H_{2}(P_{X})}\cdot\frac{\log n}{n}.
\]
The fallacy of such an argument is that it ignores that an overlap
is detected only if the probability of the matching substring is sufficiently
small, as expressed in the condition 
\[
\Gamma_{+}(t)=\frac{1}{P_{X_{1}^{t}}\left(x_{1}^{t}(2)\right)}\geq n_{\ell}
\]
in the MAP detector (\ref{eq: optimal detection rule noiseless memoryless}).
The root of this condition is that the likelihood of the two reads
at the given (false) overlap should overcome the strong prior probability
that the two reads do not overlap at all. As we have discussed above,
the typical length of strings which satisfy this condition is the
minimal reliably detectable overlap given by $t^{*}\sim\frac{\log n}{H_{1}(P_{X})}$,
and this introduces dependence on $H_{1}(P_{X_{1}})$. Nonetheless,
from the discussion at this level it is not obvious which of the two
entropy values is the relevant one. Next, we will outline the proof
of Theorem \ref{thm: exponent upper bound noiseless}, and this will
clarify this issue. 

\paragraph*{The main ideas of the proof}

The proof of Theorem \ref{thm: exponent upper bound noiseless} is
fairly long and involved, so we next outline its steps. The proof
carefully analyzes the different types of error probability in the
decomposition (\ref{eq: error decomposition model tight upper bound})
for the detector (\ref{eq: optimal detection rule noiseless memoryless})
of Prop. \ref{prop: Bayes estimator noiseless memoryless}. In the
case of memory, this detector is sub-optimal, and its error probability
is thus an upper bound on the Bayesian error probability of the MAP
detector. 

Let us begin by considering $p_{\text{type-I}}(\hat{T})$, that is,
the error conditioned on $T=0$. Recall that the detector $\hat{T}$
never decides $0<|\hat{T}|<t_{{\scriptscriptstyle \text{MDO}}}$,
where $t_{{\scriptscriptstyle \text{MDO}}}\sim\frac{\log n}{{\cal H_{-\infty}}(\mathbf{X})}$,
because for $0<|T|<t_{{\scriptscriptstyle \text{MDO}}}$, the posterior
evidence of any substring common to both reads is not strong enough
compared to the prior knowledge that there is no overlap between the
reads. This has two implications. First, the error probability is
at least the probability that $0<|T|<t_{{\scriptscriptstyle \text{MDO}}}$,
which is $\frac{t_{{\scriptscriptstyle \text{MDO}}}(n)}{n}=\Omega(\frac{\log n}{n})$.
Second, the detector will never err from $T=0$ to $0<|\hat{T}|<t_{{\scriptscriptstyle \text{MDO}}}$,
only to longer overlaps (either positive or negative). As it turns
out, however, this property alone does not suffice to tightly bound
$p_{\text{type-I}}(\hat{T})$. So, let us generalize this observation
and consider, instead of $t_{{\scriptscriptstyle \text{MDO}}}\sim\frac{\log n}{{\cal H_{-\infty}}(\mathbf{X})}$,
a general effective minimal overlap value, given by $t(h)\sim\frac{\log n}{h}$
where $h<{\cal H_{-\infty}}(\mathbf{X})$, and aim to find the minimal
$h$ which leads to the tightest upper bound on $p_{\text{type-I}}(\hat{T})$.
Conditioned on $T=0$, according to the detector of Prop. \ref{prop: Bayes estimator noiseless memoryless},
an error will occur whenever two non-overlapping reads of length longer
than $t(h)$ match, and their probability is smaller than $1/n_{\ell}$.
We then take a union bound over all possible lengths $t(h)\leq t\leq\ell=\beta\log n$,
so there are $\Theta(\log n)$ such possibilities. To discuss the
analysis of one of these possibilities, that is, an error from $T=0$
to some $\hat{T}=t>t(h)$, let us assume for simplicity that $\mathbf{X}$
is memoryless. By ignoring here several subtle technical details,
simple algebraic manipulations show that 
\begin{equation}
\P\left[\hat{T}=t\mid T=0\right]\lessapprox\sum_{x_{1}^{t}\in{\cal X}^{\otimes t}}P_{X_{1}^{t}}^{2}\left(x_{1}^{t}\right)\cdot\I\left[\frac{1}{P_{X_{1}^{t}}(x_{1}^{t})}\geq n\right].\label{eq: specific type I error probability noiseless proof outline}
\end{equation}
If we can guarantee that this probability is $o(\frac{1}{n})$, then
after the union bound over all the candidates $t(h)\leq t\leq\ell$
($\Theta(\log n)$ terms) we obtain that 
\[
p_{\text{type-I}}(\hat{T})=o\left(\frac{\log n}{n}\right).
\]
Since, as explained above, the error probability is $\Omega(\frac{\log n}{n})$,
this means that $p_{\text{type-I}}(\hat{T})$ has a negligible effect
on the error probability. Thus, let us explore what is the minimal
value of $h$ required to make the right-hand side (r.h.s.) of (\ref{eq: specific type I error probability noiseless proof outline})
$o(\frac{1}{n})$. Evidently, if we trivially upper bound the indicator
by $1$ we get 
\[
\sum_{x_{1}^{t}\in{\cal X}^{\otimes t}}P_{X_{1}^{t}}^{2}(x_{1}^{t})=\exp_{|{\cal X}|}\left[-tH_{2}(P_{X_{1}})\right]\leq\exp_{|{\cal X}|}\left[-\frac{H_{2}(P_{X_{1}})}{h}\log n\right]=n^{-\frac{H_{2}(P_{X_{1}})}{h}}
\]
and so $h<H_{2}(P_{X_{1}})$ suffices. However, the indicator in the
sum (\ref{eq: specific type I error probability noiseless proof outline})
is $\I[P_{X_{1}^{t}}(x_{1}^{t})\leq1/n]$, and retains only small
probabilities in the summation. Thus, one may conjecture that taking
this indicator into account may substantially reduce the value of
the sum below $n^{-\frac{H_{2}(P_{X_{1}})}{h}}$ (as obtained when
the indicator is trivially bounded by $1$). The proof of Theorem
\ref{thm: exponent upper bound noiseless} begins with Lemma \ref{lem: partial probability power sum upper bound}
(Appendix \ref{sec:Proofs-noiseless}), which uses a Chernoff-style
bound on the indicator function to show that this is indeed the case,
and that the sum in (\ref{eq: specific type I error probability noiseless proof outline})
is essentially bounded by
\[
\exp_{|{\cal X}|}\left[-\frac{H_{1}(P_{X_{1}})}{h}\log n\right]=n^{-\frac{H_{1}(P_{X_{1}})}{h}}.
\]
This allows to show that (\ref{eq: specific type I error probability noiseless proof outline})
is $o(\frac{1}{n})$, for all $h<H_{1}(P_{X_{1}})$, which improves
the previous sufficient condition $h<H_{2}(P_{X_{1}})$ (recall that
the monotonicity property of the R\'{e}nyi entropy assures that $H_{2}(P_{X_{1}})\leq H_{1}(P_{X_{1}})$).
Lemma \ref{lem: partial probability power sum upper bound} shows
this property in greater generality, for any strongly mixing process
$\mathbf{X}$, with condition given by the entropy rate ${\cal H}_{1}(\mathbf{X})$.
This justifies that $t^{*}\sim\frac{\log n}{{\cal H}_{1}(\mathbf{X})}$
is the \emph{effective, or reliable,} minimal detectable overlap. 

Given that the type-I error probability is a negligible $o(\frac{\log n}{n})$,
we now turn to consider the type-II error probability. Due to the
symmetry between positive and negative overlaps, let us upper bound
the contribution of $T>0$, given by
\[
\sum_{t\in[\ell]}\P[T=t]\cdot p_{\text{type-II}}(\hat{T};t)=\frac{1}{n}\sum_{t\in[\ell]}p_{\text{type-II}}(\hat{T};t).
\]
We may then double it to include also the contribution of $T<0$.
Now, if $0<T<t_{{\scriptscriptstyle \text{MDO}}}$, then the conditional
error probability is $1$. However, for the type-I error probability
to be negligible it is required to avoid deciding $T<t^{*}$, we upper
bound the conditional error probability by $1$ for all $0<T<t^{*}$.
The total contribution of such events to the error probability is
thus $\frac{t^{*}}{n}\sim\frac{1}{{\cal H}_{1}(\mathbf{X})}\frac{\log n}{n}$.
This is the dominating term in the bound, and multiplying it by $2$,
to account for negative overlaps, leads to the bound of Theorem \ref{thm: exponent upper bound noiseless}.
Our previous refinement of the upper bound in (\ref{eq: specific type I error probability noiseless proof outline})
becomes crucial here, because had our necessary condition was $h<{\cal H}_{2}(\mathbf{X})$,
then this term was larger $\sim\frac{1}{{\cal H}_{2}(\mathbf{X})}\frac{\log n}{n}$. 

\begin{figure}
\begin{centering}
\includegraphics[scale=1.25]{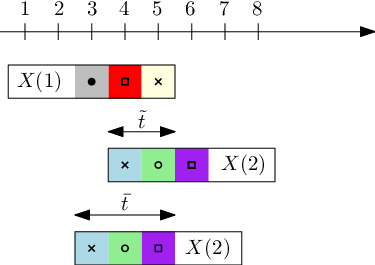}
\par\end{centering}
\caption{A case in which $X(1)$ and $X(2)$ have both overlap of length $\tilde{t}=2$
and $\overline{t}=3$ for $\ell=5$. We show $X(1)$ and the two possible
alignments of $X(2)$ to $X(1)$. By relating the two possible alignments
we obtain relations between the symbols of $X(1)$. For example, the
overlap of length $\tilde{t}$ (top) implies that the sub-sequence
marked in red (square) in $X(1)$ equals the sub-sequence marked in
blue (x) in $X(2)$. The overlap of length $\overline{t}$ (bottom)
implies that the sub-sequence marked in grey (disk) in $X(1)$ also
equals the sub-sequence marked in blue (x) in $X(2)$. Both together
imply that the sub-sequence marked in red (square) in $X(1)$ equals
the sub-sequence marked in grey (disk) in $X(1)$. Using similar arguments,
it stems that $X_{I(1)+2}^{I(1)+4}=X_{I(1)+3}^{I(1)+5}$, a repetition
of length $\overline{t}=3$ (here it is ``tail-biting'', but this
is not the case for larger differences $\overline{t}-\tilde{t}$).
\label{fig: two matchings implies repetition}}
\end{figure}

It is thus remain to bound $p_{\text{type-II}}(\hat{T};t)$ for $T=t>t^{*}$.
Specifically, there are $\ell-t^{*}=\Theta(\log n)$ such terms, and
so if we show that each of them is $o_{n}(1)$, then after multiplying
by the prior $\P[T=t]=\frac{1}{n}$, their overall contribution to
the error probability will be negligible $o(\frac{\log n}{n})$. An
error from $T=t>t^{*}$ may be to one of three possibilities: First,
to $\hat{T}=0$, second, to a longer overlap $\hat{T}>T$, and third,
to a longer negative overlap. Let us begin with the first case, which
is somewhat dual to the type-I error probability discussed above.
In the memoryless case, it can be shown to be upper bounded as
\begin{equation}
\P\left[\hat{T}=0\mid T=t\right]\lessapprox\sum_{x_{1}^{t}\in{\cal X}^{\otimes t}}P_{X_{1}^{t}}(x_{1}^{t})\cdot\I\left[\frac{1}{P_{X_{1}^{t}}(x^{t})}\leq n\right],\label{eq: specific type II to zero error probability noiseless proof outline}
\end{equation}
which is similar to the upper bound in (\ref{eq: specific type I error probability noiseless proof outline}),
albeit the summed probability is not squared here, and the inequality
in the indicator is reversed, that is, the sum retains large probabilities.
An additional result of Lemma \ref{lem: partial probability power sum upper bound},
proved similarly, is that the same condition $t>t^{*}\sim\frac{\log n}{{\cal H}_{1}(\mathbf{X})}$
suffices to make this upper bound $o_{n}(1)$ (which is less demanding
than the $o(\frac{1}{n})$ derived above for (\ref{eq: specific type I error probability noiseless proof outline})).
As for (\ref{eq: specific type I error probability noiseless proof outline}),
Lemma \ref{lem: partial probability power sum upper bound} shows
this result for general processes, possibly with memory. The second
and third cases, to wit, an error to $\hat{T}>T$ or to $\hat{T}<-T$
can be similarly bounded, and so we only discuss the second case.
Specifically, let us focus on an error from $T=\tilde{t}>t^{*}$ to
$\hat{T}=\overline{t}>\tilde{t}$. According to the detection rule
of $\hat{T}$ ((\ref{eq: optimal detection rule noiseless memoryless})
in Prop. \ref{prop: Bayes estimator noiseless memoryless}), such
an error occurs when both $X_{\ell-\overline{t}+1}^{\ell}(1)=X_{1}^{\overline{t}}(2)$
and $1/P_{X_{1}^{\overline{t}}}(X_{1}^{\overline{t}}(2))\geq n_{\ell}$
occur. For an upper bound, we may ignore the second condition, and
only upper bound the probability of the first one. This event implies
that while the true overlap is $\tilde{t}$, there is also a matching
for a longer overlap $\overline{t}$. It is not difficult to verify
that for this to occur, it must be that the process $X_{1}^{n}$ contains
a \emph{repetition} of length $\overline{t}$, that is, an event of
the form $\{X_{1}^{\overline{t}}=X_{1+s}^{\overline{t}+s}\}$ for
some $s\in\mathbb{N}_{+}$. See Fig. \ref{fig: two matchings implies repetition}
for an illustration. 
 Lemma \ref{lem: matching probability} establishes that if the process
is aperiodic and strongly mixing then the probability of any such
repetition is $o(\frac{1}{\log n})$. After taking a union bound over
all possible $\overline{t}>\tilde{t}$ (there are $\Theta(\log n)$
such terms), the resulting contribution to the conditional error probability
is $\P[\hat{T}>T\mid T=\tilde{t}]=o(\frac{1}{n})$. Taking a union
bound over all possible $T=\tilde{t}$ (again, there are $\Theta(\log n)$
such terms) results a total contribution to the type-II error probability,
and thus to the Bayesian error probability of $o(\frac{\log n}{n})$,
which is negligible compared to the $\frac{1}{{\cal H}_{1}(\mathbf{X})}\frac{\log n}{n}$
contribution from $T<t^{*}$. This completes the upper bounds of the
various types of error probability. 

To conclude the description of the proof's main ideas, we next shortly
outline the proof of Lemma \ref{lem: matching probability}, which
upper bounds the probability of the repetition $\{X_{1}^{\overline{t}}=X_{1+s}^{\overline{t}+s}\}$.
The motivation for the proof technique is that the process $\mathbf{X}$
is strongly mixing, and so events which are at time points that are
sufficiently far apart from one another are approximately independent,
even if the process has memory. Thus, we may consider repetition at
jumps of $\tau\equiv\tau(n)$ time points, for (slowly) increasing
sequence $\tau(n)$, that is, the event 
\begin{equation}
\left\{ X_{1}=X_{1+s}\right\} \cap\left\{ X_{1+\tau}=X_{1+\tau+s}\right\} \cap\cdots.\label{eq: proof outline intersection of pair mathching events}
\end{equation}
Nonetheless, the probability of this event cannot be upper bounded
as if all the symbols $\{X_{i}\}$ involved in this event are approximately
independent, since for a general value of $s$, the index of the end
symbol in the matching of a pair might be closer than $\tau$ to the
start index of a different pair. For example, $1+s$ might be close
by less than $\tau$ time points from some $1+j\tau$, $j\in\mathbb{N}_{+}$.
In other words, the time indices of the symbols in the events defining
(\ref{eq: proof outline intersection of pair mathching events}) are
intertwined, and not trivially separated. One technical challenge
of the proof is to adequately dilute pair matching events from (\ref{eq: proof outline intersection of pair mathching events}),
in order to assure a separation of $\tau$ time indices between any
two symbols. 

\paragraph*{The trade-off between the two types of error}

Evidently, the optimal Bayesian trade-off is as follows: To make the
type-I error probability a negligible $o(\frac{\log n}{n})$, the
MAP detector rarely decides on overlaps shorter than $\sim\frac{\log n}{{\cal H}_{1}(\mathbf{X})}$,
and in turn, when such short overlaps actually occur the detector
err. This is the dominating error event. Now, suppose that we would
like to further reduce the contribution of the type-I error probability,
to make it decay faster than $o(\frac{\log n}{n})$. This can be achieved
by modifying the detection rule (\ref{eq: optimal detection rule noiseless memoryless})
of Prop. \ref{prop: Bayes estimator noiseless memoryless} by replacing
$n_{\ell}\sim n$ with $n^{\mu}$ for some $\mu>1$. It can be shown
that this reduces the decay rate of the type-I error probability to
$o(\frac{\log n}{n^{\mu}})$, and increases the Bayesian error probability
to $\sim\frac{\mu}{{\cal H}_{1}(\mathbf{X})}\frac{\log n}{n}$, that
is, by a factor of $\mu$. A short outline of the details appears
in Appendix \ref{sec:Proofs-noiseless}. By contrast, the effect of
reducing $\mu$ below $1$ will make the type-I error probability
dominating, by detecting short, unreliable, overlaps. When $\mu<1$,
a characterization of the dependence of the type-II contribution to
the error probability as a function of the increased type-I error
probability is more delicate, since we still need to assure that $p_{\text{type-I}}(\hat{T})\sim\frac{\varphi_{\mu}}{{\cal H}_{1}(\mathbf{X})}\frac{\log n}{n}$
for some $\varphi_{\mu}<1$ to be in the interesting regime of $P_{\text{error}}^{*}=\Theta(\frac{\log n}{n})$.
\footnote{Bounds on the type-I error probability of the form $o(\frac{\log n}{n^{q}})$
for $q<1$ are most likely easier to derive, but are already rather
large, and less interesting given that the prior probability of $T\neq0$
is $\Theta(\frac{\log n}{n})$.} We thus refrain from pursuing the analysis of this case. 

Interestingly, the trade-off involved in our upper bound on the Bayesian
error probability does \emph{not} balance between the contributions
of the type-I and type-II error probabilities to the total error probability
(while taking into account the prior probability). Indeed, the contribution
of the type-I error probability is $o(\frac{\log n}{n})$ whereas
that of the type-II error probability is $\Theta(\frac{\log n}{n})$.
As explained in the introduction, this is because the setting here
is, loosely speaking, should be described as a collection of $2\ell-1$
binary detection problems. 

We next discuss the various assumptions of Theorem \ref{thm: exponent upper bound noiseless}. 

\paragraph*{The finite minus-infinity-order R\'{e}nyi entropy rate assumption}

This assumption implies a lower bound on the minimal error probability
of the process. Specifically, for any $\epsilon>0$, there exists
$s_{0}(\epsilon)$ such that for all $s>s_{0}$ it holds that 
\[
\min_{\tilde{x}_{1}^{s}\in\supp(P_{X_{1}^{s}})}P_{X_{1}^{s}}(\tilde{x}_{1}^{s})\geq\exp_{|{\cal X}|}\left[-s[{\cal H}_{-\infty}(\mathbf{X})-\epsilon]\right].
\]
Thus, the minimal  probability of a sequence of length $s$ from the
process decays exponentially, but not faster. Since the R\'{e}nyi
entropy is monotonically decreasing in its order, this guarantees
that ${\cal H}_{\alpha}(\mathbf{X})<\infty$ for any order $\alpha\in\mathbb{R}$.
This assumption is rather mild, but of technical nature and so it
is not obvious that it cannot be relaxed. Its origin is in the analysis
of the type-I error probability, and can be explained as follows:
In this analysis, $T=0$ is assumed, and so the reads do not overlap.
However, it is possible that they are close, or even adjacent, i.e.,
have start indices $I(1)=1$ and $I(2)+\ell+1$. When the process
$\mathbf{X}$ has memory, such reads, e.g., $X(1)=X_{1}^{\ell}$ and
$X(2)=X_{\ell+1}^{2\ell}$ may be highly correlated, even though $T=0$.
In those cases, the analysis of the type-I error, which is based on
upper bounding the probability of matching events such as $X_{\ell-t+1}^{\ell}(1)=X_{1}^{t}(2)$,
should be done under the true joint probability of the suffix/prefix
of the reads. We circumvent this challenging aspect by upper bound
the probability of the event $\{X_{\ell-t+1}^{\ell}(1)=X_{1}^{t}(2)\}$,
with the probability of a slightly shorter match $\{X_{\ell-t+1+\tau}^{\ell}(1)=X_{\tau}^{t}(2)\}$
for some $\tau\equiv\tau(n)$. This results a distance of $\tau$
time points between the symbols defining this event. Now, if we choose
$\tau(n)$ to increase slowly with $n$ (it turns out that $\tau(n)=\lceil\sqrt{\log n}\rceil$
is a valid choice), then the strongly mixing property of $\mathbf{X}$
implies that the symbols in $X_{\ell-t+1+\tau}^{\ell}(1)$ are approximately
independent from the symbols in $X_{\tau}^{t}(2)$, and this leads
to tractable analysis. Nonetheless, the error event also includes
the condition $\frac{1}{P_{X_{1}^{t}}(X_{1}^{t}(2))}\geq n_{\ell}$
but in order to agree with the reduced match length, we would like
to modify this condition to $1/P_{X_{\tau}^{t}}(X_{\tau}^{t}(2))\geq n_{\ell}$,
that is, for a read shorter by $\tau$ symbols. The assumption on
${\cal H}_{-\infty}(\mathbf{X})$ guarantees that this modification
has a mild effect on the probability, and thus this modification is
inconsequential. 

\paragraph*{The moderately strongly mixing assumption}

It is obvious that some type of mixing property is required for reliable
overlap detection. Indeed, if the process may output long repetitions
of the same symbol, of typical length $\Omega(\log n)$, then it is
impossible to reliably detect the overlap in case the two reads belong
to that segment in the process. Our assumption is fairly mild, because
for a process to be strongly mixing, it is only required that $d(s)\to0$
as $s\to\infty$. One common class of processes mixes with an \emph{exponential}
decay rate, that is $d(s)=\gamma^{s}$ for some $\gamma<1$ (otherwise
it is only sub-exponential), for which $\sum\gamma^{s}=\frac{1}{1-\gamma}<\infty$.
However, our assumption is much more lenient because it suffices that
$d(s)=o(\frac{1}{s})$ for $\sum_{s=0}^{\infty}d(s)<\infty$ to converge. 

\paragraph*{The aperiodicity assumption}

The probability ${\cal R}(\mathbf{X})=\P[X_{1}=X_{1+s}]$ is the \emph{recurrence
probability} at time delay $s$. Now, if there exists an $s^{*}>0$
such that $\P[X_{1}=X_{1+s^{*}}]=1$, then the process is periodic
with a period $s^{*}$. An aperiodicity property is clearly also necessary
for reliable overlap detection, because if the process has period
$s^{*}$ then it is impossible to distinguish between overlaps $T$
and $T+ms^{*}$, for any $m\in\mathbb{N}_{+}$. We assume that the
supremum over all $s$ is bounded away from $1$, but under the strong
mixing assumption, it is satisfied naturally for all $s$ sufficiently
large. Indeed, let $p_{\text{min}}:=\min_{x\in{\cal X}}P_{X_{1}}(x)$,
for which we may assume, w.l.o.g., that $p_{\text{min}}>0$ (otherwise
we may remove the irrelevant symbol from the alphabet), and anyhow
holds under the assumption ${\cal H}_{-\infty}(\mathbf{X})<\infty$.
It holds that $d(s)\to0$ as $s\to\infty$ and so
\begin{align}
\P\left[X_{1}=X_{1+s}\right] & =\sum_{x\in{\cal X}}\P[X_{1}=x]\P[X_{1+s}=x\mid X_{1}=x]\\
 & \leq\sum_{x\in{\cal X}}\P[X_{1}=x]\left(\P[X_{1+s}=x]+d(s)\right)\\
 & \leq\left(1+\frac{d(s)}{p_{\text{min}}}\right)\sum_{x\in{\cal X}}P_{X_{1}}^{2}[X_{1}=x]\\
 & =\left(1+\frac{d(s)}{p_{\text{min}}}\right)|{\cal X}|^{-H_{2}(P_{X_{1}})}\\
 & <1,
\end{align}
for all $s>s_{0}$, where $s_{0}$ is determined by $(\{d(s)\},p_{\text{min}},H_{2}(P_{X_{1}}))$. 

\paragraph*{The parameters determining the convergence to asymptotic limits}

The convergence of the upper bound to its asymptotic limit depends
on the problem parameters: The normalized length of the read $\beta$,
the minus-infinity-order R\'{e}nyi entropy ${\cal H}_{-\infty}(\mathbf{X})$,
and the related quantity $p_{\text{min}}$, the decay rate of the
strong mixing coefficient $d(s)$ and on the recurrence probability
${\cal R}(\mathbf{X})$. 

We conclude with two simple examples that illustrate the result of
Theorem \ref{thm: exponent upper bound noiseless}. 
\begin{example}[Memoryless sources]
\label{exa: Memoryless noiseless}If $\mathbf{X}$ is memoryless
process, then the assumptions of Theorem \ref{thm: exponent upper bound noiseless}
trivially hold (${\cal H}_{-\infty}(\mathbf{X})=-\log p_{\text{min}}<\infty$,
$d(s)\equiv0$ for all $s\in\mathbb{N}$, and ${\cal R}(\mathbf{X})=|{\cal X}|^{-H_{2}(P_{X_{1}})}<1$).
Also, the entropy rate exists, and so 
\[
P_{\text{error}}^{*}\leq2\left[1+o_{n}(1)\right]\cdot\left\{ \beta\wedge\frac{1}{H_{1}(P_{X_{1}})}\right\} \cdot\frac{\log n}{n}.
\]
\end{example}
\begin{example}[First-order Markov Process (Markov chains)]
\label{exa: Markov noiseless}If $\mathbf{X}$ is a first-order Markov
process, which forms an irreducible and aperiodic Markov chain, then
it has a unique stationary distribution (e.g., \cite[Sec. 1.5]{levin2017markov}).
Let us denote its transition matrix (Markov kernel) by
\[
K(x,x'):=\P\left[X_{i}=x'\mid X_{i-1}=x\right]
\]
for any $x,x'\in{\cal X}$ and $i\in\mathbb{Z}$. As a simple and
concrete example, let us focus on the case
\[
K(x,x')=\begin{cases}
1-(|{\cal X}|-1)\epsilon, & x=x'\\
\epsilon, & x\neq x'
\end{cases}
\]
which has a uniform stationary distribution $P_{X_{1}}(x)=\frac{1}{|{\cal X}|}$
for all $x\in{\cal X}$.

We next discuss the various assumptions of Theorem \ref{thm: exponent upper bound noiseless}.
First, from the Markov property
\begin{align}
{\cal H}_{-\infty}(\mathbf{X}) & =\lim_{t\to\infty}\frac{1}{t}\max_{x_{1}^{t}\in\supp(P_{X_{1}^{t}})}\log\frac{1}{P_{X_{1}^{t}}(x_{1}^{t})}\\
 & =\lim_{t\to\infty}\frac{1}{t}\max_{x_{1}^{t}\in\supp(P_{X_{1}^{t}})}\log\frac{1}{P_{X_{1}}(x_{1})\prod_{i=2}^{t}K(x_{i-1},x_{i})}\\
 & =\lim_{t\to\infty}\frac{1}{t-1}\max_{x_{1}^{t}\in\supp(P_{X_{1}^{t}})}\sum_{i=2}^{t}\log\frac{1}{K(x_{i-1},x_{i})}.
\end{align}
This value is upper bounded by the maximal possible sum of $\log\frac{1}{K(x_{i-1},x_{i})}$
along a cycle (i.e., a path that starts and end at the same symbol)
on the state space graph of the chain (for which each node is $x\in{\cal X}$).
In our concrete example, it holds that 
\[
{\cal H_{-\infty}}(\mathbf{X})=\begin{cases}
-\log\left[\left(1-(|{\cal X}|-1)\epsilon\right)\right], & \epsilon\geq\frac{1}{|{\cal X}|}\\
-\log(\epsilon), & \epsilon<\frac{1}{|{\cal X}|}
\end{cases},
\]
which indeed satisfies the condition ${\cal H_{-\infty}}(\mathbf{X})<\infty$.
Second, for a given delay $s\in\mathbb{N}_{+}$ and length $t$, Markovity
implies that 
\begin{align}
 & \left|\P[X_{i+s+1}^{i+s+t}\mid X_{-\infty}^{i+1}]-\P[X_{i+s+1}^{i+s+t}]\right|\nonumber \\
 & =\left|\P[X_{i+s+1}^{i+s+t}\mid X_{i+1}]-\P[X_{i+s+1}^{i+s+t}]\right|\\
 & =\left|\P[X_{i+s+1}\mid X_{i+1}]-\P[X_{i+s+1}]\right|\cdot\prod_{j=2}^{t}\P[X_{i+s+j}\mid X_{i+s+j-1}]\\
 & \leq\left|\P[X_{i+s+1}\mid X_{i+1}]-\P[X_{i+s+1}]\right|\\
 & \leq\max_{x\in{\cal X}}\|K^{s}(x,\cdot)-P_{X_{1}}\|_{\text{TV}},
\end{align}
where the latter is commonly take as the definition of the strong
mixing coefficients of a Markov chain. In our concrete example, it
is not difficult to verify (See details on Appendix \ref{sec:Proofs-noiseless})
that $d(s)\leq e^{-|{\cal X}|\epsilon\cdot s}$, which clearly implies
that the chain is strongly mixing. Third, aperiodicity holds by assumption.
Finally, the error probability is determined by the entropy rate,
easily verified to be given by
\[
{\cal H}_{1}(\mathbf{X})=\frac{1}{|{\cal X}|}\log\frac{\epsilon}{1-(|{\cal X}|-1)\epsilon}+\log\frac{1}{\epsilon}.
\]
\end{example}
We next turn to derive a lower bound, and prove that the bound of
Theorem \ref{thm: exponent upper bound noiseless} is tight (under
proper conditions). 
\begin{thm}
\label{thm: exponent lower bound noiseless}Consider the noiseless
setting (Def. \ref{def: Noiseless setting}). Assume that for some
$\eta>0$, it holds that 
\[
d(s)\leq\frac{1}{s^{\beta{\cal H}_{-\infty}(\mathbf{X})+\eta}},
\]
for all $s\in\mathbb{N}_{+}$, and that the entropy rate ${\cal H}_{1}(\mathbf{X})$
exists. Then 
\begin{equation}
P_{\text{\emph{error}}}^{*}\geq2\left[1+o_{n}(1)\right]\cdot\left\{ \beta\wedge\frac{1}{{\cal H}_{1}(\mathbf{X})}\right\} \cdot\frac{\log n}{n}.\label{eq: lower bound on the error probability noiseless}
\end{equation}
\end{thm}
Theorem \ref{thm: exponent lower bound noiseless} thus establishes
that the bound in Theorem \ref{thm: exponent upper bound noiseless}
is tight. 

\paragraph*{The main ideas of the proof}

In its core, the proof relies on the fact that the dominant error
event is when $|T|\leq t^{*}\sim\frac{\log n}{{\cal H}_{1}(\mathbf{X})}$,
in which the probability of the overlapping part of the read $X_{1}^{T}(2)$
is not sufficiently small to assure that this overlap will actually
be detected, i.e., $1/P_{X_{1}^{T}(2)}(X_{1}^{T}(2))\leq n_{\ell}$,
and so $\hat{T}=0$ is erroneously decided. To prove this rigorously,
we lower bound the error probability of the MAP detector by the error
probability of a \emph{genie-aided detector}, whose operation can
be described as follows, with some simplifications of technical details:
If $T>t^{*}\sim\frac{\log n}{{\cal H}_{1}(\mathbf{X})}$, then the
genie reveals the detector the true value of $T$, and so there is
no error. If $0<|T|\leq t^{*}$, the genie only reveals to the detector
that the overlap is either $T=0$ or $T=t^{*}$. If the overlap is
$T=0$, then the genie reveals to the detector that the true value
is either $T=0$ or some random $T=t$, chosen uniformly in $[t^{*}]$.
This reduces the operation of the genie-aided detector to that of
$[t^{*}]$ \emph{binary} detectors. The average error probability
of each of these binary detectors can be lower bounded similarly,
so we may focus on one of them. As usual, this lower bound is based
on comparing the likelihood ratio to the ratio of the prior probabilities.
In turn, this requires evaluating both $f(y_{1}^{\ell}(1),y_{1}^{\ell}(2);t)$
and $f(y_{1}^{\ell}(1),y_{1}^{\ell}(2);0)$. As we have seen in the
proof of Prop. \ref{prop: Bayes estimator noiseless memoryless},
both likelihoods are challenging to handle in case of memory. For
$T=t\neq0$, we utilize the asymptotic equipartition property (AEP)
theorem of Shannon--McMillan--Breiman (e.g., \cite[Th. 16.8.1]{cover2012elements})
to show that, with high probability, the likelihood 
\[
f(y_{1}^{\ell}(1),y_{1}^{\ell}(2);t)\leq e^{o_{n}(1)\cdot\log n}\cdot\I\left[x_{\ell-t+1}^{\ell}(1)=x_{1}^{t}(2)\right]\cdot P_{X_{1}^{\ell}}\left(x_{1}^{\ell}(1)\right)\cdot P_{X_{1}^{\ell}}\left(x_{1}^{\ell}(2)\right)\cdot\Gamma_{+}(t),
\]
where the r.h.s. is the likelihood given for a memoryless processes,
or more generally, for any finite-order Markov chain, as developed
in (\ref{eq: likelihood noiseless positive overlap}) (proof of Prop.
\ref{prop: Bayes estimator noiseless memoryless}), multiplied by
a $e^{o_{n}(1)\cdot\log n}$ factor. For $T=0$, we argued that for
ergodic processes (\ref{eq: approximate likelihood for T=00003D0 when the process has memory})
the reads are approximately independent, and the likelihood is a multiplication
of the likelihood of each read. Lemma \ref{lem: asymptotic independence condition on T=00003D0}
shows that under the stronger mixing conditions imposed by Theorem
\ref{thm: exponent lower bound noiseless}, the approximate relation
(\ref{eq: approximate likelihood for T=00003D0 when the process has memory})
is actually an accurate asymptotic lower bound 
\[
f\left(y_{1}^{\ell}(1),y_{1}^{\ell}(2);0\right)\geq\left[1-o_{n}(1)\right]\cdot P_{X_{1}^{\ell}}\left(x_{1}^{\ell}(1)\right)\cdot P_{X_{1}^{\ell}}\left(x_{1}^{\ell}(2)\right).
\]
With these asymptotic bounds, we can upper bound the likelihood ratio,
and prove a lower bound on the error probability of this binary detector.
It should be mentioned that this lower bound requires the almost sure
convergence of $-\frac{1}{t}\log P_{X_{1}^{t}}(x_{1}^{t})$ to the
entropy rate ${\cal H}_{1}(\mathbf{X})$, as assured by the Shannon--McMillan--Breiman
AEP, and thus also the existence of the entropy rate ${\cal H}_{1}(\mathbf{X})$. 

\paragraph*{The strongly mixing assumption}

Compared to Theorem \ref{thm: exponent upper bound noiseless}, which
only assumed $d(s)=o(\frac{1}{s})$, the lower bound makes a slightly
stronger assumption that $d(s)$ is polynomially decaying with power
strictly larger than $\beta{\cal H}_{-\infty}(\mathbf{X})$. Since
for many processes the strong mixing coefficient decays much faster,
this difference is rather minor. Nonetheless, pinpointing the exact
dependence on the strong mixing coefficient remains an elusive open
problem. We mention that the source of this assumption is Lemma \ref{lem: asymptotic independence condition on T=00003D0},
and there it is used to show that $d(\tau)$ is negligible compared
to $P_{X_{1}^{\ell}}(x_{1}^{\ell}(1))\cdot P_{X_{1}^{\ell}}(x_{1}^{\ell}(2))$
whenever $\tau\equiv\tau(n)=\frac{n}{\log n}$. 

\section{Overlap Detection for Memoryless Processes and Noisy Memoryless Reading
Kernels \label{sec:Alignment-Detection-for noisy}}

In this section, we address the noisy setting of Definition \ref{def: Noisy setting}.
For this analysis, we make the following additional notation conventions.
We denote by $P_{Y}$ the $Y$-marginal of $P_{X}\otimes P_{Y\mid X}$,
and let $\tilde{Y}$ be such that $P_{\tilde{Y}\mid X}=P_{Y\mid X}$
and the Markov chain relation $Y-X-\tilde{Y}$ holds. We then denote
by $P_{Y\tilde{Y}}$ the $(Y,\tilde{Y})$-marginal, that is, 
\[
P_{Y\tilde{Y}}(y,\tilde{y})=\sum_{x\in{\cal X}}P_{X}(x)\cdot P_{Y\mid X}(y\mid x)P_{Y\mid X}(\tilde{y}\mid x)
\]
for any $(y,\tilde{y})\in{\cal Y}^{\otimes2}$ (hence $P_{Y}=P_{\tilde{Y}}$
holds). Furthermore, we denote the \emph{likelihood ratio} between
$y\in{\cal Y}$ and $\tilde{y}\in{\cal Y}$ as 
\begin{equation}
\lambda(y,\tilde{y}):=\frac{P_{Y\tilde{Y}}(y,\tilde{y})}{P_{Y}(y)P_{\tilde{Y}}(\tilde{y})}.\label{eq: information density}
\end{equation}
We may note that 
\[
\lambda_{\text{max}}:=\max_{(y,\tilde{y})\in{\cal Y}^{\otimes2}}\lambda(y,\tilde{y})\in[1,\infty),
\]
and, in addition, since $P_{Y\tilde{Y}}\ll P_{Y}\otimes P_{Y}$ then
\[
\lambda(Y,\tilde{Y})\geq\lambda_{\text{min}}:=\min\left\{ \lambda(y,\tilde{y})\colon(y,\tilde{y})\in\supp(P_{Y\tilde{Y}})\right\} >0
\]
with probability $1$. 

\subsection{The MAP Detection Rule \label{subsec:The-MAP-Detection noisy}}

Our first step in deriving the MAP detection rule is to obtain an
expression for the likelihood function. 
\begin{lem}
\label{lem: likelihood for noisy reads}Consider the noisy setting
(Def. \ref{def: Noisy setting}). Then, the likelihood of overlap
$t\in[\ell]$ given reads $y_{1}^{\ell}(1)$ and $y_{1}^{\ell}(2)$
is
\begin{equation}
f\left(y_{1}^{\ell}(1),y_{1}^{\ell}(2);t\right)\propto\Gamma_{+}(t):=\prod_{i=1}^{t}\lambda\left(y_{\ell-t+i}(1),y_{i}(2)\right),\label{eq: likelihood for noisy reads positive}
\end{equation}
and for $t\in-[\ell-1]$ 
\begin{equation}
f\left(y_{1}^{\ell}(1),y_{1}^{\ell}(2);t\right)\propto\Gamma_{-}(t):=\prod_{i=1}^{t}\lambda\left((y_{\ell-t+i}(2),y_{i}(1)\right),\label{eq: likelihood for noisy reads negative}
\end{equation}
and where we set $\Gamma_{\pm}(0)=1$. 
\end{lem}
Inspecting the likelihood in (\ref{eq: likelihood for noisy reads positive}),
the function $t\mapsto\Gamma_{+}(t)$ is not necessarily monotonic
\emph{increasing} with $t$. Thus, unlike the noiseless case, any
potential overlap $t\in[\ell]$ should be considered, as each has
possibly non-zero likelihood, and an overlap is possible even if the
overlapping noisy parts do not match. Nonetheless, the detector still
has a rather simple form, as follows:
\begin{prop}
\label{Bayes estimator shotgun IID noise}Consider the noisy setting
(Def. \ref{def: Noisy setting}). Let $\Gamma_{\pm}(t)$ be as defined
in Lemma \ref{lem: likelihood for noisy reads}, and let the most
likely positive and negative overlap be 
\begin{align}
t_{+} & :=\argmax_{t\in[\ell]}\Gamma_{+}(t)\\
t_{-} & :=\argmax_{t\in-[\ell-1]}\Gamma_{-}(t).
\end{align}
Then, the MAP detector is as in (\ref{eq: optimal detection rule noiseless memoryless}).
Furthermore, let the minimum detectable overlap be denoted as
\[
t_{{\scriptscriptstyle \text{\emph{MDO}}}}:=\frac{\log(n_{\ell})}{\log\lambda_{\text{\emph{max}}}}.
\]
Then, $0<|t|\leq t_{{\scriptscriptstyle \text{\emph{MDO}}}}$ will
never be detected by the MAP detector.
\end{prop}
The proof of Prop. \ref{Bayes estimator shotgun IID noise} is similar
to the proof of Prop. \ref{prop: Bayes estimator noiseless memoryless},
and thus omitted. 

\subsection{Error Probability Analysis \label{subsec:Error-Probability-Analysis noisy}}
\begin{thm}
\label{thm: exponent upper bound noisy}For the noisy setting (Def.
\ref{def: Noisy setting})
\[
P_{\text{\emph{error}}}^{*}\leq2\left[1+o_{n}(1)\right]\cdot\left\{ \beta\wedge\frac{1}{I(Y;\tilde{Y})}\right\} \cdot\frac{\log n}{n}.
\]
\end{thm}

\paragraph*{The dependence of the error on the mutual information}

When the reading kernel is noisy, the mutual information $I(Y;\tilde{Y})$
replaces the role of the Shannon entropy rate (and indeed reduces
to it when the reading kernel is noiseless). It emerges as the key
measure because it is the expected log-likelihood ratio between the
likelihood that the overlapping part of the reads are from separated
locations in the sequence and the likelihood that this part are from
the same location.

\paragraph*{The main ideas of the proof}

As in the proof of Theorem \ref{thm: exponent upper bound noiseless},
we carefully analyze the probabilities of the various types of error.
To simplify the proof and the description here, we only focus on errors
from and to positive overlaps. Let us begin by considering the type-I
error. Since $\hat{T}=t<t_{{\scriptscriptstyle \text{MDO}}}$ is never
detected, we may take a union bound of the $\Theta(\log n)$ possibilities
to err from $T=0$ to $t_{{\scriptscriptstyle \text{MDO}}}<\hat{T}=t<\ell$.
For each one of these possibilities, it can be shown that according
to the detection rule of Prop. \ref{Bayes estimator shotgun IID noise},
the error probability is upper bounded by 
\begin{equation}
\P\left[\sum_{i=1}^{t}\log\lambda\left(Y_{\ell-t+i}(1),Y_{i}(2)\right)\geq\log n_{\ell}\,\middle\vert\,T=0\right].\label{eq: noisy proof outline type I upper bound}
\end{equation}
This is a sum of $t$ independent r.v.'s, where the expected value
of each (conditioned on $T=0$) is $I(Y;\tilde{Y})$. Hence, we expect
that such probability will vanish only if $t>t^{*}\sim\frac{\log n}{I(Y;\tilde{Y})}$.
Following the same arguments as in the proof of Theorem \ref{thm: exponent upper bound noiseless},
the contribution of the type-II error probability is $\frac{1}{I(Y;\tilde{Y})}\frac{\log n}{n}$.
However, to establish that the type-I error probability is negligible,
we need to prove that it is $o(\frac{\log n}{n})$, and so taking
into account the union bound over $\Theta(\log n)$ terms, we need
to prove that the probability in (\ref{eq: noisy proof outline type I upper bound})
is $o(\frac{1}{n})$. To this end, if we deploy the standard Chernoff
bound, then we only obtain $O(\frac{1}{n})$, which does not suffice
for our purpose. Therefore, we deploy a stronger large-deviations
bound, similar to that of Bahadur--Rao \cite{Bahadur1960on} \cite[Th. 3.7.4.]{dembo2009large},
which bounds this probability with the correct pre-exponent. In our
case, this pre-exponent contributes a factor of $\frac{1}{\sqrt{t}}=\Theta(\frac{1}{\sqrt{\log n}})$
to the upper bound, which shows that the overall type-I error probability
is $O(\frac{\sqrt{\log n}}{n})=o(\frac{\log n}{n})$. The type-II
error is similarly bounded using Chernoff's bounding technique. The
upper bound on the error probability from $T>0$ to $T=0$ involves
the analysis of a probability similar to (\ref{eq: noisy proof outline type I upper bound}),
albeit with a reverse sign, and under the condition $T=t>0$. Such
probability is analyzed in Lemma \ref{lem: Chernoff analysis of crossing events}.
The error probability for a wrong detection, that is from $T=t$ to
a different $\hat{T}\neq0$, is more complicated compared to the noiseless
setting because there is also a possibility to underestimate the true
overlap. As in the noiseless setting, the conditioning on $T>0$ results
dependencies between the terms involved in the sum of log-likelihoods.
This is addressed in Lemma \ref{lem: crossing probability correlated log-likelihoods}. 

\paragraph*{The trade-off between the two types of error}

As in the noiseless setting, it can be verified from the proof of
Theorem \ref{thm: exponent upper bound noisy}, that if $n_{\ell}$
is replaced in the detector by $n^{\mu}$ for some $\mu>0$, then
the type-I error would be $p_{\text{type-I}}(\hat{T})=o(\frac{1}{n^{\mu}})$
whereas the contribution of the type-II error probability would increase
by a factor of $\mu$, that is, given by $[1+o_{n}(1)]\frac{\mu}{I(Y;\tilde{Y})}\cdot\frac{\log n}{n}$. 

\paragraph*{The parameters determining the convergence to asymptotic limits}

Naturally, here the convergence to the asymptotic limit depends on
the source distribution $P_{X}$, the noise kernel $P_{Y|X}$, and
the scaled read length $\beta$. 

We conclude with a matching lower bound. 
\begin{thm}
\label{thm:lower bound on Bayesian error}For the noisy setting (Def.
\ref{def: Noisy setting}),
\[
P_{\text{\emph{error}}}^{*}\geq2\left[1-o_{n}(1))\right]\cdot\left\{ \beta\wedge\frac{1}{I(Y;\tilde{Y})}\right\} \cdot\frac{\log n}{n}.
\]
\end{thm}
The proof of Theorem \ref{thm:lower bound on Bayesian error} follows
similar lines to that of the lower bound in the noiseless reading
kernel case. 

\section{Conclusion \label{sec:Conclusion}}

We considered the problem of detecting the overlap between two short
reads of length $\ell=\beta\log n$ drawn from two random locations
in a long sequence of length $n$, taken from a process $\mathbf{X}$.
We considered the optimal Bayesian MAP detector, and provided tight
asymptotic expressions for its error probability, both in the case
the reading kernel is noiseless, allowing $\mathbf{X}$ to be stationary
and ergodic, as well as in the case the reading kernel is noisy and
memoryless, while restricting $\mathbf{X}$ to be memoryless. We next
discuss a few possible future research directions. 

First, it is plausible to conjecture that in the general case, in
which both the process has memory and the reading kernel is noisy
(perhaps also with memory), the error probability is given by $[1+o_{n}(1)]\cdot[\beta\wedge(1/{\cal I}(\mathbf{Y},\tilde{\mathbf{Y}}))]\cdot\frac{\log n}{n}$,
where, here, ${\cal I}(\mathbf{Y},\tilde{\mathbf{Y}})$ is the mutual-information
rate between two reads satisfying $\mathbf{Y}-\mathbf{X}-\mathbf{\tilde{Y}}$
(that is, two independent noisy observations $\mathbf{Y},\mathbf{\tilde{Y}}$
of the same process $\mathbf{X}$). One of the challenges in proving
such a result, beyond the additional anticipated technical effort,
is that even if $\mathbf{X}$ is a first-order Markov process and
the reading kernel is memoryless, each read is a \emph{hidden} Markov
process, for which the likelihood more complicated than the cases
we considered. In a similar spirit, it would be interesting to consider
noisy reading channels which introduce synchronization errors, such
as deletions and insertions, and to characterize the error probability
of the optimal detector. 

Second, it would be interesting to utilize techniques similar to the
ones used here in order to derive the fundamental limits of the perfect
reconstruction of the set of contigs (islands in the parlance of \cite{lander1988genomic})
from reads, whenever the reads do not fully cover the sequence. This
is challenging, because without any prior assumption on the coverage,
there are $\Theta(m^{2}\cdot\frac{\log n}{n})$ pairs of reads on
the average that only have a short overlap, and the detection of this
overlap is likely to either not to be detected, or erroneously detected.
The problem is even more challenging to analyze in the noisy setting,
because in this case multiple reads of the same symbol may improve
the de-noising of that symbol, and this must be taken into account
in the analysis of the optimal detector. 

Third, following \cite{shomorony2021sketching}, another open problem
is related to detection of the alignment based on \emph{sketches}
of the read. In this setting, each of the reads $Y_{1}^{\ell}(1)$
and $Y_{1}^{\ell}(2)$ is separately compressed, or sketched, to $Z_{1}^{B}(1)$
and $Z_{1}^{B}(2)$. The detector is then required to detect the correct
overlap based on the sketches $(Z_{1}^{B}(1),Z_{1}^{B}(2))$. The
question is what is the optimal error probability, where here the
sketching operation can also be optimized. Clearly, for $B=0$ the
error probability is $\beta\frac{\log n}{n}$, whereas for $B\geq H(Y_{1}^{\ell})$
the error probability, is as in our setting. The exact dependence
of the Bayesian error probability on $B$ is interesting to characterize. 

Fourth, a variation of our problem may address soft-detection, in
which the detector outputs the posterior probability of the overlap,
and the performance is measure by the logarithmic loss functions,
thus given by $H(T\mid Y_{1}^{\ell}(1),Y_{1}^{\ell}(2))$. It is then
of interest to tightly characterize the asymptotic value of this equivocation.
It would also be interesting to embed the posterior of the overlap
into assembly algorithms that merge reads that approximately overlap
\cite{shomorony2016information,kamath2017hinge}. 

Fifth, following \cite{mohajer2013reference,weinberger2024fundamental},
we may consider the overlap detection problem in case that there is
a \emph{reference sequence} that is correlated with the main sequence,
and aids the detection task. It is then of interest to characterize
how the asymptotic error probability improves as a function of the
amount of dependence between the sequence and the reference sequence. 

\appendices{\numberwithin{equation}{section}}

\section{Proofs and Details for Section \ref{sec:Alignment-Detection-for noiseless}
\label{sec:Proofs-noiseless}}

\subsection{Proof of Theorem \ref{thm: exponent upper bound noiseless}}

We begin with the following lemma, which states bounds that will be
instrumental in the analysis of the probability of error events from
$T=0$ to $\hat{T}\neq0$ (the bound (\ref{eq: partial second-order Renyi entropy upper bound})),
and from $T\neq0$ to $\hat{T}=0$ (the bound (\ref{eq: partial probability sum upper bound})). 
\begin{lem}
\label{lem: partial probability power sum upper bound}Let $\mathbf{X}:=\{X_{i}\}_{i\in\mathbb{Z}}$
be a stationary ergodic process for which ${\cal H}_{1}(\mathbf{X})<\infty$
. Let $\rho\in(0,1)$ be given and $\{\tau(n)\}_{n\in\mathbb{N}}$
be a sequence of non-negative integers such that $\tau(n)=\omega_{n}(1)$
and $\tau(n)=o(\log n)$. Then, for any $\eta>0$ 
\begin{equation}
\max_{t>\frac{(1+\eta)}{{\cal H}_{1}(\mathbf{X})}\log n}\sum_{x_{1}^{t}\in{\cal X}^{\otimes t}}P_{X_{1}^{t}}^{2}(x_{1}^{t})\cdot\I\left[\frac{1}{P_{X_{1}^{t}}(x^{t})}\geq\rho^{\tau-1}\cdot n_{\ell}\right]=o\left(\frac{1}{n}\right),\label{eq: partial second-order Renyi entropy upper bound}
\end{equation}
and
\begin{equation}
\max_{t>\frac{(1+\eta)}{{\cal H}_{1}(\mathbf{X})}\log n}\sum_{x_{1}^{t}\in{\cal X}^{\otimes t}}P_{X_{1}^{t}}(x_{1}^{t})\cdot\I\left[\frac{1}{P_{X_{1}^{t}}(x^{t})}\leq n_{\ell}\right]=o_{n}(1).\label{eq: partial probability sum upper bound}
\end{equation}
The constants involved in the asymptotic expressions depend on $\eta$. 
\end{lem}
\begin{IEEEproof}
Let any $t\geq\frac{(1+\eta)\log n}{{\cal H}_{1}(\mathbf{X})}$ be
given. We first prove (\ref{eq: partial second-order Renyi entropy upper bound}).
As $n_{\ell}=n-(2\ell-1)$ and $\tau\equiv\tau(n)=\omega_{n}(1)$,
it holds that for all $n$ sufficiently large (depending only on $(\rho,\beta)$)
that
\begin{align}
\rho^{\tau-1}\cdot n_{\ell} & =\rho^{\tau-1}\cdot\left(n-(2\ell-1)\right)\\
 & \geq\rho^{\tau-1}\left(1-\frac{2\beta\log n}{n}\right)\cdot n\\
 & \geq\left(\frac{\rho}{2}\right)^{\tau}\cdot n.\label{eq: approximate log of 1 over n}
\end{align}
In addition, since ${\cal H}_{2-r}(\mathbf{X})\uparrow{\cal H}_{1}(\mathbf{X})$
as $r\uparrow1$ (monotonicity of R\'{e}nyi entropy rate \cite{rached2001renyi}),
there exists $r_{0}(\eta)\in(0,1)$ and $\epsilon_{0}(\eta)>0$ for
which
\begin{equation}
\frac{(1+\eta)\left({\cal H}_{2-r_{0}}(\mathbf{X})-\epsilon_{0}\right)}{{\cal H}_{1}(\mathbf{X})}>1,\label{eq: ratio between almost Shannon entropy and entropy}
\end{equation}
such that $r_{0}(\eta)\to1$ and $\epsilon_{0}(\eta)\to0$ as $\eta\to0$.
So, for all $n$ sufficiently large
\begin{align}
 & \sum_{x_{1}^{t}\in{\cal X}^{\otimes t}}P_{X_{1}^{t}}^{2}(x_{1}^{t})\cdot\I\left[\frac{1}{P_{X_{1}^{t}}(x^{t})}\geq\rho^{\tau-1}\cdot n_{\ell}\right]\\
 & \trre[\leq,a]\sum_{x_{1}^{t}\in{\cal X}^{\otimes t}}P_{X_{1}^{t}}^{2}(x^{t})\cdot\I\left[\frac{1}{P_{X_{1}^{t}}(x^{t})}\geq\left(\frac{\rho}{2}\right)^{\tau}\cdot n\right]\\
 & \trre[=,b]\sum_{x_{1}^{t}\in{\cal X}^{\otimes t}}P_{X_{1}^{t}}^{2}(x^{t})\cdot\I\left[\log P_{X_{1}^{t}}(x^{t})\leq-(1+\kappa)\log n\right]\\
 & \trre[\leq,c]\sum_{x_{1}^{t}\in{\cal X}^{\otimes t}}P_{X_{1}^{t}}^{2}(x^{t})\cdot\min_{r\geq0}e^{r\left[-(1+\kappa)\log n-\log P_{X_{1}^{t}}(x^{t})\right]}\\
 & \leq\frac{1}{n^{(1+\kappa)r_{0}}}\sum_{x^{t}\in{\cal X}^{\otimes t}}P_{X_{1}^{t}}^{2-r_{0}}(x^{t})\\
 & =\frac{1}{n^{(1+\kappa)r_{0}}}|{\cal X}|^{-(1-r_{0})H_{2-r_{0}}(P_{X_{1}^{t}})}\\
 & =\exp_{n}\left(-(1+\kappa)r_{0}-\frac{(1-r_{0})H_{2-r_{0}}(P_{X_{1}^{t}})}{\log n}\right)\\
 & \trre[\le,d]\exp_{n}\left(-(1+\kappa)r_{0}-(1-r_{0})\frac{(1+\eta)}{{\cal H}_{1}(\mathbf{X})}\left({\cal H}_{2-r_{0}}(\mathbf{X})-\epsilon_{0}\right)\right)\\
 & \trre[=,e]o\left(\frac{1}{n}\right),
\end{align}
where $(a)$ follows from (\ref{eq: approximate log of 1 over n}),
$(b)$ follows by denoting 
\[
\kappa\equiv\kappa(n):=\frac{\tau(n)}{\log n}\log\frac{\rho}{2},
\]
$(c)$ follows by $\I[a>0]\leq\min_{r\geq0}e^{ra}$, and $(d)$ holds
for all $n$ sufficiently large, since for $\epsilon_{0}>0$ there
exists $n_{0}(\epsilon_{0})$ such that for all $n\ge n_{0}$
\[
H_{2-r_{0}}(P_{X_{1}^{t}})\geq t({\cal H}_{2-r_{0}}(\mathbf{X})-\epsilon)\geq\frac{(1+\eta)}{{\cal {\cal H}}_{1}(\mathbf{X})}\left({\cal H}_{2-r_{0}}(\mathbf{X})-\epsilon_{0}\right)\log(n).
\]
Finally, $(e)$ follows since (\ref{eq: ratio between almost Shannon entropy and entropy})
and $\kappa\equiv\kappa(n)=o_{n}(1)$ (which holds by the assumption
$\tau_{n}=o(\log n)$) imply that 
\[
\left(1+\kappa(n)\right)r_{0}+(1-r_{0})\frac{(1+\eta)}{{\cal H}_{1}(\mathbf{X})}\left(\underline{{\cal H}}_{2-r_{0}}(\mathbf{X})-\epsilon_{0}\right)>1
\]
for all $n$ sufficiently large (which depends on $(\rho,\{\tau(n)\})$).
Since these constants do not depend on $t$, the result (\ref{eq: partial second-order Renyi entropy upper bound})
is obtained.

We turn to prove (\ref{eq: partial probability sum upper bound}).
For any $r\in(0,1)$ and $\epsilon>0$, there exists $n_{0}(\epsilon)$
such that for all $n\geq n_{0}(\epsilon)$
\begin{align}
 & \sum_{x_{1}^{t}\in{\cal X}^{\otimes t}}P_{X_{1}^{t}}(x_{1}^{t})\cdot\I\left[\frac{1}{P_{X_{1}^{t}}(x^{t})}\leq n_{\ell}\right]\nonumber \\
 & =\P\left[P_{X_{1}^{t}}^{1-r}(X_{1}^{t})\geq\frac{1}{n_{\ell}^{1-r}}\right]\\
 & \trre[\leq,a]n^{1-r}\cdot\E\left[P_{X_{1}^{t}}^{1-r}\left(X_{1}^{t}(1)\right)\right]\\
 & =n^{1-r}\sum_{x_{1}^{t}\in{\cal X}^{\otimes t}}P_{X_{1}^{t}}^{2-r}(x_{1}^{t})\\
 & \leq n_{\ell}^{1-r}\cdot\exp_{|{\cal X}|}\left[-(1-r)t({\cal H}_{2-r}(\mathbf{X})-\epsilon)\right]\\
 & =\exp_{|{\cal X}|}\left[-(1-r)\left((1+\eta)\left(\frac{{\cal H}_{2-r}(\mathbf{X})-\epsilon}{{\cal H}_{1}(\mathbf{X})}\right)-1\right)\log n\right]\\
 & \trre[=,b]o_{n}(1),
\end{align}
where $(a)$ follows from Markov's inequality and $n_{\ell}\leq n$,
and $(b)$ follows from (\ref{eq: ratio between almost Shannon entropy and entropy}),
by setting $r=r_{0}(\eta)$ and $\epsilon=\epsilon_{0}(\eta)$. 
\end{IEEEproof}
We continue with the following lemma, which will be used to analyze
error events from a non-zero overlap $T=t$ to a longer overlap $\hat{T}>t$.
\begin{lem}
\label{lem: matching probability}Let $\mathbf{X}:=\{X_{i}\}_{i\in\mathbb{Z}}$
be a stationary strongly mixing process with mixing coefficients $\{d(s)\}_{s\in\mathbb{N}}$,
which satisfies $\sum_{s=0}^{\infty}d(s)<\infty$, and is also aperiodic
in the sense that 
\[
{\cal R}(\mathbf{X}):=\sup_{s\in\mathbb{N}_{+}}\P\left[X_{1}=X_{1+s}\right]<1.
\]
Then,
\[
\sup_{1\leq s\leq\ell}\,\max_{\frac{(1+\eta)}{{\cal H}_{1}(\mathbf{X})}\log n\leq t\leq\beta\log n}\P\left[X_{1}^{t}=X_{1+s}^{t+s}\right]=o\left(\frac{1}{\log n}\right).
\]
The constants depend on $p_{\text{\emph{min}}}:=\min_{x\in\supp({\cal X})}P_{X_{1}}(x)$. 
\end{lem}
To prove Lemma \ref{lem: matching probability}, we will be assisted
by the following lemma. Recall first Definition \ref{def: strong mixing coefficient}
of the strong mixing coefficient, which compares between the probability
of an event and its probability conditioned on a distant past event.
The next lemma upper bounds a probability conditioned on both past
and \emph{future }events, which are sufficiently distant apart. 
\begin{lem}
\label{lem: probability condining on both past and future}Let $\mathbf{X}:=\{X_{i}\}_{i\in\mathbb{Z}}$
be a stationary strongly mixing process with strong mixing coefficients
$\{d(s)\}_{s\in\mathbb{N}}$. Let $\underline{r},\overline{r}\in\mathbb{N}_{+}$
and let 
\[
t_{-\underline{r}}<t_{-\underline{r}+1}<\cdots t_{-1}<t_{0}<t_{1}<\cdots<t_{\overline{r}}
\]
 be a given sequence of $\underline{r}+\overline{r}+1$ times points.
Also let $p_{\text{\emph{min}}}:=\min_{x\in\supp({\cal X})}P_{X_{1}}(x)>0$
and 
\[
d_{\text{\emph{min}}}:=\min_{0\leq i\leq\overline{r}-1}\left|t_{i+1}-t_{i}\right|\wedge\min_{0\leq i\leq\underline{r}-1}\left|t_{-i}-t_{i-1}\right|,
\]
as well as 
\[
\overline{d}:=\frac{d_{\text{\emph{min}}}}{p_{\text{\emph{min}}}}.
\]
If $\overline{d}\leq\frac{1}{2}$ then 
\[
\P\left[X_{t_{0}}\mid X_{-t_{1}},\ldots,X_{-t_{\underline{r}}},X_{t_{1}},\ldots,X_{t_{\overline{r}}}\right]\leq\P[X_{t_{0}}]\cdot e^{2\overline{d}(\overline{r}+1)}.
\]
\end{lem}
\begin{IEEEproof}
To ease the notation, let us denote $V_{i}:=X_{t_{i}}$. Then, 
\begin{align}
 & \P\left[V_{0}\mid V_{-\underline{r}},\ldots,V_{-1},V_{1},\ldots,V_{\overline{r}}\right]\nonumber \\
 & \trre[=,a]\frac{\P\left[V_{-\underline{r}},\ldots,V_{-1},V_{0},V_{1},\ldots,V_{\overline{r}}\right]}{\P\left[V_{-\underline{r}},\ldots,V_{-1},V_{1},\ldots,V_{\overline{r}}\right]}\\
 & \trre[=,b]\frac{\P\left[V_{-\underline{r}},\ldots,V_{-1}\right]\P\left[V_{0}\mid V_{-\underline{r}},\ldots,V_{-1}\right]\P\left[V_{1},\ldots,V_{\overline{r}}\mid V_{0},V_{-\underline{r}},\ldots,V_{-1}\right]}{\P\left[V_{-\underline{r}},\ldots,V_{-1}\right]\P\left[V_{1},\ldots,V_{\overline{r}}\mid V_{-\underline{r}},\ldots,V_{-1}\right]}\\
 & =\frac{\P\left[V_{0}\mid V_{-\underline{r}},\ldots,V_{-1}\right]\P\left[V_{1},\ldots,V_{\overline{r}}\mid V_{0},V_{-\underline{r}},\ldots,V_{-1}\right]}{\P\left[V_{1},\ldots,V_{\overline{r}}\mid V_{-\underline{r}},\ldots,V_{-1}\right]}\\
 & =\P\left[V_{0}\mid V_{-\underline{r}},\ldots,V_{-1}\right]\cdot\prod_{i=1}^{\overline{r}}\frac{\P\left[V_{i}\mid V_{i-1},,\ldots,V_{1},V_{0},V_{-\underline{r}},\ldots,V_{-1}\right]}{\P\left[V_{i}\mid V_{i-1},,\ldots,V_{1},V_{-\underline{r}},\ldots,V_{-1}\right]}\\
 & \trre[\le,c]\left(\P[V_{0}]+d_{\text{min}}\right)\cdot\prod_{i=1}^{\overline{r}}\left(\frac{\P[V_{i}]+d_{\text{min}}}{\P[V_{i}]-d_{\text{min}}}\right)\\
 & \trre[\le,d]\P[V_{0}]\left(1+\overline{d}\right)\cdot\prod_{i=1}^{\overline{r}}\left(1+2\overline{d}\right)\\
 & \leq\P[V_{0}]\left(1+2\overline{d}\right)^{\overline{r}+1}\\
 & \trre[\le,e]\P[V_{0}]\cdot e^{2\overline{d}(\overline{r}+1)},
\end{align}
where $(a)$ and $(b)$ follow from Bayes rule, $(c)$ follows from
the definition of the strong mixing coefficient $d(s)$ and its assumed
monotonicity, as well as the assumption $\P[V_{i}]\geq p_{\text{min}}\geq2d_{\text{min}}$,
$(d)$ follows since for $0<a\leq\frac{1}{2}$ it holds that $\frac{1}{1-a}\leq1+2a$
and using this for $a=\frac{d_{\text{min}}}{\P[V_{1}]}\leq\frac{1}{2}$,
and $(e)$ from $1+a\leq e^{a}$ for $a\in\mathbb{R}$. 
\end{IEEEproof}
Using Lemma \ref{lem: probability condining on both past and future}
we may now prove Lemma \ref{lem: matching probability}.
\begin{IEEEproof}[Proof of Lemma \ref{lem: matching probability}]
Let arbitrary $t\ge t^{*}(n)=\frac{(1+\eta)}{{\cal {\cal H}}_{1}(\mathbf{X})}\log n$
and $1\leq s\leq\ell$ be given. We let $p_{\text{min}}:=\min_{x\in\supp({\cal X})}P_{X_{1}}(x)$.
Since $\sum_{s=0}^{\infty}d(s)<\infty$ it holds that $d(s)=O(\frac{1}{s\cdot\log s})$.
Thus, there must exists $\tau\equiv\tau(n)\in\mathbb{N}_{+}$ which
satisfies that 
\[
k\equiv k(n)=\left\lfloor \frac{t^{*}(n)+1}{\tau(n)}\right\rfloor =\omega(\log\log n),
\]
and
\[
d(\tau(n))\cdot k(n)=o_{n}(1)
\]
(as an extreme illustrative example, if $d(s)=\frac{1}{s\cdot\log s}$
then we can take $\tau(n)=\frac{\log n}{(\log\log n)^{2}}$ so that
\begin{align}
d(\tau(n))\cdot k(n) & =O\left(\frac{1}{\tau(n)\cdot\log\tau(n)}\cdot\frac{\log n}{\tau(n)}\right)\\
 & =O\left(\frac{(\log\log n)^{3}}{\log n}\right)
\end{align}
and $k(n)=\Omega((\log\log n))^{2}$). Assume further for simplicity
that $k$ is even, otherwise we may substitute $k$ by $k-1$. 

We consider two cases. First, we consider the case $1\leq s\leq4\tau$
. The idea is to focus on a subset of matching pairs from $X_{1}^{t}=X_{1+s}^{t+s}$,
which are at sufficiently distant time indices. We have 
\begin{align}
 & \P\left[X_{1}^{t}=X_{1+s}^{t+s}\right]\nonumber \\
 & \trre[\leq,a]\P\left[\left\{ X_{1}=X_{1+s}\right\} \cap\left\{ X_{1+8\tau}=X_{1+8\tau+s}\right\} \cap\left\{ X_{1+16\tau}=X_{1+16\tau+s}\right\} \cap\cdots\cap\left\{ X_{1+k\tau}=X_{1+k\tau+s}\right\} \right]\\
 & =\P\left[X_{1}=X_{1+s}\right]\cdot\prod_{i=1}^{k/8}\P\left[X_{1+8i\tau}=X_{1+8i\tau+s}\,\middle\vert\,\bigcap_{j=0}^{i-1}\left\{ X_{1+8j\tau}=X_{1+8j\tau+s}\right\} \right]\\
 & \leq\prod_{i=1}^{k/8}\P\left[X_{1+8i\tau}=X_{1+8i\tau+s}\,\middle\vert\,\bigcap_{j=0}^{i-1}\left\{ X_{1+8j\tau}=X_{1+8j\tau+s}\right\} \right]\\
 & \trre[\leq,b]\prod_{i=1}^{k/8}\left(\P\left[X_{1+8i\tau}=X_{1+8i\tau+s}\right]+d(8\tau-s)\right)\\
 & \trre[\leq,c]\prod_{i=1}^{k/8}\left(\P\left[X_{1+8i\tau}=X_{1+8i\tau+s}\right]+d(4\tau)\right)\\
 & \trre[=,d]\left(\P\left[X_{1}=X_{1+s}\right]+d(4\tau)\right)^{k/8}\\
 & \trre[\leq,e]\left(\P\left[X_{1}=X_{1+s}\right]+o_{n}(1)\right)^{k/8}\\
 & \trre[=,f]o\left(\frac{1}{\log n}\right),\label{eq: matching probability Markov low delay 1}
\end{align}
where $(a)$ follows by only taking into account symbol pair matching
events at distant $8\tau$ apart, $(b)$ follows from the definition
of the strong mixing coefficient, and since the events $\{X_{1+2i\tau}=X_{1+2i\tau+s}\}$
and $\bigcap_{j=0}^{i-1}\{X_{1+2j\tau}=X_{1+2j\tau+s}\}$ are $8\tau-s$
time points apart (obtained at the symbols and $X_{1+2(i-1)\tau+s}$
and $X_{1+2i\tau}$, $(c)$ follows from the assumed monotonicity
of $d(s)$ w.l.o.g., and since $s\le4\tau$, $(d)$ follows from stationarity,
$(e)$ follows since $d(\tau(n))=o_{n}(1)$, and $(f)$ follows since
by assumption $\sup_{s\in\mathbb{N}_{+}}\P[X_{1}=X_{1+s}]<1$ and
as $k\equiv k(n)=\omega(\log\log n)$. 

Second, we consider the case $4\tau\leq s\leq\ell$. Let $m=\lfloor\frac{s}{\tau}\rfloor$,
where $m\geq4$. The idea is again to focus on a subset of matching
pairs from the event $\{X_{1}^{t}=X_{1+s}^{t+s}\}$. In principle,
as in the previous case, we would like to choose them to be $\tau$
apart (or a multiple of $\tau$), to allow for approximate independence
between these events. A natural candidate for a loosening of the event
$\overline{{\cal E}}:=\{X_{1}^{t}=X_{1+s}^{t+s}\}$ is the subset
\begin{equation}
{\cal E}:=\bigcap_{i=0}^{k}{\cal E}_{i}\label{eq: pair matching at distant tau}
\end{equation}
where 
\[
{\cal E}_{i}:=\left\{ X_{1+i\tau}=X_{1+i\tau+s}\right\} .
\]
However, for the case $4\tau\leq s\leq\ell$, this does not guarantee
that the symbols involved in the definition of ${\cal E}$ are $\tau$
apart. Indeed, suppose, for example, that $m_{0}=\frac{s}{\tau}$
is integer. Then, the end symbol in the matching of ${\cal E}_{0}$,
i.e., $X_{1+s}$ is equal to the start symbol in the matching of ${\cal E}_{m_{0}}$,
to wit, $X_{1+m_{0}\tau}$, and thus they are definitely not $\tau$
symbols apart. To circumvent this, we will modify the event (\ref{eq: pair matching at distant tau})
by diluting some of the events ${\cal E}_{i}$, and consider the event
\begin{equation}
\tilde{{\cal E}}:=\bigcap_{i\in{\cal I}}\left\{ X_{1+i\tau}=X_{1+i\tau+s}\right\} \subset{\cal E},\label{eq: pair matching at distant tau modified}
\end{equation}
where ${\cal I}\subset\{0\}\cup[k]$ is a retained set of indices.
Specifically, we would like to guarantee \emph{any} two symbols that
are a part of the definition the even $\tilde{{\cal E}}$ to wit,
$\{X_{1+i\tau},X_{1+i\tau+s}\}_{i\in{\cal I}}$ are at least $\tau$
time points apart. This can be seen to be achieved by the following
choice. First, we include the events $\{{\cal E}_{j}\}_{j\in\{0,1,\ldots,m-1\}}$,
but exclude ${\cal E}_{m}$ and ${\cal E}_{m+1}$, as their start
symbol indices, i.e., $1+m\tau,1+(m+1)\tau$, resp., may be too close
to the final symbol index $1+s$ of ${\cal E}_{0}$. For a similar
reason, we exclude ${\cal E}_{m+2}$ (closeness to ${\cal E}_{1}$),
and so on. The index of the smallest event ${\cal E}_{i}$ that can
be included while satisfying the requirements, to wit, whose start
symbol index is assured to be more than $\tau$ apart from the final
symbol index of ${\cal E}_{m-1}$, is ${\cal E}_{2m+1}$. Similarly
to the inclusion of $\{{\cal E}_{i}\}_{\{0,1,\ldots,m-1\}}$, we may
also include all the $m-2$ following events, that is, $\{{\cal E}_{j}\}_{j\in\{2m+2,\ldots,3m-1\}}$.
See Fig. \ref{fig:Time-indices-of-pairs} for an illustration. 
\begin{figure}
\begin{centering}
\includegraphics[scale=1.25]{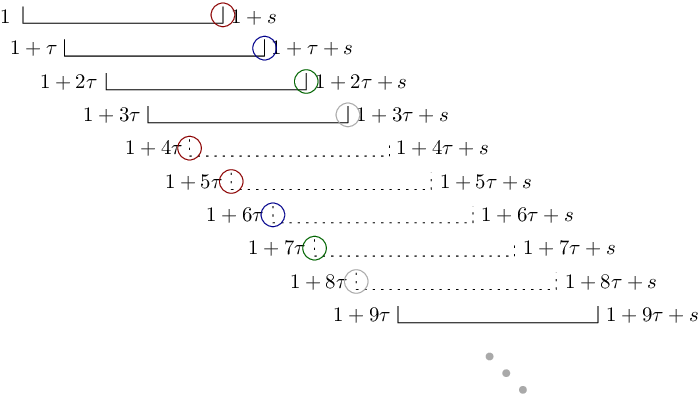}
\par\end{centering}
\caption{Time indices of pair matching of the modified event (\ref{eq: pair matching at distant tau modified}).
Here $m=4$, and the symbols $X_{1+4\tau}$ and $X_{1+5\tau}$ are
less than $\tau$ time points from $X_{1+s}$, thus their corresponding
events are removed from the event in (\ref{eq: pair matching at distant tau})
(appearing in dash points in the figure). Symbols with time indices
that are less than $\tau$ time points are marked by a circle of a
similar color. \label{fig:Time-indices-of-pairs}}
\end{figure}
 Generalizing the above considerations, traversing from $i=0$ to
$i=k$, we allow inclusion of a sequence of at least $m-1$ events,
followed by exclusion of the following $m+1$ events, inclusion of
at least $m-1$ events, and so on. The cardinality of the retained
set of indices is thus at least $|{\cal I}|\geq\frac{k}{3}$. With
this in mind, we bound 
\begin{align}
 & \P\left[X_{1}^{t}=X_{1+s}^{t+s}\right]\nonumber \\
 & \trre[\leq,a]\P\left[\bigcap_{i\in{\cal I}}\left\{ X_{1+i\tau}=X_{1+i\tau+s}\right\} \right]\\
 & \trre[=,b]\prod_{i\in{\cal I}}\P\left[X_{1+i\tau}=X_{1+i\tau+s}\,\middle\vert\,\bigcap_{j\in{\cal I}\colon j\leq i-1}\left\{ X_{1+j\tau}=X_{1+j\tau+s}\right\} \right]\\
 & \trre[=,c]\prod_{i\in{\cal I}}\sum_{x\in{\cal X}}\P\left[X_{1+i\tau+s}=x\,\middle\vert\,\bigcap_{j\in{\cal I}\colon j\leq i-1}\left\{ X_{1+j\tau}=X_{1+j\tau+s}\right\} \right]\times\nonumber \\
 & \hphantom{==\prod_{i\in{\cal I}}\sum_{x\in{\cal X}}}\P\left[X_{1+i\tau}=x\,\middle\vert\,\bigcap_{j\in{\cal I}\colon j\leq i-1}\left\{ X_{1+j\tau}=X_{1+j\tau+s}\right\} \cap\left\{ X_{1+i\tau+s}=x\right\} \right]\\
 & \trre[\leq,d]\prod_{i\in{\cal I}}\sum_{x\in{\cal X}}\P\left[X_{1+i\tau+s}=x\right]\P\left[X_{1+i\tau}=x\right]e^{\frac{d(\tau)}{p_{\text{min}}}(2k+1)}\\
 & \trre[=,e]\prod_{i\in{\cal I}}\exp_{|{\cal X}|}\left[\log(e)\cdot\frac{d(\tau)}{p_{\text{min}}}(2k+1)-H_{2}(P_{X_{1}})\right]\\
 & \trre[\leq,f]\exp_{|{\cal X}|}\left[-\frac{k}{4}\left(H_{2}(P_{X_{1}})-\log(e)\cdot\frac{d(\tau)}{p_{\text{min}}}(2k+1)\right)\right]\\
 & \trre[\leq,g]\exp_{|{\cal X}|}\left[-\frac{k}{8}\cdot H_{2}(P_{X_{1}})\right]\\
 & \trre[=,g]o\left(\frac{1}{\log n}\right),\label{eq: matching probability Markov low delay 2}
\end{align}
where $(a)$ follows by considering the probability ${\cal E}\subset{\cal \overline{E}}$
in (\ref{eq: pair matching at distant tau modified}), $(b)$ follows
from Bayes rule, $(c)$ follows again from Bayes rule, $(d)$ follows
due to the following bounds on the two probabilities: The first probability
is bounded as 
\begin{align}
\P\left[X_{1+i\tau+s}=x\,\middle\vert\,\bigcap_{j\in{\cal I}\colon j\leq i-1}\left\{ X_{1+j\tau}=X_{1+j\tau+s}\right\} \right] & \leq\P\left[X_{1+i\tau+s}=x\right]+d(\tau)\\
 & \leq\P\left[X_{1+i\tau+s}=x\right]\left[1+\frac{d(\tau)}{p_{\text{min}}}\right]\\
 & \leq\P\left[X_{1+i\tau+s}=x\right]\cdot e^{\frac{d(\tau)}{p_{\text{min}}}},
\end{align}
since the intersection $\bigcap_{j\in{\cal I}\colon j\leq i-1}\{X_{1+j\tau}=X_{1+j\tau+s}\}\in\sigma(X_{-\infty}^{1+(i-1)\tau+s})$,
that is, it is at distance of $\tau$ times points from $X_{1+i\tau+s}$.
The second probability is bounded as 
\begin{align}
 & \P\left[X_{1+i\tau}=x\,\middle\vert\,\bigcap_{j\in{\cal I}\colon j\leq i-1}\left\{ X_{1+j\tau}=X_{1+j\tau+s}\right\} \cap\left\{ X_{1+i\tau+s}=x\right\} \right]\nonumber \\
 & \leq\P\left[X_{1+i\tau}=x\right]e^{2\frac{d(\tau)}{p_{\text{min}}}|{\cal I}|}\\
 & \leq\P\left[X_{1+i\tau}=x\right]e^{2\frac{d(\tau)}{p_{\text{min}}}k},
\end{align}
using Lemma \ref{lem: probability condining on both past and future},
as the conditioning event involve symbols which are at least $\tau$
time points apart from $X_{1+i\tau}$ (either to the past or the future),
$(e)$ follows from stationarity 
\[
\sum_{x\in{\cal X}}\P\left[X_{1+i\tau+s}=x\right]\P\left[X_{1+i\tau}=x\right]=\sum_{x\in{\cal X}}P_{X_{1}}^{2}(x)=|{\cal X}|^{-H_{2}(P_{X_{1}})},
\]
and $1+a\leq e^{a}=|{\cal X}|^{\log(e)\cdot a}$, $(f)$ follows since
there are at least $|{\cal I}|\ge\frac{k}{3}$ terms in the product,
and $(g)$ follows since $d(\tau(n))\cdot k(n)=o_{n}(1)$ while $k(n)=\omega(\log\log n)$. 

Combining the two cases (\ref{eq: matching probability Markov low delay 1})
and (\ref{eq: matching probability Markov low delay 2}), and noting
that the bounds do not depend on $t$ and $s$, proves the claim of
the lemma. 
\end{IEEEproof}
With Lemma \ref{lem: partial probability power sum upper bound} and
Lemma \ref{lem: matching probability} at hand, may now prove Theorem
\ref{thm: exponent upper bound noiseless}. 
\begin{IEEEproof}[Proof of Theorem \ref{thm: exponent upper bound noiseless}]
We first consider the case that $\beta\leq\frac{1}{{\cal H}_{1}(\mathbf{X})}$,
which is simple to analyze. Indeed, consider a trivial detector that
always outputs $T=0$. This detector will err if $T\neq0$, which
happens with probability $\frac{2\ell-1}{n}=\frac{2\beta\log n-1}{n}$.
Hence, 
\[
P_{\text{error}}^{*}\leq\frac{2\ell}{n}=2\beta\cdot\frac{\log n}{n}.
\]
The rest of the proof is devoted to the case that $\beta>\frac{1}{{\cal H}_{1}(\mathbf{X})}$
and to prove the bound
\[
P_{\text{error}}^{*}\leq[1+o_{n}(1)]\cdot\frac{2}{{\cal H}_{1}(\mathbf{X})}\cdot\frac{\log n}{n},
\]
under the assumptions of the theorem, thus establishing the upper
bound on the error probability in (\ref{eq: upper bound on the error probability noiseless}).
To this end, let 
\[
t^{*}:=\frac{\log n}{{\cal H}_{1}(\mathbf{X})},
\]
let $\eta>0$ be given and set 
\[
t_{\eta}^{*}:=\lceil(1+2\eta)t^{*}\rceil=\left\lceil (1+2\eta)\frac{\log n}{{\cal H}_{1}(\mathbf{X})}\right\rceil ,
\]
where $\eta$ is also small enough so that $t_{\eta}^{*}\leq\ell$
(we will eventually take $\eta\downarrow0$). The idea is to use the
detector of Prop. \ref{prop: Bayes estimator noiseless memoryless},
which is possibly sub-optimal when $\mathbf{X}$ has memory. To slightly
simplify the analysis, we modify this detector further, and analyze
the detector
\begin{equation}
\overline{T}:=\begin{cases}
\hat{T}, & |\hat{T}|>t_{\eta}^{*}\\
0, & |\hat{T}|\leq t_{\eta}^{*}
\end{cases}.\label{eq: sub-optimal detector}
\end{equation}
The error probability of this sub-optimal detector is clearly an upper
bound on the error probability of the MAP detector. 

Let us decompose the error probability $\overline{T}$ as
\begin{equation}
P_{\text{error}}(\overline{T})\leq\sum_{t\in{\cal T}}P_{T}(t)\cdot\P\left[\overline{T}\neq T\mid T=t\right],\label{eq: decomposition of Bayesian error}
\end{equation}
so the summation over $t\in{\cal T}$ can be partitioned into the
following five index sets: $-\{[\ell-1]\backslash[t_{\eta}^{*}]\}$,
$\{-[t_{\eta}^{*}]\}$, $\{0\}$, $[t_{\eta}^{*}]$, and $[\ell]\backslash[t_{\eta}^{*}]$.
We next separately analyze each of them. 

\uline{Analysis of the type-I error probability:} 

We begin with the set $\{0\}$, that is, analyzing the type-I error
probability -- the probability of a wrong detection in case $T=0$
to $\{\overline{T}\neq0\}=\{\overline{T}>0\}\cup\{\overline{T}<0\}$.
For $\overline{T}>0$, we bound for all $n$ sufficiently large,
\begin{align}
 & \P\left[\overline{T}>0\mid T=0\right]\nonumber \\
 & \trre[=,a]\P\left[\bigcup_{t\in[\ell]\backslash[t_{\eta}^{*}]}\left\{ X_{\ell-t+1}^{\ell}(1)=X_{1}^{t}(2)\right\} \cap\left\{ \frac{1}{P_{X_{1}^{t}}\left(X_{1}^{t}(2)\right)}\geq n_{\ell}\right\} \,\middle\vert\,T=0\right]\\
 & \trre[\leq,b]\sum_{t\in[\ell]\backslash[t_{\eta}^{*}]}\P\left[\left\{ X_{\ell-t+1}^{\ell}(1)=X_{1}^{t}(2)\right\} \cap\left\{ \frac{1}{P_{X_{1}^{t}}\left(X_{1}^{t}(2)\right)}\geq n_{\ell}\right\} \,\middle\vert\,T=0\right]\\
 & \leq\sum_{t\in[\ell]\backslash[t_{\eta}^{*}]}\sum_{k=-(n-1)}^{n-1}\P\left[I(1)-I(2)=k\mid T=0\right]\times\nonumber \\
 & \hphantom{=\sum_{t\in[\ell]\backslash[t_{\eta}^{*}]}\sum_{k=-n}^{n}}\P\left[\left\{ X_{\ell-t+I(1)}^{\ell+I(1)-1}=X_{I(2)}^{t+I(2)-1}\right\} \cap\left\{ \frac{1}{P_{X_{I(2)}^{t+I(2)-1}}\left(X_{I(2)}^{t+I(2)-1}\right)}\geq n_{\ell}\right\} \,\middle\vert\,I(1)-I(2)=k,T=0\right]\\
 & \trre[=,c]\sum_{t\in[\ell]\backslash[t_{\eta}^{*}]}\sum_{k=-n}^{n}\P\left[I(1)-I(2)=k\mid T=0\right]\P\left[\left\{ X_{\ell-t+k+1}^{\ell+k}=X_{1}^{t}\right\} \cap\left\{ \frac{1}{P_{X_{1}^{t}}\left(X_{1}^{t}\right)}\geq n_{\ell}\right\} \right]\\
 & \trre[\leq,d]\frac{1}{n}\sum_{t\in[\ell]\backslash[t_{\eta}^{*}]}\sum_{k=-n}^{-\ell}\P\left[\left\{ X_{\ell-t+k+1}^{\ell+k}=X_{1}^{t}\right\} \cap\left\{ \frac{1}{P_{X_{1}^{t}}\left(X_{1}^{t}\right)}\geq n_{\ell}\right\} \right]\nonumber \\
 & \hphantom{===}+\frac{1}{n}\sum_{t\in[\ell]\backslash[t_{\eta}^{*}]}\sum_{k=\ell}^{n}\P\left[\left\{ X_{\ell-t+k+1}^{\ell+k}=X_{1}^{t}\right\} \cap\left\{ \frac{1}{P_{X_{1}^{t}}\left(X_{1}^{t}\right)}\geq n_{\ell}\right\} \right],\label{eq: Simple case cond err prob T zero first}
\end{align}
where $(a)$ follows from the detection rule of Prop. \ref{prop: Bayes estimator noiseless memoryless}
which states that an error occurs if there is a matching overlapping
fragment of length larger than $t_{{\scriptscriptstyle \text{MDO}}}$,
and it has small enough probability, as well as the modification in
(\ref{eq: sub-optimal detector}), which results that an error to
$0<\overline{T}\leq t_{\eta}^{*}$ is impossible, $(b)$ follows from
the union bound, $(c)$ follows from stationarity, and $(d)$ follows
since given $T=0$, $\P[I(1)-I(2)=k\mid T=0]>0$ only if either $\ell\le k\leq n$
or $-n\le k\leq-\ell$, and otherwise it follows from the conditional
distribution of $I(2)$ given $I(1)$ in (\ref{eq: I(2) conditioned on I(1)})
that 
\[
\P\left[I(1)-I(2)=k\mid T=0\right]\leq\frac{1}{n}.
\]
Let us continue by upper bounding the first double summation in (\ref{eq: Simple case cond err prob T zero first}),
and the second double summation can be similarly upper bounded. To
this end, let $\{\tau(n)\}_{n\in\mathbb{N}}$ be a monotonic non-decreasing
sequence of integers such that $\tau(n)=\omega_{n}(1)$ and $\tau(n)=o(\log n)$.
For example, $\tau(n)=\lceil\sqrt{\log n}\rceil$ is a suitable choice.
For brevity, let us next denote $\tau\equiv\tau(n)$. We now bound
the first double sum in (\ref{eq: Simple case cond err prob T zero first})
as
\begin{align}
 & \frac{1}{n}\sum_{t\in[\ell]\backslash[t_{\eta}^{*}]}\sum_{k=-n}^{-\ell}\P\left[\left\{ X_{\ell-t+k+1}^{\ell+k}=X_{1}^{t}\right\} \cap\left\{ \frac{1}{P_{X_{1}^{t}}\left(X_{1}^{t}\right)}\geq n_{\ell}\right\} \right]\nonumber \\
 & \trre[\leq,a]\frac{1}{n}\sum_{t\in[\ell]\backslash[t_{\eta}^{*}]}\sum_{k=-n}^{-\ell}\P\left[\left\{ X_{\ell-t+k+\tau}^{\ell+k}=X_{\tau}^{t}\right\} \cap\left\{ \frac{1}{P_{X_{1}^{t}}\left(X_{1}^{t}\right)}\geq n_{\ell}\right\} \right]\\
 & \trre[\leq,b]\frac{1}{n}\sum_{t\in[\ell]\backslash[t_{\eta}^{*}]}\sum_{k=-n}^{-\ell}\P\left[\left\{ X_{\ell-t+k+\tau}^{\ell+k}=X_{\tau}^{t}\right\} \cap\left\{ \frac{1}{P_{X_{\tau}^{t}}\left(x_{\tau}^{t}\right)}\geq\rho^{\tau-1}\cdot n_{\ell}\right\} \right]\\
 & =\frac{1}{n}\sum_{t\in[\ell]\backslash[t_{\eta}^{*}]}\sum_{k=-n}^{-\ell}\sum_{x_{\tau}^{t}\in{\cal X}^{t-\tau+1}}P_{X_{\ell-t+k+\tau}^{\ell+k}}\left(x_{\tau}^{t}\right)\P\left[X_{\tau}^{t}=x_{\tau}^{t}\mid X_{\ell-t+k+\tau}^{\ell+k}=x_{\tau}^{t}\right]\cdot\I\left[\frac{1}{P_{X_{\tau}^{t}}\left(x_{\tau}^{t}\right)}\geq\rho^{\tau-1}\cdot n_{\ell}\right]\\
 & \trre[\leq,c]\frac{1}{n}\sum_{t\in[\ell]\backslash[t_{\eta}^{*}]}\sum_{k=-n}^{-\ell}\sum_{x_{1}^{t-\tau+1}\in{\cal X}^{t-\tau+1}}P_{X_{1}^{t-\tau+1}}^{2}\left(x_{1}^{t-\tau+1}\right)\I\left[\frac{1}{P_{X_{1}^{t-\tau+1}}(x_{1}^{t-\tau+1})}\geq\rho^{\tau-1}\cdot n_{\ell}\right]\nonumber \\
 & \hphantom{==}+\frac{1}{n}\sum_{t\in[\ell]\backslash[t_{\eta}^{*}]}\sum_{k=-n}^{-\ell}\sum_{x_{1}^{t-\tau+1}\in{\cal X}^{t-\tau+1}}P_{X_{1}^{t-\tau+1}}\left(x_{1}^{t-\tau+1}\right)d\left(\tau-\ell-k\right)\cdot\I\left[\frac{1}{P_{X_{1}^{t-\tau+1}}(x_{1}^{t-\tau+1})}\geq\rho^{\tau-1}\cdot n_{\ell}\right],\label{eq: Simple case cond err prob T zero second}
\end{align}
where $(a)$ follows by slightly shortening the required match length
by $\tau+1$ symbols, $(b)$ follows from the assumption ${\cal H}_{-\infty}(\mathbf{X})<\infty$,
which assures that for any $\rho<|{\cal X}|^{-{\cal H}_{-\infty}(\mathbf{X})}<1$
there exists $n_{0}$ such that for all $n\geq n_{0},$ 
\begin{align}
\frac{1}{P_{X_{1}^{t}}\left(x_{1}^{t}\right)} & \leq\frac{1}{P_{X_{\tau}^{t}}\left(x_{\tau}^{t}\right)\cdot\min_{\tilde{x}_{1}^{\tau-1}\in\supp\left[P_{X_{1}^{\tau-1}}\right]}P_{X_{1}^{\tau-1}}(\tilde{x}_{1}^{\tau-1})}\\
 & \leq\frac{1}{P_{X_{\tau}^{t}}\left(x_{\tau}^{t}\right)}\left(\frac{1}{\rho}\right)^{\tau-1},
\end{align}
$(c)$ follows from the definition of the strong mixing coefficient
(Definition \ref{def: strong mixing coefficient}), which implies
that 
\[
\P\left[X_{\tau}^{t}=x_{\tau}^{t}\mid X_{\ell-t+k+\tau}^{\ell+k}=x_{\tau}^{t}\right]\leq\P\left[X_{\tau}^{t}=x_{\tau}^{t}\right]+d\left(\tau-\ell-k\right)
\]
and utilizing stationarity to replace $P_{X_{\ell-t+k+1+\tau}^{\ell+k}}=P_{X_{\tau}^{t}}=P_{X_{1}^{t-\tau+1}}$.
We now consider each of the triple summations in (\ref{eq: Simple case cond err prob T zero second}).
The first triple summation is upper bounded as
\begin{align}
 & \frac{1}{n}\sum_{t\in[\ell]\backslash[t_{\eta}^{*}]}\sum_{k=-n}^{-\ell}\sum_{x_{1}^{t-\tau+1}\in{\cal X}^{t-\tau+1}}P_{X_{1}^{t-\tau+1}}^{2}\left(x_{1}^{t-\tau+1}\right)\I\left[\frac{1}{P_{X_{1}^{t-\tau+1}}(x_{1}^{t-\tau+1})}\geq\rho^{\tau-1}\cdot n_{\ell}\right]\nonumber \\
 & \leq\ell\cdot\max_{t_{\eta}^{*}\le t\le\ell}\sum_{x_{1}^{t-\tau+1}\in{\cal X}^{t-\tau+1}}P_{X_{1}^{t-\tau+1}}^{2}\left(x_{1}^{t-\tau+1}\right)\I\left[\frac{1}{P_{X_{1}^{t-\tau+1}}(x_{1}^{t-\tau+1})}\geq\rho^{\tau-1}\cdot n_{\ell}\right]\\
 & \trre[=,a]\ell\cdot o\left(\frac{1}{n}\right)\\
 & \trre[=,b]o\left(\frac{\log n}{n}\right),\label{eq: Simple case cond err prob T zero third A}
\end{align}
where $(a)$ follows from (\ref{eq: partial second-order Renyi entropy upper bound})
in Lemma \ref{lem: partial probability power sum upper bound}, by
first replacing $t\leftrightarrow t-\tau+1$ and noting that for all
$t\geq t_{\eta}^{*}=(1+2\eta)t^{*}$ the assumption $\tau=o(\log n)=o(t_{\eta}^{*})$
implies that 
\begin{align}
t-\tau+1 & \geq(1+2\eta)t_{\eta}^{*}-\tau+1\\
 & \geq(1+\eta)t_{\eta}^{*}\\
 & =\frac{(1+\eta)}{{\cal {\cal H}}_{1}(\mathbf{X})}\log n,
\end{align}
and $(b)$ follows since $\ell=\beta\log n$. The second triple summation
in (\ref{eq: Simple case cond err prob T zero second}) is upper bounded
as 
\begin{align}
 & \frac{1}{n}\sum_{t\in[\ell]\backslash[t_{\eta}^{*}]}\sum_{k=-n}^{-\ell}\sum_{x_{1}^{t-\tau+1}\in{\cal X}^{t-\tau+1}}P_{X_{1}^{t-\tau+1}}\left(x_{1}^{t-\tau+1}\right)d\left(\tau-\ell-k\right)\cdot\I\left[\frac{1}{P_{X_{1}^{t-\tau+1}}(x_{1}^{t-\tau+1})}\geq\rho^{\tau-1}\cdot n_{\ell}\right]\nonumber \\
 & \leq\frac{1}{n}\sum_{t\in[\ell]\backslash[t_{\eta}^{*}]}\sum_{k=-n}^{-\ell}d\left(\tau-\ell-k\right)\sum_{x_{1}^{t-\tau+1}\in{\cal X}^{t-\tau+1}}P_{X_{1}^{t-\tau+1}}\left(x_{1}^{t-\tau+1}\right)\\
 & =\frac{1}{n}\sum_{t\in[\ell]\backslash[t_{\eta}^{*}]}\sum_{k=-n}^{-\ell}d\left(\tau-\ell-k\right)\\
 & \leq\frac{\ell}{n}\cdot\sum_{s=\tau}^{\infty}d(s)\\
 & \trre[=,a]\frac{\ell}{n}\cdot o_{n}(1)\\
 & \trre[=,b]o\left(\frac{\log n}{n}\right),\label{eq: Simple case cond err prob T zero third B}
\end{align}
where $(a)$ follows since the assumption $\sum_{s=0}^{\infty}d(s)<\infty$
of the theorem, combined with $\tau\equiv\tau(n)=\omega_{n}(1)$ implies
that $\sum_{s=\tau(n)}^{\infty}d(s)\to0$ as $n\to\infty$, and $(b)$
follows again since $\ell=\beta\log n$. Substituting (\ref{eq: Simple case cond err prob T zero third A})
and (\ref{eq: Simple case cond err prob T zero third B}) into (\ref{eq: Simple case cond err prob T zero second})
results that the first double summation in (\ref{eq: Simple case cond err prob T zero first})
is 
\[
\frac{1}{n}\sum_{t\in[\ell]\backslash[t_{\eta}^{*}]}\sum_{k=-n}^{-\ell}\P\left[\left\{ X_{\ell-t+k+1}^{\ell+k}=X_{1}^{t}\right\} \cap\left\{ \frac{1}{P_{X_{1}^{t}}\left(X_{1}^{t}\right)}\geq n_{\ell}\right\} \right]=o\left(\frac{\log n}{n}\right).
\]
Similarly, the second double summation in (\ref{eq: Simple case cond err prob T zero first})
is also $o(\frac{\log n}{n})$, and so (\ref{eq: Simple case cond err prob T zero first})
implies that 
\[
\P\left[\overline{T}>0\mid T=0\right]=o\left(\frac{\log n}{n}\right).
\]
It can be proved using analogous arguments that
\[
\P\left[\overline{T}<0\mid T=0\right]=o\left(\frac{\log n}{n}\right)
\]
and so the type-I error probability is upper bounded as
\begin{equation}
\P\left[\overline{T}\neq0\mid T=0\right]=\P\left[\overline{T}>0\mid T=0\right]+\P\left[\overline{T}<0\mid T=0\right]=o\left(\frac{\log n}{n}\right).\label{eq: type I error probability upper bound noiseless}
\end{equation}

\uline{Analysis of the type-II error probability from} $T\in\{-[t_{\eta}^{*}]\}\cup[t_{\eta}^{*}]$:

In this case, we trivially bound the conditional error probability
by $1$ and use $P_{T}(t)=\frac{1}{n}$ to obtain that the contribution
of these sets to the error probability in (\ref{eq: decomposition of Bayesian error})
is upper bounded by 
\begin{equation}
\sum_{t\in\{-[t_{\eta}^{*}]\}\cup[t_{\eta}^{*}]}P_{T}(t)\cdot\P\left[\overline{T}\neq T\mid T=t\right]\leq\frac{2t_{\eta}^{*}}{n}.\label{eq: Simple case cond err prob T small overlaps bound}
\end{equation}

\uline{Analysis of the type-II error probability from} $T\in[\ell]\backslash[t_{\eta}^{*}]$:

In this case, an error may occur in one of three events: (1) An error
to $\overline{T}=0$, that is, deciding that the reads do not overlap.
(2) An error to a longer positive overlap $\overline{T}\in[\ell]\backslash[\tilde{t}]$.
(3) An error to a longer negative overlap $\overline{T}\in-\{[\ell-1]\backslash[t_{\eta}^{*}]\}$.
We next upper bound the probability of each of these error events,
and then use a union bound to upper bound $\P[\overline{T}\neq T\mid T=\tilde{t}]$. 

First, the error probability of an error from $T=\tilde{t}$ to $\overline{T}=0$,
is upper bounded as
\begin{align}
\P\left[\overline{T}=0\mid T=\tilde{t}\right] & \trre[\leq,a]\P\left[\frac{1}{P_{X_{1}^{\tilde{t}}}\left(X_{1}^{\tilde{t}}(1)\right)}\leq n_{\ell}\right]\\
 & =\sum_{x_{1}^{t}\in{\cal X}^{\otimes t}}P_{X_{1}^{t}}(x_{1}^{t})\cdot\I\left[\frac{1}{P_{X_{1}^{\tilde{t}}}\left(X_{1}^{\tilde{t}}(1)\right)}\leq n_{\ell}\right]\\
 & =o_{n}(1),\label{eq: noiseless error from positive to zero}
\end{align}
where $(a)$ follows since an error will occur only if $P_{X_{1}^{\tilde{t}}}^{-1}(X_{1}^{\tilde{t}}(1))\leq n_{\ell}$
because the reads are noiseless and so $X_{\ell-\tilde{t}+1}^{\ell}(1)=X^{\tilde{t}}(2)$
occurs with probability $1$, and $(b)$ follows from (\ref{eq: partial probability sum upper bound})
in Lemma \ref{lem: partial probability power sum upper bound} since
$\tilde{t}\geq t_{\eta}^{*}\geq\frac{(1+\eta)}{{\cal H}_{1}(\mathbf{X})}\log n$.
Second, the probability of an error from $T=\tilde{t}$ to $\overline{T}\in[\ell]\backslash[\tilde{t}]$
is upper bounded from the union bound as 
\begin{align}
\P\left[\overline{T}\in[\ell]\backslash[\tilde{t}]\mid T=\tilde{t}\right] & \leq\sum_{\overline{t}=\tilde{t}+1}^{\ell}\P\left[\overline{T}=\overline{t}\mid T=\tilde{t}\right]\\
 & \leq\ell\cdot\max_{\tilde{t}+1\leq\overline{t}\leq\ell}\P\left[\overline{T}=\overline{t}\mid T=\tilde{t}\right].\label{eq: noiseless error from positive to longer positive first}
\end{align}
Now, for such $\overline{t}$ which satisfies $\ell\geq\overline{t}\geq\tilde{t}+1$,
\begin{align}
 & \P\left[\overline{T}=\overline{t}\mid T=\tilde{t}\right]\nonumber \\
 & \trre[=,a]\P\left[\left\{ X_{\ell-\overline{t}+1}^{\ell}(1)=X_{1}^{\overline{t}}(2)\right\} \cap\left\{ \frac{1}{P_{X_{1}^{\overline{t}}}\left(X_{1}^{\overline{t}}(2)\right)}\geq n_{\ell}\right\} \,\middle\vert\,T=\tilde{t}\right]\\
 & \trre[\le,b]\P\left[X_{\ell-\overline{t}+1}^{\ell}(1)=X_{1}^{\overline{t}}(2)\mid T=\tilde{t}\right]\\
 & =\P\left[X_{\ell-\overline{t}+1+I(1)}^{\ell+I(1)}=X_{1+I(2)}^{\overline{t}+I(2)}\mid T=\tilde{t}\right]\\
 & \trre[=,c]\P\left[X_{\ell-\overline{t}+1+I(1)}^{\ell+I(1)}=X_{1+I(1)+\ell-\tilde{t}}^{\overline{t}+I(1)+\ell-\tilde{t}}\mid T=\tilde{t}\right]\\
 & \trre[=,d]\P\left[X_{1}^{\overline{t}}=X_{1+\overline{t}-\tilde{t}}^{2\overline{t}-\tilde{t}}\mid T=\tilde{t}\right]\\
 & \trre[=,e]\P\left[X_{1}^{\overline{t}}=X_{1+\overline{t}-\tilde{t}}^{2\overline{t}-\tilde{t}}\right]\\
 & \trre[=,f]o\left(\frac{1}{\log n}\right),\label{eq: noiseless error from positive to longer positive second}
\end{align}
where $(a)$ follows from the definition of the detection rule $\overline{T}$,
$(b)$ follows by relaxing the condition $\frac{1}{P_{X_{1}^{\overline{t}}}(X_{1}^{\overline{t}}(2))}\geq n_{\ell}$,
$(c)$ follows since for $T=\tilde{t}>0$ it holds that $I(2)=\ell-\tilde{t}-I(1)$,
$(d)$ follows from stationarity, $(e)$ follows from independence
of $T$ and $\{X_{i}\}_{i\in\mathbb{Z}}$, and $(f)$ follows from
Lemma \ref{lem: matching probability}, with the notation change $t\leftarrow\overline{t}$
and $s\leftarrow\overline{t}-\tilde{t}$. Substituting (\ref{eq: noiseless error from positive to longer positive second})
into (\ref{eq: noiseless error from positive to longer positive first})
implies that
\begin{equation}
\P\left[\overline{T}\in[\ell]\backslash[\tilde{t}]\mid T=\tilde{t}\right]=o_{n}(1).\label{eq: noiseless error from positive to longer positive third}
\end{equation}

Third, similarly to (\ref{eq: noiseless error from positive to longer positive first}),
the probability of an error from $T=\tilde{t}$ to $\overline{T}\in-\{[\ell-1]\backslash[\tilde{t}]\}$
is upper bounded from the union bound as
\begin{align}
\P\left[\overline{T}\in-\{[\ell-1]\backslash[\tilde{t}]\}\mid T=\tilde{t}\right] & \leq\sum_{\overline{t}=\tilde{t}+1}^{\ell}\P\left[\overline{T}=-\overline{t}\mid T=\tilde{t}\right]\\
 & \leq\ell\cdot\max_{\tilde{t}+1\leq\overline{t}\leq\ell}\P\left[\overline{T}=-\overline{t}\mid T=\tilde{t}\right].\label{eq: noiseless error from positive to longer negative first}
\end{align}
Now, for such $\ell\geq-\overline{t}\geq\tilde{t}+1$ we may analyze
similarly to (\ref{eq: noiseless error from positive to longer positive second}),
by replacing the role of $X(1)$ and $X(2)$, as:
\begin{align}
 & \P\left[\overline{T}=-\overline{t}\mid T=\tilde{t}\right]\nonumber \\
 & =\P\left[\left\{ X_{\ell-\overline{t}+1}^{\ell}(2)=X_{1}^{\overline{t}}(1)\right\} \cap\left\{ \frac{1}{P_{X_{1}^{\overline{t}}}\left(X_{1}^{\overline{t}}(1)\right)}\geq n_{\ell}\right\} \,\middle\vert\,T=\tilde{t}\right]\\
 & \leq\P\left[X_{\ell-\overline{t}+1}^{\ell}(2)=X_{1}^{\overline{t}}(1)\mid T=\tilde{t}\right]\\
 & =\P\left[X_{\ell-\overline{t}+1+I(2)}^{\ell+I(2)}=X_{1+I(1)}^{\overline{t}+I(1)}\mid T=\tilde{t}\right]\\
 & =\P\left[X_{\ell-\overline{t}+1+I(2)}^{\ell+I(2)}=X_{1+\ell-\tilde{t}-I(2)}^{\overline{t}+\ell-\tilde{t}-I(2)}\mid T=\tilde{t}\right]\\
 & =\P\left[X_{1+\tilde{t}-\overline{t}+2I(2)}^{\tilde{t}+2I(2)}=X_{1}^{\overline{t}}\mid T=\tilde{t}\right]\\
 & =\P\left[X_{1+\tilde{t}-\overline{t}+2I(2)}^{\tilde{t}+2I(2)}=X_{1}^{\overline{t}}\right]\\
 & =o\left(\frac{1}{\log n}\right),\label{eq: noiseless error from positive to longer negative second}
\end{align}
where the transitions follow from the same arguments as in (\ref{eq: noiseless error from positive to longer positive second}),
by replacing the role of $X(1)$ and $X(2)$ (while $I(2)=\ell-\tilde{t}-I(1)$
still holds). Substituting (\ref{eq: noiseless error from positive to longer negative second})
into (\ref{eq: noiseless error from positive to longer negative first})
implies that
\begin{equation}
\P\left[t_{\eta}^{*}\in-\{[\ell-1]\backslash[\tilde{t}]\}\mid T=\tilde{t}\right]=o_{n}(1).\label{eq: noiseless error from positive to longer negative third}
\end{equation}
Combining all three conditional error probabilities (\ref{eq: noiseless error from positive to zero})
(\ref{eq: noiseless error from positive to longer positive third})
and (\ref{eq: noiseless error from positive to longer negative third})
we obtain 
\[
\P\left[\overline{T}\neq T\mid T=\tilde{t}\right]=o_{n}(1).
\]
Hence, also 
\begin{equation}
\sum_{\tilde{t}\in[\ell]\backslash[t_{\eta}^{*}]}P_{T}(t)\cdot\P\left[\overline{T}\neq T\mid T=t\right]=o\left(\frac{\ell}{n}\right)=o\left(\frac{\log n}{n}\right).\label{eq: Simple case cond err prob T large positive}
\end{equation}
This completes the analysis of the contribution of $\tilde{t}\in[\ell]\backslash[t_{\text{\ensuremath{\eta}}}]$
to the error probability in the decomposition (\ref{eq: decomposition of Bayesian error}). 

\uline{Analysis of the type-II error probability from} $T=\tilde{t}\in-\{[\ell-1]\backslash[t_{\eta}^{*}]\}$:

This case can be analyzed in a complete analogous manner to the previous
case, and results, similarly to (\ref{eq: Simple case cond err prob T large positive}),
the upper bound
\begin{equation}
\sum_{\tilde{t}\in-\{[\ell-1]\backslash[t_{\eta}^{*}]\}}P_{T}(t)\cdot\P\left[\overline{T}\neq T\mid T=t\right]=o\left(\frac{\log n}{n}\right).\label{eq: Simple case cond err prob T large negative}
\end{equation}

\uline{Combining the different error probabilities}:

Substituting (\ref{eq: type I error probability upper bound noiseless}),
(\ref{eq: Simple case cond err prob T small overlaps bound}), (\ref{eq: Simple case cond err prob T large positive})
and (\ref{eq: Simple case cond err prob T large negative}) into (\ref{eq: decomposition of Bayesian error})
results 
\begin{align}
P_{\text{error}}^{*} & \leq P_{\text{error}}(\overline{T})\\
 & \leq\frac{2t_{\eta}^{*}}{n}+o\left(\frac{1}{n}\right)+o\left(\frac{\log n}{n}\right)\\
 & \leq2(1+2\eta)\cdot[1+o_{n}(1)]\cdot\frac{1}{{\cal H}_{1}(\mathbf{X})}\frac{\log n}{n}.
\end{align}
The proof of the upper bound (\ref{eq: upper bound on the error probability noiseless})
is then completed by taking $\eta\downarrow0$. This completes the
proof of Theorem \ref{thm: exponent upper bound noiseless}. 
\end{IEEEproof}

\subsection{Details for analysis of trade-off between two types of error \ref{exa: Markov noiseless}}

Suppose that we modify the detection rule (\ref{eq: optimal detection rule noiseless memoryless})
of Prop. \ref{prop: Bayes estimator noiseless memoryless} and make
the condition of overlap detection more stringent, with the goal of
reducing the type-I error probability to $o(\frac{\log n}{n^{q}})$
for some degree $q>1$. We can achieve this by replacing $n_{\ell}\sim n$
in the detection rule (\ref{eq: optimal detection rule noiseless memoryless})
of Prop. \ref{prop: Bayes estimator noiseless memoryless} by $n^{\mu}$
for some $\mu>1$. In turn, the minimal reliably detectable overlap
$t^{*}\sim\frac{\log n}{{\cal H}_{1}(\mathbf{X})}$ will now be replaced
by a modified parameter $t^{\#}\sim\frac{\log n}{h}$ for some $h<{\cal H}_{1}(\mathbf{X})$
which in turn will increase the dominating contribution of the type-II
error probability (misdetection of overlaps) to $\frac{1}{h}\cdot\frac{\log n}{n}$.
Our goal is thus to characterize the trade-off between $q$ and $h$
(and how they can be tuned by controlling $\mu$). The analysis of
the error probability of this modified detector, follows the same
events as in the analysis of the proof of Theorem \ref{thm: exponent upper bound noiseless}.
As the technical details can be easily discerned from that proof,
we only provide here a sketch of required modifications. Since the
contribution of error in detecting a wrong overlap to the type-II
error, that is, an error from $T=t$ to $\hat{T}=\tilde{t}$, is negligible
compared to the error from $T=t$ to $\hat{T}=0$ as long as $|t|>t^{*}$,
and since now $|t|>t^{\#}>t^{*}$ it remains to evaluate the error
probability from $T=0$ to $\hat{T}=t$ for some $t>t^{\#}$ and vice-versa.

First, the probability of an error from $T=0$ to $\hat{T}=t>0$,
previously bounded in (\ref{eq: specific type I error probability noiseless proof outline}),
will now be bounded as
\[
\P\left[\hat{T}=t\mid T=0\right]\lessapprox\sum_{x_{1}^{t}\in{\cal X}^{\otimes t}}P_{X_{1}^{t}}^{2}\left(x_{1}^{t}\right)\I\left[\frac{1}{P_{X_{1}^{t}}(x_{1}^{t})}\geq n^{\mu}\right].
\]
Following the proof of Lemma \ref{lem: partial probability power sum upper bound},
and ignoring terms which eventually vanish, this bound decays as $o(\frac{1}{n^{q}})$
where 
\[
q=\mu r-(r-1)\frac{{\cal H}_{2-r}(\mathbf{X})}{h}
\]
for some choice of $r>0$. Second, the probability of an error from
$T=t$ to $\hat{T}=0$, which was previously bounded in (\ref{eq: specific type II to zero error probability noiseless proof outline}),
will now be bounded as
\[
\P\left[\hat{T}=0\mid T=t\right]\lessapprox\sum_{x_{1}^{t}\in{\cal X}^{\otimes t}}P_{X_{1}^{t}}\left(x_{1}^{t}\right)\I\left[\frac{1}{P_{X_{1}^{t}}(x_{1}^{t})}\le n^{\mu}\right],
\]
and this is $o_{n}(1)$ only if 
\[
\left(1-r\right)\left[\frac{{\cal H}_{2-r}(\mathbf{X})}{h}-\mu\right]>0
\]
for some choice of $r\in(0,1)$. This last condition means that $\mu\leq\frac{{\cal H}_{2-r}(\mathbf{X})}{h}$.
Let us assume that the same parameter $r$ is used for both types
of error. Then, by choosing $\mu$ to its maximal value $\frac{{\cal H}_{2-r}(\mathbf{X})}{h}$,
the maximal degree of the type-I error upper bound is 
\[
q=\frac{{\cal H}_{2-r}(\mathbf{X})}{h}r-(r-1)\frac{{\cal H}_{2-r}(\mathbf{X})}{h}=\frac{{\cal H}_{2-r}(\mathbf{X})}{h},
\]
which could be at most $\mu$ for an optimized choice of $r$. Consequently,
we deduce that for any given choice of $\mu$ in the modified detector,
it holds that the contribution of the type-I error to the error probability
is 
\[
P_{T}(0)\cdot p_{\text{type-I}}(\hat{T})=o\left(\frac{\log n}{n^{\mu}}\right)
\]
whereas the contribution of the type-II error probability is 
\[
\sum_{t\in\{-[\ell-1]\}\cup[\ell]}\P[T=t]\cdot p_{\text{type-II}}(\hat{T};t)\sim\frac{\mu}{{\cal H}_{1}(\mathbf{X})}\frac{\log n}{n}.
\]
In words, the cost of improving the decay rate of the type-I error
probability to $o(\frac{\log n}{n^{\mu}})$ results an increase of
a factor of $\mu$ in the total error probability.

\subsection{Details for Example \ref{exa: Markov noiseless}}

We next provide details on the computation of the strong mixing coefficient
for the Markov example (Example \ref{exa: Markov noiseless}). We
may write the Markov kernel in a matrix form 
\[
K=\epsilon\cdot\boldsymbol{1}\boldsymbol{1}^{\top}+(1-|{\cal X}|\epsilon)\cdot\boldsymbol{I},
\]
where $\boldsymbol{1}^{\top}:=(1,1,\ldots,1)\in\mathbb{R}^{|{\cal X}|}$
is the all ones vector (here, of dimension $|{\cal X}|$), and $\boldsymbol{I}\in\mathbb{R}^{|{\cal X}|\times|{\cal X}|}$
is the identity matrix. Then, for an integer $s>1$, it is easy to
verify that 
\[
K^{s}=a_{s}\cdot\boldsymbol{1}\boldsymbol{1}^{\top}+b_{s}\cdot\boldsymbol{I},
\]
where $a_{1}=\epsilon$, $b_{1}=(1-|{\cal X}|\epsilon)$, and the
recursive formulas hold:
\begin{align}
a_{s+1} & =a_{s}+b_{s}\epsilon\\
b_{s+1} & =b_{s}\cdot\left(1-|{\cal X}|\epsilon\right).
\end{align}
Thus, $b_{s}=(1-|{\cal X}|\epsilon)^{s}$ and
\[
a_{s}=\epsilon\cdot\sum_{i=0}^{s-1}\left(1-|{\cal X}|\epsilon\right)^{i}=\epsilon\cdot\frac{1-\left(1-|{\cal X}|\epsilon\right)^{s}}{|{\cal X}|\epsilon}=\frac{1}{|{\cal X}|}\left[1-\left(1-|{\cal X}|\epsilon\right)^{s}\right].
\]
Due to the symmetry between the letters, the mixing coefficient is
determined as the total variation distance between $\Pi_{X}$ and
any of the rows of $K^{s}$, say the first, given by $(a_{s}+b_{s},a_{s},a_{s},\ldots,a_{s})$.
So, since $a_{s}\leq\frac{1}{|{\cal X}|}$ it must hold that $a_{s}+b_{s}\geq\frac{1}{|{\cal X}|}$.
Hence, 
\begin{align}
d(s) & =\max_{x\in{\cal X}}\left\Vert K^{s}(x,\cdot)-P_{X_{1}}\right\Vert _{\text{TV}}\\
 & =\frac{1}{2}\left(a_{s}+b_{s}-\frac{1}{|{\cal X}|}\right)+\frac{1}{2}(|{\cal X}|-1)\left(\frac{1}{|{\cal X}|}-a_{s}\right)\\
 & =\frac{|{\cal X}|-1}{|{\cal X}|}\left(1-|{\cal X}|\epsilon\right)^{s}\\
 & \leq\frac{|{\cal X}|-1}{|{\cal X}|}\cdot e^{-|{\cal X}|\epsilon\cdot s}.
\end{align}

\subsection{Proof of Theorem \ref{thm: exponent lower bound noiseless}}

We next turn to the proof of Theorem \ref{thm: exponent lower bound noiseless}.
To this end, recall that in case the process has memory, the likelihoods
for both $T=t\neq0$ and $T=0$ do not take the simple form of the
memoryless case (Prop. \ref{prop: Bayes estimator noiseless memoryless}).
Nonetheless, we show that these likelihoods are sufficiently tight
approximations. We show this separately for the case $T=t\ne0$ and
then for $T=0$. To this end, will use the following known result
from ergodic theory, which is an easy consequence of the AEP (see,
for example, \cite[Lemma 3.4.1.]{gray2011entropy}). We provide for
completeness the simple proof. 
\begin{lem}
\label{lem: asymptotic independence lemma AEP}Let $\boldsymbol{X}$
be a stationary and ergodic process. Let $m\in\mathbb{N}_{+}$ and
$k\equiv k(m)$ be such that $k(m)=\omega_{m}(1)$. Then, 
\[
\frac{1}{m+k}\log\frac{P_{X_{1}^{m+k}}(X_{1}^{m+k})}{P_{X_{1}^{k}}(X_{1}^{k})\cdot P_{X_{k+1}^{m+k}}(X_{k+1}^{m+k})}\to0
\]
as $m\to\infty$, with probability $1$. 
\end{lem}
\begin{IEEEproof}
The AEP for ergodic sources \cite[Th. 16.8.1]{cover2012elements}
implies that 
\[
\frac{1}{m+k}\log P_{X_{1}^{m+k}}(X_{1}^{m+k})\to-{\cal H}_{1}(\mathbf{X})
\]
and 
\begin{align}
 & \frac{1}{m+k}\log\frac{1}{P_{X_{1}^{k}}(X_{1}^{k})\cdot P_{X_{k+1}^{m+k}}(X_{k+1}^{m+k})}\nonumber \\
 & =\frac{k}{m+k}\cdot\frac{1}{k}\log\frac{1}{P_{X_{1}^{k}}(X_{1}^{k})}+\frac{m}{m+k}\cdot\frac{1}{m}\log\frac{1}{P_{X_{k+1}^{m+k}}(X_{k+1}^{m+k})}\\
 & \to\frac{k}{m+k}{\cal H}_{1}(\mathbf{X})+\frac{m}{m+k}{\cal H}_{1}(\mathbf{X})\\
 & ={\cal H}_{1}(\mathbf{X}),
\end{align}
as $m\to\infty$, with probability $1$. The result then follows by
summing both limits. 
\end{IEEEproof}
\begin{lem}
\label{lem: asymptotic independence condition on T=00003Dt}Consider
the noiseless setting (Def. \ref{def: Noiseless setting}). Let $0<\theta_{0}<\theta_{1}<\beta$
be given. For any $q\in(0,1)$ and $\nu>0$ there exists $n_{0}(\nu,q)$
such that for all $n\geq n_{0}$ it holds with probability $1-q$
that 
\[
\max_{\theta_{0}\log n\leq t\leq\theta_{1}\log n}\log\frac{f\left(X_{1}^{\ell}(1),X_{1}^{\ell}(2);t\right)}{\I\left[X_{\ell-t+1}^{\ell}(1)=X_{1}^{t}(2)\right]\cdot P_{X_{1}^{\ell}}\left(X_{1}^{\ell}(1)\right)\cdot P_{X_{1}^{\ell}}\left(X_{1}^{\ell}(2)\right)\cdot\Gamma_{+}(t)}\leq\nu\cdot\log n.
\]
\end{lem}
\begin{IEEEproof}
For $T=t$, the likelihood was derived in Lemma \ref{lem: likelihood for noisless reads},
and is given by
\begin{equation}
f\left(x_{1}^{\ell}(1),x_{1}^{\ell}(2);t\right)=\I\left[x_{\ell-t+1}^{\ell}(1)=x_{1}^{t}(2)\right]\cdot P_{X_{1}^{\ell}}\left(x_{1}^{\ell}(1)\right)\cdot\frac{P_{X_{1}^{\ell}\mid X_{-(\ell-t-1)}^{0}}\left(x_{1}^{\ell}(2)\mid x_{1}^{\ell-t}(1)\right)}{P_{X_{1}^{t}\mid X_{-(\ell-t-1)}^{0}}(x_{1}^{t}(2)\mid x_{1}^{\ell-t}(1))}.\label{eq: likelihood T =00003Dt in proof of asymptotic term}
\end{equation}
For clarity, we will next omit the arguments and use the abbreviated
notation $P_{X_{1}^{\ell}\mid X_{-(\ell-t-1)}^{0}}\equiv\P[X_{1}^{\ell}\mid X_{-(\ell-t-1)}^{0}].$
Our goal is to show that the ratio in (\ref{eq: likelihood T =00003Dt in proof of asymptotic term}),
given by 
\[
\frac{\P[X_{1}^{\ell}\mid X_{-(\ell-t-1)}^{0}]}{\P[X_{1}^{t}\mid X_{-(\ell-t-1)}^{0}]},
\]
is tightly upper bounded by $\P[X_{1}^{\ell}]/\P[X_{1}^{t}]$, up
to a sufficiently small multiplicative term. We begin with the probability
on the numerator. Let us write 
\[
\P[X_{1}^{\ell}\mid X_{-(\ell-t-1)}^{0}]=\P[X_{1}^{\ell}]\cdot\frac{\P[X_{-(\ell-t-1)}^{\ell}]}{\P[X_{-(\ell-t-1)}^{0}]\cdot\P[X_{1}^{\ell}]}.
\]
To bound the ratio appearing in this term, we use Lemma \ref{lem: asymptotic independence lemma AEP}
that states that if $m\equiv m(n)\to\infty$ and $k\equiv k(n)\to\infty$
as $n\to\infty$ then 
\begin{equation}
\frac{1}{m+k}\log\frac{\P[X_{1}^{m+k}]}{\P[X_{1}^{k}]\cdot\P[X_{k+1}^{m+k}]}\to0,\label{eq: probability of m plus k vs k times k plus one to m plus k}
\end{equation}
as $n\to\infty$ with probability $1$. Let $(\Omega,{\cal F},P)$
be the probability space underlying the process $\boldsymbol{X}\equiv\boldsymbol{X}(\omega)$,
where $\omega\in\Omega$. In principle, the rate of the convergence
to the limit $0$ in (\ref{eq: probability of m plus k vs k times k plus one to m plus k})
may depend on $\omega\in\Omega$. However, by Egorov's theorem (e.g.,
\cite[p. 79, Ex. 5]{chung2000course}), for any $q\in(0,1)$ there
exists a set ${\cal B}_{1}\in{\cal F}$ such that $\P[{\cal B}_{1}]\leq q/2$
and the convergence is \emph{uniform} on ${\cal B}_{1}^{c}$. Thus,
for any $\nu>0$ there exists $n_{0}(\nu,q)$ and such that 
\[
\sup_{\omega\in{\cal B}^{c}}\frac{1}{m+k}\log\frac{\P[X_{1}^{m+k}(\omega)]}{\P[X_{1}^{k}(\omega)]\cdot\P[X_{k+1}^{m+k}(\omega)]}\leq\nu
\]
for all $n\geq n_{0}$ sufficiently large so that $m(n)\geq m(n_{0})$
and $k(n)\geq k(n_{0})$. Using $m=\ell-t$ and $k=\ell$ and noting
that for any $\theta_{0}\log n\leq t\leq\theta_{1}$ it holds that
$m\equiv m(n)\to\infty$ as $n\to\infty$, we deduce that 
\[
\sup_{\omega\in{\cal B}^{c}}\max_{\theta_{0}\log n\leq t\leq\theta_{1}\log n}\frac{1}{2\ell-t}\log\frac{\P[X_{-(\ell-t-1)}^{\ell}(\omega)]}{\P[X_{-(\ell-t-1)}^{0}(\omega)]\cdot\P[X_{1}^{\ell}(\omega)]}\leq\nu.
\]
Thus, also 
\[
\sup_{\omega\in{\cal B}^{c}}\max_{\theta_{0}\log n\leq t\leq\theta_{1}\log n}\log\frac{\P[X_{-(\ell-t-1)}^{\ell}]}{\P[X_{-(\ell-t-1)}^{0}]\cdot\P[X_{1}^{\ell}]}\leq2\nu\beta\cdot\log n
\]
In a completely similar fashion, we can lower bound the probability
on the denominator, that is, $\P[X_{1}^{t}\mid X_{-(\ell-t-1)}^{0}]$,
showing uniform convergence, except on a set ${\cal B}_{2}\in{\cal F}$
which satisfy $\P[{\cal B}_{2}]\leq q/2$. The result then follows
by taking ${\cal B}={\cal B}_{1}\cup{\cal B}_{2}$ for which $\P[{\cal B}]\leq q$,
and combining the bounds for both the numerator and denominator. 
\end{IEEEproof}
Next, we consider the case $T=0$, and state the approximation bound
for this case. 
\begin{lem}
\label{lem: asymptotic independence condition on T=00003D0}Consider
the noiseless setting (Def. \ref{def: Noiseless setting}). Assume
that there exists $\eta>0$ so that
\[
d(s)\leq\frac{1}{s^{\beta{\cal H}_{-\infty}(\mathbf{X})+\eta}},
\]
 for all $s\in\mathbb{N}_{+}$. Then, the likelihood for $T=0$ is
lower bounded as
\[
f\left(x_{1}^{\ell}(1),x_{1}^{\ell}(2);0\right)\geq\xi(n)\cdot P_{X_{1}^{\ell}}\left(x_{1}^{\ell}(1)\right)P_{X_{1}^{\ell}}\left(x_{1}^{\ell}(2)\right),
\]
where $\xi(n)\to1$ and does not depend on $(x_{1}^{\ell}(1),x_{1}^{\ell}(2))$. 
\end{lem}
\begin{IEEEproof}
Under the conditions of the lemma, the likelihood is bounded as
\begin{align}
 & f\left(x_{1}^{\ell}(1),x_{1}^{\ell}(2);0\right)\nonumber \\
 & =\P\left[\left\{ X_{1}^{\ell}(1)=x_{1}^{\ell}(1)\right\} \cap\left\{ X_{1}^{\ell}(2)=x_{1}^{\ell}(2)\right\} \,\middle\vert\,T=0\right]\\
 & =\sum_{i_{1}=1}^{n}\sum_{i_{2}=1}^{n}\P\left[\left\{ I(1)=i_{1}\right\} \cap\left\{ I(2)=i_{2}\right\} \mid T=0\right]\times\nonumber \\
 & \hphantom{=\sum_{i_{1}=1}^{n}\sum_{i_{2}=1}^{n}}\P\left[\left\{ X_{1}^{\ell}(1)=x_{1}^{\ell}(1)\right\} \cap\left\{ X_{1}^{\ell}(2)=x_{1}^{\ell}(2)\right\} \,\middle\vert\,\left\{ I(1)=i_{1}\right\} \cap\left\{ I(2)=i_{2}\right\} \right]\\
 & =\frac{1}{n}\sum_{i_{1}=1}^{n}\sum_{i_{2}=1}^{n}\P\left[I(2)=i_{2}\mid T=0,I(1)=i_{1}\right]\cdot\P\left[\left\{ X_{i_{1}}^{i_{1}+\ell-1}=x_{1}^{\ell}(1)\right\} \cap\left\{ X_{i_{2}}^{i_{2}+\ell-1}=x_{1}^{\ell}(2)\right\} \right]\\
 & \trre[=,a]\frac{1}{n}\sum_{i_{1}=1}^{n}\sum_{i_{2}=1}^{n}\P\left[I(2)=i_{2}\mid T=0,I(1)=i_{1}\right]\cdot\P\left[\left\{ X_{1}^{\ell}=x_{1}^{\ell}(1)\right\} \cap\left\{ X_{i_{2}-i_{1}+1}^{i_{2}-i_{1}+\ell}=x_{1}^{\ell}(2)\right\} \right]\\
 & \trre[\geq,b]\frac{1}{n}\sum_{i_{1}=\tau(n)}^{n-\tau(n)}\sum_{i_{2}=1}^{n}\P\left[I(2)=i_{2}\mid T=0,I(1)=i_{1}\right]\cdot\P\left[\left\{ X_{1}^{\ell}=x_{1}^{\ell}(1)\right\} \cap\left\{ X_{i_{2}-i_{1}+1}^{i_{2}-i_{1}+\ell}=x_{1}^{\ell}(2)\right\} \right],\label{eq: asym likelihood noiseless memory first}
\end{align}
where $(a)$ follows from stationarity, and in $(b)$ we choose $\tau\equiv\tau(n)=2\lfloor\frac{n}{2\log n}\rfloor,$which
is an even integer. It holds that $\frac{\tau(n)}{\ell}=\frac{\tau(n)}{\beta\log n}=\omega_{n}(1)$,
and so according to the way that $I(1)$ and $I(2)$ are drawn, as
described in (\ref{eq: I(2) conditioned on I(1)}), whenever $\ell\leq I(1)\leq n-\ell+1$
it holds that $I(2)\sim\text{Uniform}[n]$. So, conditioned on $I(1)=i_{1}$
and $T=0$ for some $i_{1}\in\{\tau(n)+1,\ldots,n-\tau(n)\}$, it
holds that 
\[
\P\left[I(2)=i_{2}\mid T=0,I(1)=i_{1}\right]\sim\text{Uniform}\left[[n]\backslash\{i_{1}-\ell+1,\ldots,i_{1}+\ell-1\}]\right],
\]
which is supported over $n-(2\ell-1)=n_{\ell}$ possible time indices.
So, the inner summation in (\ref{eq: asym likelihood noiseless memory first})
is given by
\begin{align}
 & \sum_{i_{2}=1}^{n}\P\left[I(2)=i_{2}\mid T=0,I(1)=i_{1}\right]\cdot\P\left[\left\{ X_{1}^{\ell}=x_{1}^{\ell}(1)\right\} \cap\left\{ X_{i_{2}-i_{1}+1}^{i_{2}-i_{1}+\ell}=x_{1}^{\ell}(2)\right\} \right]\nonumber \\
 & =\frac{1}{n_{\ell}}\sum_{i_{2}=1}^{i_{1}-\ell}\P\left[\left\{ X_{1}^{\ell}=x_{1}^{\ell}(1)\right\} \cap\left\{ X_{i_{2}-i_{1}+1}^{i_{2}-i_{1}+\ell}=x_{1}^{\ell}(2)\right\} \right]+\nonumber \\
 & \hphantom{=\frac{1}{n_{\ell}}\sum_{i_{2}=1}^{i_{1}-\ell}}\frac{1}{n_{\ell}}\sum_{i_{2}=i_{1}+\ell}^{n}\P\left[\left\{ X_{1}^{\ell}=x_{1}^{\ell}(1)\right\} \cap\left\{ X_{i_{2}-i_{1}+1}^{i_{2}-i_{1}+\ell}=x_{1}^{\ell}(2)\right\} \right].\label{eq: asym likelihood noiseless memory second}
\end{align}
We may lower bound the first sum in the r.h.s. of (\ref{eq: asym likelihood noiseless memory second})
as 
\begin{align}
 & \frac{1}{n_{\ell}}\sum_{i_{2}=1}^{i_{1}-\ell}\P\left[\left\{ X_{1}^{\ell}=x_{1}^{\ell}(1)\right\} \cap\left\{ X_{i_{2}-i_{1}+1}^{i_{2}-i_{1}+\ell}=x_{1}^{\ell}(2)\right\} \right]\nonumber \\
 & \trre[=,a]\frac{1}{n_{\ell}}\sum_{s=\ell}^{i_{1}-1}\P\left[\left\{ X_{1}^{\ell}=x_{1}^{\ell}(1)\right\} \cap\left\{ X_{-s+1}^{-s+\ell}=x_{1}^{\ell}(2)\right\} \right]\\
 & \geq\frac{1}{n_{\ell}}\sum_{s=\ell+\tau/2}^{i_{1}-1}\P\left[\left\{ X_{1}^{\ell}=x_{1}^{\ell}(1)\right\} \cap\left\{ X_{-s+1}^{-s+\ell}=x_{1}^{\ell}(2)\right\} \right]\\
 & \trre[\geq,b]\frac{1}{n_{\ell}}\sum_{s=\ell+\tau/2}^{i_{1}-1}\P\left[X_{1}^{\ell}=x_{1}^{\ell}(1)\right]\left(\P\left[X_{-s+1}^{-s+\ell}=x_{1}^{\ell}(2)\right]-d(s-\ell)\right)\\
 & \trre[\geq,c]\frac{1}{n_{\ell}}\sum_{s=\ell+\tau/2}^{i_{1}-1}\P\left[X_{1}^{\ell}=x_{1}^{\ell}(1)\right]\left(\P\left[X_{-s+1}^{-s+\ell}=x_{1}^{\ell}(2)\right]-d(\tau/2)\right)\\
 & \trre[=,d]\frac{1}{n_{\ell}}\sum_{s=\ell+\tau/2}^{i_{1}-1}\P\left[X_{1}^{\ell}=x_{1}^{\ell}(1)\right]\left(\P\left[X_{1}^{\ell}=x_{1}^{\ell}(2)\right]-d(\tau/2)\right)\\
 & =\left(\frac{i_{1}-\ell-\tau/2}{n_{\ell}}\right)\P\left[X_{1}^{\ell}=x_{1}^{\ell}(1)\right]\left(\P\left[X_{1}^{\ell}=x_{1}^{\ell}(2)\right]-d(\tau/2)\right),
\end{align}
where $(a)$ follows from the change of variables $s=i_{1}-i_{2}$,
$(b)$ follows from the strong mixing property, $(c)$ follows from
the assumed monotonicity of $d(s)$, and $(d)$ follows from stationarity. 

Similarly, we may lower bound the second sum in the r.h.s. of (\ref{eq: asym likelihood noiseless memory second})
as 
\begin{align}
 & \frac{1}{n_{\ell}}\sum_{i_{2}=i_{1}+\ell}^{n}\P\left[\left\{ X_{1}^{\ell}=x_{1}^{\ell}(1)\right\} \cap\left\{ X_{i_{2}-i_{1}+1}^{i_{2}-i_{1}+\ell}=x_{1}^{\ell}(2)\right\} \right]\nonumber \\
 & \trre[=,a]\frac{1}{n_{\ell}}\sum_{s=\ell}^{n-i_{1}}\P\left[\left\{ X_{1}^{\ell}=x_{1}^{\ell}(1)\right\} \cap\left\{ X_{s+1}^{s+\ell}=x_{1}^{\ell}(2)\right\} \right]\\
 & \trre[\geq,b]\frac{1}{n_{\ell}}\sum_{s=\ell+\tau/2}^{n-i_{1}}\P\left[\left\{ X_{1}^{\ell}=x_{1}^{\ell}(1)\right\} \cap\left\{ X_{s+1}^{s+\ell}=x_{1}^{\ell}(2)\right\} \right]\\
 & \geq\frac{1}{n_{\ell}}\sum_{s=\ell+\tau/2}^{n-i_{1}}\P\left[X_{1}^{\ell}=x_{1}^{\ell}(1)\right]\left(\P\left[X_{-s+1}^{-s+\ell}=x_{1}^{\ell}(2)\right]-d(s-\ell)\right)\\
 & \geq\frac{1}{n_{\ell}}\sum_{s=\ell+\tau/2}^{n-i_{1}}\P\left[X_{1}^{\ell}=x_{1}^{\ell}(1)\right]\left(\P\left[X_{-s+1}^{-s+\ell}=x_{1}^{\ell}(2)\right]-d(\tau/2)\right)\\
 & \geq\left(\frac{n-i_{1}-\ell-\tau/2+1}{n_{\ell}}\right)\P\left[X_{1}^{\ell}=x_{1}^{\ell}(1)\right]\left(\P\left[X_{1}^{\ell}=x_{1}^{\ell}(2)\right]-d(\tau)\right),
\end{align}
where $(a)$ follows from the change of variables $s=i_{2}-i_{1}$,
and $(b)$ is a non-empty summation (for all $n$ sufficiently large)
since $i_{1}\leq n-\tau$. 

Summing the lower bounds on the first and second terms in the r.h.s.
of (\ref{eq: asym likelihood noiseless memory second}) leads to the
lower bound 
\begin{align}
 & \sum_{i_{2}=1}^{n}\P\left[I(2)=i_{2}\mid T=0,I(1)=i_{1}\right]\cdot\P\left[\left\{ X_{1}^{\ell}=x_{1}^{\ell}(1)\right\} \cap\left\{ X_{i_{2}-i_{1}+1}^{i_{2}-i_{1}+\ell}=x_{1}^{\ell}(2)\right\} \right]\nonumber \\
 & \geq\left(\frac{i_{1}-\ell-\tau/2}{n_{\ell}}+\frac{n-i_{1}-\ell-\tau/2+1}{n_{\ell}}\right)\cdot\P\left[X_{1}^{\ell}=x_{1}^{\ell}(1)\right]\left(\P\left[X_{1}^{\ell}=x_{1}^{\ell}(2)\right]-d(\tau/2)\right)\\
 & =\left(\frac{n-2\ell-\tau+1}{n_{\ell}}\right)\P\left[X_{1}^{\ell}=x_{1}^{\ell}(1)\right]\cdot\left(\P\left[X_{1}^{\ell}=x_{1}^{\ell}(2)\right]-d(\tau/2)\right)\\
 & =\left(\frac{n-2\ell-2\tau-1}{n_{\ell}}\right)\P\left[X_{1}^{\ell}=x_{1}^{\ell}(1)\right]\P\left[X_{1}^{\ell}=x_{1}^{\ell}(2)\right]\left(1-\frac{d(\tau/2)}{\P\left[X_{1}^{\ell}=x_{1}^{\ell}(2)\right]}\right)\\
 & \sim\P\left[X_{1}^{\ell}=x_{1}^{\ell}(1)\right]\P\left[X_{1}^{\ell}=x_{1}^{\ell}(2)\right],
\end{align}
where last step holds since
\[
\frac{n-2\ell-2\tau-1}{n_{\ell}}=1-O\left(\frac{1}{\log n}\right)\sim1,
\]
and since 
\begin{align}
 & \frac{d(\tau/2)}{\P\left[X_{1}^{\ell}=x_{1}^{\ell}(2)\right]}\nonumber \\
 & \leq\frac{d\left(\frac{n}{2\log n}\right)}{\P\left[X_{1}^{\ell}=x_{1}^{\ell}(2)\right]}\\
 & \trre[\le,i]\frac{\frac{(2\log n)^{\beta{\cal H}_{-\infty}(\mathbf{X})+\eta}}{n^{\beta{\cal H}_{-\infty}(\mathbf{X})+\eta}}}{\P\left[X_{1}^{\ell}=x_{1}^{\ell}(2)\right]}\\
 & \trre[\le,ii]\frac{\frac{(2\log n)^{\beta{\cal H}_{-\infty}(\mathbf{X})+\eta}}{n^{\beta{\cal H}_{-\infty}(\mathbf{X})+\eta}}}{n^{-\beta\left[{\cal H}_{-\infty}(\mathbf{X})+\epsilon(n)\right]}}\\
 & =\frac{(2\log n)^{\beta{\cal H}_{-\infty}(\mathbf{X})+\eta}}{n^{\beta\left[\eta-\epsilon(n)\right]}}\\
 & =o_{n}(1),
\end{align}
where, here, $(i)$ follows from the assumption of the lemma, and
$(ii)$ holds for some $\epsilon(n)=o_{n}(1)$.
\end{IEEEproof}
Before getting into the proof of Theorem \ref{thm: exponent lower bound noiseless},
we remind the reader the following lower bound on the error probability
in binary hypothesis testing. This result will also be used in the
proof of Theorem \ref{thm: exponent lower bound noiseless}. 
\begin{lem}[{Strong converse on errors in binary hypothesis testing \cite[Th. 14.9]{polyanskiy2023information}}]
\label{lem: strong converse HT}Consider a binary hypothesis testing
problem between $H_{0}:X\sim P$ and $H_{1}:X\sim Q$. Let $Z\in\{0,1\}$
be a test based on $X$. Then, for any $\eta>0$ it holds that the
two types of error must satisfy
\[
P[Z=1]+\eta\cdot Q[Z=0]\geq1-\P_{X\sim P}\left[\log\frac{\d P}{\d Q}(X)>\log\eta\right].
\]
\end{lem}
We may now prove Theorem \ref{thm: exponent lower bound noiseless}. 
\begin{IEEEproof}[Proof of Theorem \ref{thm: exponent lower bound noiseless}]
Let us first assume that $\beta>\frac{1}{{\cal H}_{1}(\mathbf{X})}$.
As we have seen in the proof of Theorem \ref{thm: exponent upper bound noiseless},
the possibility of both positive and negative overlaps simply doubles
the error probability, and only involves repeating similar arguments
for both the positive and negative overlaps. Therefore, in this proof,
as well as the rest of the proofs to follow, we slightly simplify
the setting, and assume that either no overlap or a positive overlap
are possible. That is, we assume that either $T=0$ or $T\in[\ell]$,
and that 
\[
P_{T}(t)=\begin{cases}
\frac{1-\ell}{n}, & t=0\\
\frac{1}{n}, & t\in[\ell]
\end{cases}
\]
(c.f. (\ref{eq: DNA shotgun sequencing overlap})). 

We consider the following genie-aided detector. Let $t_{0}:=\lceil\theta_{0}\log n\rceil$
and $t_{1}:=\lfloor\theta_{1}\log n\rfloor$ so that $0<\theta_{0}<\theta_{1}<\beta$,
where 
\[
\theta_{1}=\frac{1-\theta_{0}}{{\cal H}_{1}(\mathbf{X})}
\]
(we take $\theta_{0}$ as a small constant). The genie provides the
detector with side-information $S$, so that the optimal Bayesian
detector is $\hat{T}_{\text{ga}}(X_{1}^{\ell}(1),X_{1}^{\ell}(2),S)$
(with a slight abuse of notation). Let $\tilde{T}_{0}\sim\text{Uniform}([t_{1}]\backslash[t_{0}])$.
The side-information is given by 
\[
S=\begin{cases}
\{0,\tilde{T}_{0}\}, & T=0\\
\{0,T\}, & T\in[t_{1}]\backslash[t_{0}]\\
T, & \text{otherwise}
\end{cases}.
\]
That is, if $T\in\{1,2,\ldots,t_{0}\}$ or $T\in\{t_{1}+1,\ldots,\ell\}$
then the genie reveals to the detector the exact value of $T$. Otherwise,
it reveals to the detector that the overlap is either zero or a positive
overlap in $[t_{1}]\backslash[t_{0}]$, where one of these options
is the true overlap. We next lower bound the Bayesian error probability
of \emph{any} genie-aided detector. First, conditioned on the event
$T\in\{1,\ldots,t_{0}\}\cup\{t_{1}+1,\ldots,\ell\}$ an optimal genie-aided
detector will decide $\hat{T}_{\text{ga}}=S$, and obtain zero error.
Second, conditioned on the event ${\cal E}_{t}:=\{S=\{0,t\}\}$ for
some $t\in[t_{1}]\backslash[t_{0}]$, an optimal genie-aided detector
will output either $0$ or $t$. Consequently, any optimal genie-aided
detector defines a set of $t_{1}-t_{0}$ binary hypothesis testers.
Let us focus on a single index $t\in[t_{1}]\backslash[t_{0}]$. The
probability of the event ${\cal E}_{t}$ is given by
\[
\P[{\cal E}_{t}]=\frac{P_{T}(0)}{t_{1}-t_{0}}+P_{T}(t)=\frac{1-\frac{\ell}{n}}{t_{1}-t_{0}}+\frac{1}{n}=\frac{1}{n}\left(\frac{n-\ell+t_{1}-t_{0}}{t_{1}-t_{0}}\right)\sim\frac{1}{t_{1}-t_{0}}.
\]
Furthermore, the binary hypothesis testing problem between $T=0$
and $T=t$ has priors 
\[
\pi_{t}:=\P\left[T=t\mid S=\{0,t\}\right]=\frac{t_{1}-t_{0}}{n-\ell+t_{1}-t_{0}}\sim(\theta_{1}-\theta_{0})\cdot\frac{\log n}{n},
\]
and $\pi_{0}=1-\pi_{t}$. Let $\hat{D}_{t}$ be any binary test between
$T=0$ and $T=t$ under the prior $\{\pi_{0},\pi_{t}\}$. So, for
the MAP detector
\begin{align}
P_{\text{error}}^{*} & \trre[\geq,a]P_{\text{error}}(\hat{T}_{\text{ga}})\\
 & =\sum_{t=t_{0}}^{t_{1}}\P[{\cal E}_{t}]\cdot\P[\hat{T}_{\text{ga}}\neq T\mid{\cal E}_{t}]+\P\left[\bigcap_{t=t_{0}}^{t_{1}}{\cal E}_{t}^{c}\right]\cdot\P\left[\hat{T}_{\text{ga}}\neq T\,\middle\vert\,\bigcap_{t=t_{0}}^{t_{1}}{\cal E}_{t}^{c}\right]\\
 & \trre[=,b]\sum_{t=t_{0}}^{t_{1}}\P[{\cal E}_{t}]\cdot\P[\hat{T}_{\text{ga}}\neq T\mid{\cal E}_{t}]\\
 & \trre[=,c]\sum_{t=t_{0}}^{t_{1}}\P[{\cal E}_{t}]\cdot\left(\pi_{t}\P\left[\hat{D}_{t}=0\mid T=t\right]+\pi_{0}\cdot\P\left[\hat{D}_{t}=t\mid T=0\right]\right)\\
 & \trre[\sim,d]\frac{1}{t_{1}-t_{0}}\cdot\sum_{t=t_{0}}^{t_{1}}\pi_{t}\P\left[\hat{D}_{t}=0\mid T=t\right]+\pi_{0}\cdot\P\left[\hat{D}_{t}=t\mid T=0\right]\\
 & \geq\min_{t\in[t_{1}]\backslash[t_{0}]}\pi_{t}\P\left[\hat{D}_{t}=0\mid T=t\right]+\pi_{0}\cdot\P\left[\hat{D}_{t}=t\mid T=0\right]\\
 & \geq\min_{t\in[t_{1}]\backslash[t_{0}]}\pi_{t}\cdot\left(\P\left[\hat{D}_{t}=0\mid T=t\right]+\frac{\pi_{0}}{\pi_{t}}\cdot\P\left[\hat{D}_{t}=t\mid T=0\right]\right),\label{eq: lower bound on error probability via binary HT}
\end{align}
where $(a)$ follows since the Bayesian error probability of an optimal
genie-aided detector is lower than that of the MAP detector, $(b)$
holds since conditioned on $\bigcap_{t=t_{0}}^{t_{1}}{\cal E}_{t}^{c}$
it holds that $T$ is known exactly to the genie-aided detector, $(c)$
holds since, conditioned on ${\cal E}_{t}$, the genie-aided detector
is reduced to a binary hypothesis test, and $(d)$ holds since $\P[{\cal E}_{t}]\sim\frac{1}{t_{1}-t_{0}}$.

We now invoke Lemma \ref{lem: strong converse HT} (or \cite[Th. 14.9]{polyanskiy2023information}),
by setting therein $\eta=\frac{\pi_{0}}{\pi_{t}}$. Let $q\in(0,1)$
be a small given probability, and let $\nu=\theta_{0}/8$. We will
use Lemma \ref{lem: asymptotic independence condition on T=00003Dt}
which assures that there exists an event ${\cal B}$ such that $\P[{\cal B}]\leq q$
and on ${\cal B}^{c}$ it holds that 
\begin{equation}
f\left(X_{1}^{\ell}(1),X_{1}^{\ell}(2);t\right)\leq e^{\nu\log n}\cdot\I\left[X_{\ell-t+1}^{\ell}(1)=X_{1}^{t}(2)\right]\cdot P_{X_{1}^{\ell}}\left(X_{1}^{\ell}(1)\right)\cdot P_{X_{1}^{\ell}}\left(X_{1}^{\ell}(2)\right)\cdot\Gamma_{+}(t).\label{eq: the approximation T=00003Dt on the likelihood in the proof}
\end{equation}
We then have that
\begin{align}
 & \P\left[\hat{D}_{t}=0\mid T=t\right]+\frac{\pi_{0}}{\pi_{t}}\cdot\P\left[\hat{D}_{t}=t\mid T=0\right]\nonumber \\
 & \trre[\ge,a]1-\P\left[\log\frac{f\left(X_{1}^{\ell}(1),X_{1}^{\ell}(2);t\right)}{f\left(X_{1}^{\ell}(1),X_{1}^{\ell}(2);0\right)}>\log\frac{\pi_{0}}{\pi_{t}}\,\middle\vert\,T=t\right]\\
 & \trre[\ge,b]1-\P\left[\log\frac{f\left(X_{1}^{\ell}(1),X_{1}^{\ell}(2);t\right)}{\xi(n)P_{X_{1}^{\ell}}\left(X_{1}^{\ell}(1)\right)\cdot P_{X_{1}^{\ell}}\left(X_{1}^{\ell}(2)\right)}>\log\frac{\pi_{0}}{\pi_{t}}\,\middle\vert\,T=t\right]\\
 & \trre[=,c]1-\P\left[\{{\cal B}^{c}\}\cap\left\{ \log\frac{f\left(X_{1}^{\ell}(1),X_{1}^{\ell}(2);t\right)}{\xi(n)P_{X_{1}^{\ell}}\left(X_{1}^{\ell}(1)\right)\cdot P_{X_{1}^{\ell}}\left(X_{1}^{\ell}(2)\right)}>\log\frac{\pi_{0}}{\pi_{t}}\right\} \,\middle\vert\,T=t\right]\nonumber \\
 & \hphantom{===}-\P\left[\{{\cal B}\}\cap\left\{ \log\frac{f\left(X_{1}^{\ell}(1),X_{1}^{\ell}(2);t\right)}{\xi(n)P_{X_{1}^{\ell}}\left(X_{1}^{\ell}(1)\right)\cdot P_{X_{1}^{\ell}}\left(X_{1}^{\ell}(2)\right)}>\log\frac{\pi_{0}}{\pi_{t}}\right\} \,\middle\vert\,T=t\right]\\
 & \trre[\ge,d]1-q-\P\left[\log\frac{\I\left[X_{\ell-t+1}^{\ell}(1)=X_{1}^{t}(2)\right]\cdot\Gamma_{+}(t)}{\xi(n)}+\nu\log n>\log\frac{\pi_{0}}{\pi_{t}}\,\middle\vert\,T=t\right]\\
 & \trre[=,e]1-q-\P\left[\log\frac{\Gamma_{+}(t)}{\xi(n)}+\nu\log n>\log\frac{\pi_{0}}{\pi_{t}}\,\middle\vert\,T=t\right]\\
 & \trre[=,f]1-q-\P\left[\frac{1}{P_{X_{1}^{t}}\left(X_{1}^{t}\right)}>\log\xi(n)-\nu\log n+\log\frac{\pi_{0}}{\pi_{t}}\,\middle\vert\,T=t\right]\\
 & \trre[\sim,g]1-q,
\end{align}
where $(a)$ follows from Lemma \ref{lem: strong converse HT}, $(b)$
follows since under the alternative hypothesis $T=0$,\footnote{We name the two hypothesis this way in order to comply with the next
use of \cite[Th. 14.9]{polyanskiy2023information}.} Lemma \ref{lem: asymptotic independence condition on T=00003D0}
implies that 
\[
f\left(x_{1}^{\ell}(1),x_{1}^{\ell}(2);0\right)\geq\xi(n)P_{X_{1}^{\ell}}\left(x_{1}^{\ell}(1)\right)\cdot P_{X_{1}^{\ell}}\left(x_{1}^{\ell}(2)\right),
\]
$(c)$ follows by using the definition of the event ${\cal B}$, $(d)$
follows by utilizing the bound (\ref{eq: the approximation T=00003Dt on the likelihood in the proof})
on ${\cal B}$, and $\P[{\cal B}]\leq q$, $(e)$ holds since the
reading kernel is noiseless, and so under $T=t$ it holds with probability
$1$ that $X_{\ell-t+1}^{\ell}(1)=X_{1}^{t}(2)$, $(f)$ holds from
the definition of $\Gamma_{+}(t)$ and stationarity. Finally, $(g)$
follows from the following derivation, which holds for all $n$ sufficiently
large: 
\begin{align}
 & \P\left[\log\frac{1}{P_{X_{1}^{t}}\left(X_{1}^{t}\right)}>\log\xi(n)-\nu\log n+\log\frac{\pi_{0}}{\pi_{t}}\,\middle\vert\,T=t\right]\nonumber \\
 & \trre[\leq,i]\P\left[\log\frac{1}{P_{X_{1}^{t}}\left(X_{1}^{t}\right)}>\log\xi(n)+\left(1-\zeta(n)-\nu\right)\log n\right]\\
 & =\P\left[\log\frac{1}{P_{X_{1}^{t}}\left(X_{1}^{t}\right)}-\E\left[\log\frac{1}{P_{X_{1}^{t}}\left(X_{1}^{t}\right)}\right]>\log\xi(n)+\left(1-\zeta(n)-\nu\right)\log n-\E\left[\log\frac{1}{P_{X_{1}^{t}}\left(X_{1}^{t}\right)}\right]\right]\\
 & \trre[\leq,ii]\P\left[\log\frac{1}{P_{X_{1}^{t}}\left(X_{1}^{t}\right)}-\E\left[\log\frac{1}{P_{X_{1}^{t}}\left(X_{1}^{t}\right)}\right]>\left(1-2\zeta(n)-\nu\right)\log n-\E\left[\log\frac{1}{P_{X_{1}^{t}}\left(X_{1}^{t}\right)}\right]\right]\\
 & \trre[\leq,iii]\P\left[\log\frac{1}{P_{X_{1}^{t}}\left(X_{1}^{t}\right)}-\E\left[\log\frac{1}{P_{X_{1}^{t}}\left(X_{1}^{t}\right)}\right]>\frac{\theta_{0}}{8}\log n\right]\\
 & =\P\left[\frac{1}{t}\log\frac{1}{P_{X_{1}^{t}}\left(X_{1}^{t}\right)}-\frac{1}{t}\E\left[\log\frac{1}{P_{X_{1}^{t}}\left(X_{1}^{t}\right)}\right]>\frac{\theta_{0}}{8\theta_{1}}\right]\\
 & \trre[=,iv]o_{n}(1),
\end{align}
where here $(i)$ follows by setting 
\begin{align}
\log\frac{\pi_{0}}{\pi_{t}} & =\log\frac{1-\pi_{t}}{\pi_{t}}\\
 & =\log\frac{n-\ell}{t_{1}-t_{0}}\\
 & \geq\log\frac{n-\ell}{(\theta_{1}-\theta_{0})\log n}\\
 & =\log n+\log\left(1-\frac{\ell}{n}\right)-\log(\theta_{1}-\theta_{0})-\log\log(n)\\
 & \geq\log n+\log\left(1-\frac{\beta\log n}{n}\right)-\log\beta-\log\log(n)\\
 & =:\left(1-\zeta(n)\right)\log n,\label{eq: log prior ratio for lower bound proof}
\end{align}
where $\zeta(n):=\Theta(\frac{\log\log(n)}{\log n})=o_{n}(1),$$(ii)$
follows since $\log\xi(n)\to0$ whereas $(1-\zeta(n))\log n\to\infty$
as $n\to\infty$, $(iii)$ follows since for all $n$ sufficiently
large 
\begin{align}
\E\left[\log\frac{1}{P_{X_{1}^{t}}\left(X_{1}^{t}\right)}\right] & \leq\E\left[\log\frac{1}{P_{X_{1}^{t_{1}}}\left(X_{1}^{t_{1}}\right)}\right]\\
 & \leq t_{1}\cdot\left[{\cal H}_{1}(\mathbf{X})+\epsilon(n)\right]\\
 & \leq(1-\theta_{0}/2)\log n,
\end{align}
where $\epsilon(n)=o_{n}(1)$ and $\nu=\frac{\theta_{0}}{8}$, $(iv)$
follows from the AEP for ergodic sources \cite[Th. 16.8.1]{cover2012elements},
which implies that 
\[
-\frac{1}{t}\log P_{X_{1}^{t}}\left(X_{1}^{t}\right)\to{\cal H}_{1}(\mathbf{X})
\]
with probability $1$, as $t\to\infty$, and since any $t\in[t_{1}]\backslash[t_{0}]$
satisfies $t>\theta_{0}\log n\to\infty$ as $n\to\infty$ and $\frac{\theta_{0}}{8\theta_{1}}$
is a constant. 

Overall, we obtain that 
\begin{align}
P_{\text{error}}(\hat{T}) & \geq[1-q-o_{n}(1)]\pi_{t}\\
 & \sim(1-q)(\theta_{1}-\theta_{0})\frac{\log n}{n}\\
 & =(1-q)\left(\frac{1-\theta_{0}}{{\cal H}_{1}(\mathbf{X})}-\theta_{0}\right)\cdot\frac{\log n}{n}.
\end{align}
 The proof is then completed by noting that both $\theta_{0}$ and
$q$ can be made arbitrarily small, and taking $\theta_{0}\downarrow0$
and $q\downarrow0$. Finally, in case that $\beta\le\frac{1}{{\cal H}_{1}(\mathbf{X})}$
a similar proof holds with $t_{1}$ chosen as $t_{1}\sim\frac{1-\theta_{0}}{\beta}\cdot\log n$. 
\end{IEEEproof}

\section{Proofs for Section \ref{sec:Alignment-Detection-for noisy} \label{sec:Proofs-noisy}}

We begin by proving Lemma \ref{lem: likelihood for noisy reads},
which provides an exact expression for the likelihood function in
the noisy setting. 
\begin{IEEEproof}[Proof of Lemma \ref{lem: likelihood for noisy reads}]
We only prove the expression for $\Gamma_{+}(t)$, as the expression
for $\Gamma_{-}(t)$ can be obtained by replacing the role of the
two reads. Recall that the pair of sequences may be partitioned into
$3$ parts. The three parts -- the left side of the left fragment,
the common fragment, and the right side of the right fragment --
are statistically independent conditioned on $T$. Using this, we
may write the likelihood as
\begin{align}
f\left(y_{1}^{\ell}(1),y_{1}^{\ell}(2);t\right) & :=P_{Y_{1}^{\ell}(1)Y_{1}^{\ell}(2)\mid T}\left(y_{1}^{\ell}(1),y_{1}^{\ell}(2)\mid T=t\right)\\
 & =P_{Y^{\ell-t}(1)}\left(y_{1}^{\ell-t}(1)\right)\cdot P_{Y_{t+1}^{\ell}(2)}\left(y_{t+1}^{\ell}(2)\right)\nonumber \\
 & \hphantom{==}\times\P\left[Y_{\ell-t+1}^{\ell}(1)=y_{\ell-t+1}^{\ell}(1),Y_{1}^{t}(2)=y^{t}(2)\mid X_{\ell-t+1}^{\ell}(1)=X_{1}^{t}(2)\right].\label{eq: likelihood for noisy reads definition}
\end{align}
Now, recalling that $P_{Y}$ is the $Y$-marginal of $P_{X}\otimes P_{Y\mid X}$,
since both the source and the channel are memoryless and stationary,
the product of the first two terms in (\ref{eq: likelihood for noisy reads definition})
is 
\begin{equation}
P_{Y^{\ell-t}(1)}\left(y_{1}^{\ell-t}(1)\right)\cdot P_{Y_{t+1}^{\ell}(2)}\left(y_{t+1}^{\ell}(2)\right)=\prod_{i=1}^{\ell-t}P_{Y}(y_{i}(1))\prod_{i=t+1}^{\ell}P_{Y}(y_{i}(2)).\label{eq: likelihood memoryless first and second terms}
\end{equation}
The third term in (\ref{eq: likelihood for noisy reads definition})
is given by 
\begin{align}
 & \P\left[Y_{\ell-t+1}^{\ell}(1)=y_{\ell-t+1}^{\ell}(1),Y_{1}^{t}(2)=y_{1}^{t}(2)\mid X_{\ell-t+1}^{\ell}(1)=X_{1}^{t}(2)\right]\nonumber \\
 & =\prod_{i=1}^{t}P_{Y\tilde{Y}}(y_{\ell-t+i}(1),y_{i}(2)),\label{eq: likelihood memoryless third term}
\end{align}
due to the independence of the noisy channels generating the reads.
Combining (\ref{eq: likelihood memoryless first and second terms})
and (\ref{eq: likelihood memoryless third term}) results the likelihood
\begin{align}
 & f\left(y_{1}^{\ell}(1),y_{1}^{\ell}(2);t\right)\nonumber \\
 & =\prod_{i=1}^{\ell-t}P_{Y}(y_{i}(1))\prod_{i=1}^{t}P_{Y\tilde{Y}}(y_{\ell-t+i}(1),y_{i}(2))\prod_{i=t+1}^{\ell}P_{Y}(y_{i}(2))\\
 & =\prod_{i=1}^{\ell-t}P_{Y}(y_{i}(1))\prod_{i=1}^{t}P_{Y}(y_{\ell-t+i}(1))\frac{\prod_{i=1}^{t}P_{Y\tilde{Y}}(y_{\ell-t+i}(1),y_{i}(2))}{\prod_{i=1}^{t}P_{Y}(y_{\ell-t+i}(1))P_{Y}(y_{i}(2))}\prod_{i=1}^{t}P_{Y}(y_{i}(2))\prod_{i=t+1}^{\ell}P_{Y}(y_{i}(2))\\
 & =\prod_{i=1}^{\ell}P_{Y}(y_{i}(1))\times\prod_{i=1}^{t}\lambda(y_{\ell-t+i}(1),y_{i}(2))\times\prod_{i=1}^{\ell}P_{Y}(y_{i}(2)).
\end{align}
Finally, since the first term $\prod_{i=1}^{\ell}P_{Y}(y_{i}(1))$
and the last term $\prod_{i=1}^{\ell}P_{Y}(y_{i}(2))$ do not depend
on $t$, they can be omitted from the likelihood (without affecting
the maximizer), and (\ref{eq: likelihood for noisy reads positive})
thus follows.
\end{IEEEproof}

\subsection{Proof of Theorem \ref{thm: exponent upper bound noisy}}

Before we prove Theorem \ref{thm: exponent upper bound noisy}, we
state and prove a few useful lemmas. The first lemma will be used
in the analysis of the type-I error probability. It is a reminder
from \cite[Lemma 47]{polyanskiy2010channel}, and is commonly used
in\emph{ finite blocklength} information theory. This result is based
on a change-of-measure argument combined with Berry--Esseen central
limit theorem (e.g. \cite[Th. 2, Chapter XVI.5]{feller1991introduction})
to obtain strong large-deviations bounds, similar to the one of Bahadur
and Rao \cite{Bahadur1960on} \cite[Th. 3.7.4.]{dembo2009large} (which
has the exact pre-exponent). Compared to Bahadur--Rao bound, this
lemma does not have the exact pre-exponent constant, which is anyway
immaterial for our analysis. However, it is non-asymptotic, and its
applicable to both lattice and non-lattice variables. See \cite[Th. 14.5]{polyanskiy2023information}
and \cite[Th. 1.3]{tan2014asymptotic} for a discussion, and how it
is used for the higher order terms in capacity expansions. See also
\cite[Ch. 3]{merhav2024toolbox} on its relation saddle-point methods
and \cite{vazquez2018saddlepoint,merhav2023d} for two additional
applications of those techniques. This lemma will be used in the analysis
of the type-I error probability.
\begin{lem}[{From \cite[Lemma 47]{polyanskiy2010channel} (simplified for i.i.d.
r.v.'s)}]
\label{lem:truncated MGF}Let $\{Z_{i}\}_{i\in[n]}$ be i.i.d. r.v's
with $\sigma^{2}=\V[Z_{i}]>0$ and $m_{3}^{3}:=\E[|Z_{i}-\E[Z_{i}]|^{3}]<\infty$.
Then, for any $a\in\mathbb{R}$
\[
\E\left[\exp\left\{ -\sum_{i=1}^{n}Z_{i}\right\} \cdot\I\left(\sum_{i=1}^{n}Z_{i}>a\right)\right]\leq2\cdot\left(\frac{\log2}{\sqrt{2\pi}}+\frac{12m_{3}^{3}}{\sigma^{2}}\right)\frac{1}{\sigma}e^{-a}.
\]
\end{lem}
The second lemma will be used in the analysis of the type-II error
probability, and specifically, to an error event from $T\neq0$ to
$T=0$ (misdetection of an overlap). This is a large-deviations result,
which is based in Chernoff's bound. 
\begin{lem}
\label{lem: Chernoff analysis of crossing events}Let $\{(Y_{i},\tilde{Y}_{i},\overline{Y}_{i})\}_{i\in\mathbb{Z}}$
be a sequence of i.i.d. r.v.'s, distributed as $P_{Y\tilde{Y}}\otimes P_{Y}$,
and assume that $I(Y;\tilde{Y})>0$. Let $a>0$ be given. Then,
\begin{equation}
\P\left[\sum_{i=1}^{t}\log\lambda(Y_{i},\tilde{Y}_{i})\leq a\log n_{\ell}\right]\leq\exp\left[-t\cdot E_{-}^{(t)}(a,P_{XY})\right],\label{eq: Chernoff for sum of log-likelihood is less than a*log(n)}
\end{equation}
for
\[
E_{-}^{(t)}(a,P_{XY}):=\sup_{\nu\leq1}\frac{(\nu-1)a\log n_{\ell}}{t}-(\nu-1)\cdot D_{\nu}(P_{Y\tilde{Y}}\mid\mid P_{Y}\otimes P_{Y}),
\]
and there exists $E_{-}(0,P_{XY})>0$ such that
\begin{equation}
\P\left[\sum_{i=1}^{t}\log\lambda(Y_{i},\tilde{Y}_{i})\leq0\right]\leq\exp\left[-t\cdot E_{-}(0,P_{XY})\right].\label{eq: Chernoff for sum of log-likelihood is negative}
\end{equation}
Finally, there exists $E_{+}(P_{XY})>0$ such that 
\begin{equation}
\P\left[\sum_{i=1}^{t}\log\lambda(Y_{i},\overline{Y}_{i})>0\right]\leq\exp\left[-t\cdot E_{+}(P_{XY})\right].\label{eq: Chernoff for sum of log-likelihood positive}
\end{equation}
\end{lem}
\begin{IEEEproof}
It follows from Chernoff's bound that
\begin{align}
 & \log\P\left[\sum_{i=1}^{t}\log\lambda(Y_{i},\tilde{Y}_{i})\leq a\log n_{\ell}\right]\nonumber \\
 & \leq\log\inf_{\text{\ensuremath{\nu\geq}0}}\E\left[\exp\left\{ \nu a\log n_{\ell}-\nu\sum_{i=1}^{t}\log\lambda(Y_{i},\tilde{Y}_{i})\right\} \right]\\
 & \trre[=,a]\log\inf_{\nu\geq0}n_{\ell}^{\nu a}\prod_{i=1}^{t}\E\left[\frac{P_{Y}^{\nu}(Y_{i})P_{Y}^{\nu}(\tilde{Y}_{i})}{P_{Y\tilde{Y}}^{\nu}(Y_{i},\tilde{Y}_{i})}\right]\\
 & =\inf_{\nu\geq0}\nu a\log n_{\ell}+t\log\sum_{(y,\tilde{y})\in{\cal Y}^{\otimes2}}P_{Y\tilde{Y}}^{1-\nu}(y,\tilde{y})P_{Y}^{\nu}(y)P_{Y}^{\nu}(\tilde{y})\\
 & =t\cdot\left(\inf_{\nu\leq1}\frac{(1-\nu)a\log n_{\ell}}{t}+\log\sum_{(y,\tilde{y})\in{\cal Y}^{\otimes2}}P_{Y\tilde{Y}}^{\nu}(y,\tilde{y})P_{Y}^{1-\nu}(y)P_{Y}^{1-\nu}(\tilde{y})\right)\\
 & \trre[=:,b]-t\cdot E_{-}^{(t)}(a,P_{XY}),\label{eq: Chernoff bound for under crossing event noisy}
\end{align}
where $(a)$ holds since $\{(Y_{i},\tilde{Y}_{i})\}_{i=1}^{t}$ are
i.i.d., and $(b)$ utilizes the definition of the R\'{e}nyi divergence
in (\ref{eq: Renyi divergence}). This shows (\ref{eq: Chernoff for sum of log-likelihood is less than a*log(n)}).
Next, we focus on the case $a=0$, and show that the infimum, which
we denote by $E_{-}(0,P_{XY})$ is \emph{strictly} negative. To this
end, let 
\[
f(\nu):=\log\sum_{(y,\tilde{y})\in{\cal Y}^{\otimes2}}P_{Y\tilde{Y}}^{\nu}(y,\tilde{y})P_{Y}^{1-\nu}(y)P_{Y}^{1-\nu}(\tilde{y}).
\]
Then, $f(1)=0$ and
\[
f(0)=\log P_{Y}\otimes P_{Y}(\supp(P_{Y\tilde{Y}}))\leq0.
\]
If $f(0)<0$ then (\ref{eq: Chernoff for sum of log-likelihood is less than a*log(n)})
holds with $E_{-}(0,P_{XY})\geq-f(0)$. Otherwise, $f(0)=f(1)=0$,
but we next show that $f(\nu)$ is strictly convex, and so there exists
$\nu_{*}\in(0,1)$ so that $f(\nu_{*})<0$, and then (\ref{eq: Chernoff for sum of log-likelihood is less than a*log(n)})
holds with $E_{-}(0,P_{XY})\geq-f(\nu_{*})$. Indeed, let $P^{(\nu)}\propto P_{Y\tilde{Y}}^{\nu}(y,\tilde{y})P_{Y}^{1-\nu}(y)P_{Y}^{1-\nu}(\tilde{y})$
denote a tilted probability measure, and let $\E_{\nu}[\cdot]$ (resp.
$\V_{\nu}[\cdot]$) denote expectation (resp. variance) with respect
to this measure. Then, 
\begin{align}
\frac{\d}{\d\nu}f(\nu) & :=\frac{\sum_{(y,\tilde{y})\in{\cal Y}^{\otimes2}}P_{Y}(y)P_{Y}(\tilde{y})\frac{P_{Y\tilde{Y}}^{\nu}(y,\tilde{y})}{P_{Y}^{\nu}(y)P_{Y}^{\nu}(\tilde{y})}\log\frac{P_{Y\tilde{Y}}(y,\tilde{y})}{P_{Y}(y)P_{Y}(\tilde{y})}}{\sum_{(y,\tilde{y})\in{\cal Y}^{\otimes2}}P_{Y\tilde{Y}}^{\nu}(y,\tilde{y})P_{Y}^{1-\nu}(y)P_{Y}^{1-\nu}(\tilde{y})}\\
 & =\E_{\nu}\left[\log\frac{P_{Y\tilde{Y}}(y,\tilde{y})}{P_{Y}(y)P_{Y}(\tilde{y})}\right],
\end{align}
and 
\begin{align}
\frac{\d^{2}}{\d\nu^{2}}f(\nu) & :=\frac{\d}{\d\nu}\frac{\sum_{(y,\tilde{y})\in{\cal Y}^{\otimes2}}P_{Y}(y)P_{Y}(\tilde{y})\frac{P_{Y\tilde{Y}}^{\nu}(y,\tilde{y})}{P_{Y}^{\nu}(y)P_{Y}^{\nu}(\tilde{y})}\log\frac{P_{Y\tilde{Y}}(y,\tilde{y})}{P_{Y}(y)P_{Y}(\tilde{y})}}{\sum_{(y,\tilde{y})\in{\cal Y}^{\otimes2}}P_{Y\tilde{Y}}^{\nu}(y,\tilde{y})P_{Y}^{1-\nu}(y)P_{Y}^{1-\nu}(\tilde{y})}\\
 & =\frac{\sum_{(y,\tilde{y})\in{\cal Y}^{\otimes2}}P_{Y}(y)P_{Y}(\tilde{y})\frac{P_{Y\tilde{Y}}^{\nu}(y,\tilde{y})}{P_{Y}^{\nu}(y)P_{Y}^{\nu}(\tilde{y})}\log^{2}\frac{P_{Y\tilde{Y}}(y,\tilde{y})}{P_{Y}(y)P_{Y}(\tilde{y})}}{\sum_{(y,\tilde{y})\in{\cal Y}^{\otimes2}}P_{Y\tilde{Y}}^{\nu}(y,\tilde{y})P_{Y}^{1-\nu}(y)P_{Y}^{1-\nu}(\tilde{y})}\nonumber \\
 & \hphantom{==}-\frac{\left(\sum_{(y,\tilde{y})\in{\cal Y}^{\otimes2}}P_{Y}(y)P_{Y}(\tilde{y})\frac{P_{Y\tilde{Y}}^{\nu}(y,\tilde{y})}{P_{Y}^{\nu}(y)P_{Y}^{\nu}(\tilde{y})}\log\frac{P_{Y\tilde{Y}}(y,\tilde{y})}{P_{Y}(y)P_{Y}(\tilde{y})}\right)^{2}}{\left(\sum_{(y,\tilde{y})\in{\cal Y}^{\otimes2}}P_{Y\tilde{Y}}^{\nu}(y,\tilde{y})P_{Y}^{1-\nu}(y)P_{Y}^{1-\nu}(\tilde{y})\right)^{2}}\\
 & =\E_{\nu}\left[\log^{2}\frac{P_{Y\tilde{Y}}(y,\tilde{y})}{P_{Y}(y)P_{Y}(\tilde{y})}\right]-\E_{\nu}^{2}\left[\log\frac{P_{Y\tilde{Y}}(y,\tilde{y})}{P_{Y}(y)P_{Y}(\tilde{y})}\right]\\
 & =\V_{\nu}\left[\log\frac{P_{Y\tilde{Y}}(y,\tilde{y})}{P_{Y}(y)P_{Y}(\tilde{y})}\right]\\
 & >0,
\end{align}
where the last inequality holds since due to the assumption that $I(Y;\tilde{Y})>0$
it holds that $Y$ and $\tilde{Y}$ are not statistically independent,
and so $\log[P_{Y\tilde{Y}}(y,\tilde{y})/P_{Y}(y)P_{Y}(\tilde{y})]$
is not a constant under the tilted measure (note that since $P_{Y}\otimes P_{Y}\gg P_{Y\tilde{Y}}$,
the tilted measure $P^{(\nu)}$ has the same support as $P_{Y\tilde{Y}}$). 

Similarly, it follows from Chernoff's bound that
\begin{align}
\log\P\left[\sum_{i=1}^{t}\log\lambda(Y_{i},\overline{Y}_{i})>0\right] & \leq\log\inf_{\nu\geq0}\E\left[\exp\left\{ \nu\sum_{i=1}^{t}\log\lambda(Y_{i},\overline{Y}_{i})\right\} \right]\\
 & =\log\inf_{\nu\geq0}\prod_{i=1}^{t}\E\left[\frac{P_{Y\tilde{Y}}^{\nu}(Y_{i},\overline{Y}_{i})}{P_{Y}^{\nu}(Y_{i})P_{Y}^{\nu}(\overline{Y}_{i})}\right]\\
 & =t\cdot\inf_{\nu\geq0}\log\sum_{(y,\overline{y})\in{\cal Y}^{\otimes2}}P_{Y\tilde{Y}}^{\nu}(y,\overline{y})P_{Y}^{1-\nu}(y)P_{Y}^{1-\nu}(\overline{y}).\\
 & =t\cdot\inf_{\nu\le1}\log\sum_{(y,\overline{y})\in{\cal Y}^{\otimes2}}P_{Y}^{\nu}(y)P_{Y}^{\nu}(\overline{y})P_{Y\tilde{Y}}^{1-\nu}(y,\overline{y}),
\end{align}
which is similar to the previous Chernoff-based bound, albeit with
the roles of $P_{Y\tilde{Y}}$ and $P_{Y}\otimes P_{Y}$ interchanged.
Defining similarly 
\[
g(\nu):=\log\sum_{(y,\overline{y})\in{\cal Y}^{\otimes2}}P_{Y}^{\nu}(y)P_{Y}^{\nu}(\overline{y})P_{Y\tilde{Y}}^{1-\nu}(y,\overline{y}),
\]
it holds that $g(1)=0$ and
\[
g(0)=\log P_{Y\tilde{Y}}(\supp(P_{Y}\otimes P_{Y}))=0
\]
since $P_{Y}\otimes P_{Y}\gg P_{Y\tilde{Y}}$. As before $g(\nu)$
is strictly convex, and so there exists $\nu_{*}\in(0,1)$ so that
$g(\nu_{*})<0$, and then (\ref{eq: Chernoff for sum of log-likelihood positive})
holds with $E_{+}(P_{XY})\geq-g(\nu_{*})$.
\end{IEEEproof}
The next lemma will also be used to analyze the type-II error probability,
however, it will be used for the case that the true overlap is either
overestimated or underestimated. 
\begin{lem}
\label{lem: crossing probability correlated log-likelihoods}It holds
that 
\[
\P\left[\sum_{i=1}^{t}\log\lambda(Y_{k+i},\tilde{Y}_{i})>0\right]\leq2\exp\left[-(t-1)\cdot\frac{E_{+}(P_{XY})}{2}\right].
\]
\end{lem}
\begin{IEEEproof}
In general, the r.v.'s $\{\lambda(Y_{k+i},\tilde{Y}_{i})\}_{i=1}^{t}$
are possibly statistically dependent, due to the dependence between
$Y_{i}$ and $\tilde{Y}_{i}$ for any given $i$. Following \cite[Lemma 16]{motahari2013information},
we note that $[t]$ can be partitioned into two sets $I_{1},I_{2}$
of almost equal size ($I_{1}\cup I_{2}=[t]$, $I_{1}\cap I_{2}=\emptyset$),
for which the $Y_{i}$'s and $\tilde{Y}_{i}$'s involved in each sum
are all statistically independent. Specifically, iteratively construct
$I_{1}$ as follows: Initializing $I_{1}^{(0)}=\emptyset$, at the
$r$th iteration, set 
\[
I_{1}^{(r)}\leftarrow I_{1}^{(r-1)}\cup\left\{ \{r,2k+r,\ldots\}\cap[t]\right\} 
\]
and 
\[
I_{2}^{(r)}\leftarrow I_{2}^{(r-1)}\cup\left\{ \{k+r,3k+r,\ldots\}\cap[t]\right\} .
\]
The construction halts at the first iteration $r_{*}$ for which $|I_{1}^{(r_{*})}|\in\{t/2-1,t/2\}$. 

For simplicity, let us assume that $t$ is odd, and so $|I_{1}|=(t-1)/2$
and $|I_{2}|=(t+1)/2$. The case of even $t$ is similar. Then, 
\begin{align}
\P\left[\sum_{i=1}^{t}\log\lambda(Y_{k+i},\tilde{Y}_{i})>0\right] & =\P\left[\sum_{i\in I_{1}}\log\lambda(Y_{k+i},\tilde{Y}_{i})+\sum_{i\in I_{2}}\log\lambda(Y_{k+i},\tilde{Y}_{i})>0\right]\\
 & \trre[\leq,a]\P\left[\sum_{i\in I_{1}}\log\lambda(Y_{k+i},\tilde{Y}_{i})>0\right]+\P\left[\sum_{i\in I_{2}}\log\lambda(Y_{k+i},\tilde{Y}_{i})>0\right]\\
 & \trre[=,b]\P\left[\sum_{i\in I_{1}}\log\lambda(Y_{i},\overline{Y}_{i})>0\right]+\P\left[\sum_{i\in I_{2}}\log\lambda(Y_{i},\overline{Y}_{i})>0\right]\\
 & \trre[\leq,c]2\exp\left[-(t-1)\cdot\frac{E_{+}(P_{XY})}{2}\right]
\end{align}
where $(a)$ follows from the union bound (at least one of the sums
must be positive), $(b)$ follows since within $I_{1}$, $\{Y_{k+i},\tilde{Y}_{i}\}_{i\in I_{1}}$
is a collection of $t$ statistically independent r.v.'s, all distributed
according to $P_{Y}$, and similarly for $I_{2}$, and $(c)$ follows
from Lemma \ref{lem: Chernoff analysis of crossing events}.
\end{IEEEproof}
Using the above lemmas, we may now prove Theorem \ref{thm: exponent upper bound noisy},
\begin{IEEEproof}[Proof of Theorem \ref{thm: exponent upper bound noisy}]
As can be discerned from the proof of Theorem \ref{thm: exponent upper bound noiseless},
and as discussed in the beginning of the proof of Theorem \ref{thm: exponent lower bound noiseless},
the analysis of positive and negative overlaps is analogous. Thus,
in order to simplify the analysis, we only consider positive overlaps,
and then double the error to account for negative overlaps, at the
end of the proof. As also discussed in the proof of Theorem \ref{thm: exponent upper bound noiseless},
the case $\beta<\frac{1}{I(Y;\tilde{Y})}$ is trivial, since this
is the error of a detector that never decides on an overlap. We thus
focus on the opposite case. With a slight abuse of notation, we let
$\{(Y_{i},\tilde{Y}_{i},\overline{Y}_{i})\}_{i\in\mathbb{Z}}$ be
a sequence of i.i.d. r.v.'s, distributed as $P_{Y\tilde{Y}}\otimes P_{Y}$.
We also let $\hat{T}\equiv\hat{T}_{\text{MAP}}$ be the optimal MAP
detector.

\uline{Analysis of the type-I error probability:} 

We begin by analyzing $p_{\text{type-I}}(\hat{T})$. By the union
bound and Prop. \ref{lem: likelihood for noisy reads}
\begin{align}
p_{\text{type-I}}(\hat{T}) & =\P\left[\hat{T}>0\mid T=0\right]\\
 & =\P\left[\bigcup_{t=t_{{\scriptscriptstyle \text{MDO}}}}^{\ell}\left\{ \sum_{i=1}^{t}\log\lambda\left(Y_{\ell-t+i}(1),Y_{i}(2)\right)\geq\log n_{\ell}\right\} \,\middle\vert\,T=0\right]\\
 & \leq\sum_{t=t_{{\scriptscriptstyle \text{MDO}}}}^{\ell}\P\left[\sum_{i=1}^{t}\log\lambda\left(Y_{\ell-t+i}(1),Y_{i}(2)\right)\geq\log n_{\ell}\,\middle\vert\,T=0\right].\label{eq: type I error first derivation}
\end{align}
Fix any $t\geq t_{{\scriptscriptstyle \text{MDO}}}$. To bound each
of the probabilities in (\ref{eq: type I error first derivation}),
we will use the truncated MGF bound of Lemma \ref{lem:truncated MGF}.
To this end, let 
\[
m_{3}^{3}:=\E\left[\left|\log\lambda(Y_{i},\tilde{Y}_{i})-\E[\log\lambda(Y_{i},\tilde{Y}_{i})]\right|^{3}\right]<\log\lambda_{\text{max}}^{3}<\infty,
\]
and let $\sigma^{2}:=\V[\log\lambda(Y_{i},\tilde{Y}_{i})]$, where
$\sigma^{2}>0$ since the channel $P_{Y\mid X}$ is not completely
noiseless and so $Y$ and $\tilde{Y}$ are not independent (as $\frac{1}{I(Y;\tilde{Y})}\leq\beta<\infty$).
Hence, it does not hold that $P_{Y\tilde{Y}}(y,\tilde{y})=P_{Y}(y)P_{Y}(\tilde{y})$
with probability $1$. In addition, conditioned on $T=0$, it holds
that $\{(Y_{\ell-t+i}(1),Y_{i}(2))\}_{i\in[t]}\eqd\{(Y_{i},\overline{Y}_{i})\}_{i\in[t]}$.
So, for any fixed $t\geq t_{{\scriptscriptstyle \text{MDO}}}$, any
of the probabilities in the sum (\ref{eq: type I error first derivation})
is bounded as
\begin{align}
 & \P\left[\sum_{i=1}^{t}\log\lambda(Y_{i},\overline{Y}_{i})\geq\log n_{\ell}\right]\nonumber \\
 & =\sum_{(y^{t},\overline{y}^{t})\in[{\cal Y}^{\otimes2}]^{t}}\prod_{i=1}^{t}P_{Y}(y_{i})P_{Y}(\overline{y}_{i})\cdot\I\left[\sum_{i=1}^{t}\log\lambda(y_{i},\overline{y}_{i})\geq\log n_{\ell}\right]\\
 & \trre[=,a]\sum_{(y^{t},\overline{y}^{t})\in[{\cal Y}^{\otimes2}]^{t}}P_{Y_{1}^{t}\tilde{Y}_{1}^{t}}(y^{t},\overline{y}^{t})\cdot\exp\left[-\sum_{i=1}^{t}\log\lambda(y_{i},\overline{y}_{i})\right]\I\left[\sum_{i=1}^{t}\log\lambda(y_{i},\overline{y}_{i})\geq\log n_{\ell}\right]\\
 & =\E\left[\exp\left[-\sum_{i=1}^{t}\log\lambda(Y_{i},\tilde{Y}_{i})\right]\I\left[\sum_{i=1}^{t}\log\lambda(Y_{i},\tilde{Y}_{i})\geq\log n_{\ell}\right]\right]\\
 & \trre[\leq,b]2\left(\frac{\log2}{\sqrt{2\pi\sigma^{2}}}+\frac{12m_{3}^{3}}{\sigma^{3}}\right)\frac{\exp(-\log n_{\ell})}{\sqrt{t}}\\
 & \leq\tilde{a}(P_{XY},\beta)\cdot\frac{1}{\sqrt{\log n}}\frac{1}{n_{\ell}},
\end{align}
for some constant $\tilde{a}(P_{XY},\beta)$ that does not depend
on $n$, where $(a)$ follows since 
\[
\exp\left[-\sum_{i=1}^{t}\log\lambda(y_{i},\overline{y}_{i})\right]=\prod_{i=1}^{t}\frac{P_{Y}(y_{i})P_{Y}(\overline{y}_{i})}{P_{Y\tilde{Y}}(y_{i},\overline{y}_{i})},
\]
and $(b)$ follows from Lemma \ref{lem:truncated MGF}. Plugging this
back to (\ref{eq: type I error first derivation}) results for $\ell=\beta\log n$
\begin{align}
p_{\text{type-I}}(\hat{T}) & \leq(\ell-t_{{\scriptscriptstyle \text{MDO}}})\tilde{a}(P_{XY},\beta)\cdot\frac{1}{\sqrt{\log n}}\frac{1}{n}\\
 & \leq\ell\tilde{a}(P_{XY},\beta)\cdot\frac{1}{\sqrt{\log n}}\frac{1}{n}\\
 & =a(P_{XY},\beta)\frac{\sqrt{\log n}}{n},
\end{align}
with $a(P_{XY},\beta)=\beta\cdot\tilde{a}(P_{XY},\beta)$, as claimed. 

\uline{Analysis of the type-II error probability:} 

We next turn to analyze $p_{\text{type-II}}(\hat{T};t)$. Let $\epsilon>0$
be fixed. To this end, we will use the Chernoff-bound based analysis
carried in Lemma \ref{lem: Chernoff analysis of crossing events}.
It is shown therein that for any $t=\theta\log n_{\ell}$ with $\frac{1}{\log\lambda_{\text{max}}}\leq\theta\leq\beta$
(this domain is not empty since $\frac{1}{\log\lambda_{\text{max}}}\leq\frac{1}{I(Y;\tilde{Y})}\leq\beta$)
it holds that
\begin{equation}
\P\left[\sum_{i=1}^{t}\log\lambda(Y_{i},\tilde{Y}_{i})\leq\log n_{\ell}\right]\leq\exp\left[-t\cdot E_{-}^{(t)}(1,P_{XY})\right],\label{eq: Chernoff for sum of log-likelihood is less than a*log(n) proof}
\end{equation}
where
\[
E_{-}^{(t)}(1,P_{XY}):=\sup_{\nu\leq1}(\nu-1)\cdot\left[\frac{\log n_{\ell}}{t}-D_{\nu}(P_{Y\tilde{Y}}\mid\mid P_{Y}\otimes P_{Y})\right].
\]
So, the negative log-probability in (\ref{eq: Chernoff for sum of log-likelihood is less than a*log(n) proof})
is lower bounded by $\log n_{\ell}\cdot A(\theta)$ where
\[
A(\theta):=\sup_{\text{\ensuremath{\nu\leq1}}}f(\nu):=\sup_{\text{\ensuremath{\nu\leq1}}}(\nu-1)-(\nu-1)\theta\cdot D_{\nu}(P_{Y\tilde{Y}}\mid\mid P_{Y}\otimes P_{Y}),
\]
and where $f(\nu)$ was implicitly defined. In addition, $\theta\to A(\theta)$
is monotonically non-decreasing. Indeed, this will follow if we show
that the supremum in $A(\theta)$ is always achieved at $\nu\in[0,1]$
since\footnote{Anyway, we may always restrict $\nu\in[0,1]$ in the Chernoff bound
(\ref{eq: Chernoff bound for under crossing event noisy}) of Lemma
\ref{lem: Chernoff analysis of crossing events}. We show here that
this does not compromises the bound. } 
\[
(\nu-1)D_{\nu}(P_{Y\tilde{Y}}\mid\mid P_{Y}\otimes P_{Y})\leq0
\]
for all $\nu\in[0,1]$ (which holds since $D_{\nu}(P_{Y\tilde{Y}}\mid\mid P_{Y}\otimes P_{Y})\geq0$
for $\nu\ge0$ \cite[Th. 8]{van2014renyi}). To show this property,
note that $f(1)=0$ and, as in the proof of Lemma \ref{lem: Chernoff analysis of crossing events},
\[
\left.\frac{\partial}{\partial\nu}f(\nu)\right|_{\nu=0}=\theta\cdot D_{1}(P_{Y}\otimes P_{Y}\mid\mid P_{Y\tilde{Y}})>0,
\]
so that $\nu\mapsto f(\nu)$ is increasing on $\nu=0$ and 
\[
\frac{\partial^{2}}{\partial\nu^{2}}f(\nu)=-\theta\cdot\V_{\nu}\left[\log\frac{P_{Y\tilde{Y}}(y,\tilde{y})}{P_{Y}(y)P_{Y}(\tilde{y})}\right]<0,
\]
so that $\nu\mapsto f(\nu)$ is strictly concave. From the above three
properties, its supremum over $\nu\leq1$ occurs in $[0,1]$. Accordingly,
we may define $\theta_{\epsilon}^{*}$ as the minimal value of $\theta$
such that $A(\theta)\geq\epsilon$, that is, 
\begin{align}
\theta_{\epsilon}^{*}\equiv\theta_{\epsilon}^{*}(P_{XY}) & :=\min\left\{ \theta\in\left(\frac{1}{\log\lambda_{\text{max}}},\beta\right)\colon A(\theta)\geq\epsilon\right\} \\
 & =\min_{0\leq\nu\leq1}\frac{1-\nu-\epsilon}{(1-\nu)\cdot D_{\nu}(P_{Y\tilde{Y}}\mid\mid P_{Y}\otimes P_{Y})},
\end{align}
where the last equality follows from the following consideration:
If $\theta\geq\theta_{\epsilon}^{*}$, then there exists $0\leq\nu\leq1$
such that
\[
(\nu-1)-(\nu-1)\theta\cdot D_{\nu}(P_{Y\tilde{Y}}\mid\mid P_{Y}\otimes P_{Y})\geq\epsilon
\]
which is equivalent to
\[
\theta\geq\frac{1-\nu-\epsilon}{(1-\nu)D_{\nu}(P_{Y\tilde{Y}}\mid\mid P_{Y}\otimes P_{Y})}
\]
as $D_{\nu}(P_{Y\tilde{Y}}\mid\mid P_{Y}\otimes P_{Y})>0$ for $\nu\geq0$
\cite[Th. 8]{van2014renyi}, and so $\theta_{\epsilon}^{*}$ is as
claimed. 

Let $t_{\epsilon}^{*}:=\lceil\theta_{\epsilon}^{*}\log n_{\ell}\rceil$.
Then, if $t\leq t_{\epsilon}^{*}$ we trivially bound the error by
$1$. For $t>t_{\epsilon}^{*}$ we decompose the error probability
into three terms, as
\begin{align}
p_{\text{type-II}}(\hat{T}_{\text{MAP}};t) & =\P\left[\hat{T}\neq t\mid T=t\right]\\
 & =\P\left[\hat{T}=0\mid T=t\right]+\sum_{k=1}^{t}\P\left[\hat{T}=t-k\mid T=t\right]+\sum_{k=1}^{\ell-t}\P\left[\hat{T}=t+k\mid T=t\right],\label{eq: error conditioned on positive t decomposition}
\end{align}
that is, an error to $\hat{T}=0$, an error that underestimate $T$
(but not zero), and an error that overestimate $T$. We next analyze
each of these terms separately. 

First, for an error to $\hat{T}=0$, it holds that
\begin{align}
 & \P\left[\hat{T}=0\mid T=t\right]\nonumber \\
 & \trre[=,a]\P\left[\sum_{i=1}^{t}\log\lambda(Y_{\ell-t+i}(1),Y_{i}(2))\leq\log n_{\ell}\,\middle\vert\,T=t\right]\\
 & \trre[=,b]\P\left[\sum_{i=1}^{t}\log\lambda(Y_{i},\tilde{Y}_{i})\leq\log n_{\ell}\right]\\
 & \trre[\leq,c]\exp\left[-A(\theta)\cdot\log n_{\ell}\right]\\
 & \trre[\leq,d]\exp\left[-\epsilon\log n_{\ell}\right]\\
 & =\exp\left[-\epsilon\log(n-\ell)\right]\\
 & =\frac{1}{(n-\ell)^{\epsilon}}\\
 & =\frac{1}{n^{\epsilon}}\cdot\frac{1}{[1-\frac{\beta\log n}{n}]{}^{\epsilon}}\\
 & \sim\frac{1}{n^{\epsilon}},\label{eq: error conditioned on t to zero}
\end{align}
where $(a)$ follows from Prop. \ref{lem: likelihood for noisy reads},
$(b)$ follows since conditioned on $T=t$, it holds that $Y_{\ell-t+i}(1)-X_{\ell-t+i}-Y_{i}(2)$
for $i\in[t]$ and that $\{Y_{\ell-t+i}(1)-X_{\ell-t+i}-Y_{i}(2)\}_{i\in[t]}$
are i.i.d., hence $\{Y_{\ell-t+i}(1),Y_{i}(2)\}_{i\in[t]}\eqd\{(Y_{i},\tilde{Y}_{i})\}_{i\in[t]}$,
$(c)$\textbf{ }follows from the statements made in the discussion
after (\ref{eq: Chernoff for sum of log-likelihood is less than a*log(n) proof}),
and $(d)$ follows since $\theta\to A(\theta)$ is monotonic non-decreasing,
and since $\theta\geq\theta_{\epsilon}^{*}$. 

Second, for an error to $\hat{T}=t-k$ for some $k\in\{1,\ldots,t-t_{{\scriptscriptstyle \text{MDO}}}\}$,
it holds that 
\begin{align}
 & \P\left[\hat{T}=t-k\mid T=t\right]\nonumber \\
 & \trre[\leq,a]\P\left[\log\frac{\prod_{i=1}^{t-k}\lambda(Y_{\ell-t+k+i}(1),Y_{i}(2))}{\prod_{i=1}^{t}\lambda(Y_{\ell-t+i}(1),Y_{i}(2))}\geq0\,\middle\vert\,T=t\right]\\
 & \trre[=,b]\P\left[\log\frac{\prod_{i=1}^{t-k}\lambda(Y_{k+i},\tilde{Y}_{i})}{\prod_{i=1}^{t}\lambda(Y_{i},\tilde{Y}_{i})}\geq0\right]\\
 & =\P\left[\sum_{i=1}^{t-k}\log\lambda(Y_{k+i},\tilde{Y}_{i})\geq\sum_{i=1}^{t}\log\lambda(Y_{i},\tilde{Y}_{i})\right]\\
 & \trre[\leq,c]\P\left[\left\{ \sum_{i=1}^{t-k}\log\lambda(Y_{k+i},\tilde{Y}_{i})>0\right\} \cup\left\{ \sum_{i=1}^{t}\log\lambda(Y_{i},\tilde{Y}_{i})<0\right\} \right]\\
 & \trre[\leq,d]\P\left[\sum_{i=1}^{t-k}\log\lambda(Y_{k+i},\tilde{Y}_{i})>0\right]+\P\left[\sum_{i=1}^{t}\log\lambda(Y_{i},\tilde{Y}_{i})<0\right]\\
 & \trre[=,e]\P\left[\sum_{i=1}^{t-k}\log\lambda(Y_{i},\overline{Y}_{i})>0\right]+\P\left[\sum_{i=1}^{t}\log\lambda(Y_{i},\tilde{Y}_{i})<0\right]\\
 & \trre[\leq,f]\exp\left[-\frac{(t-k-1)}{2}\cdot E_{+}\right]+\exp\left[-t\cdot E_{0}\right]\\
 & \trre[\leq,g]\exp\left[-\frac{E_{+}}{2\log\lambda_{\text{max}}}\log n_{\ell}\right]+\exp\left[-\frac{E_{0}}{\log\lambda_{\text{max}}}\log n_{\ell}\right]\\
 & \trre[\leq,h]\frac{1}{n^{\eta(P_{XY},\beta)}},\label{eq: error conditioned on t underestimate}
\end{align}
where $(a)$ follows from Prop. \ref{lem: likelihood for noisy reads},
$(b)$ follows since, conditioned on $T=t$, it holds that 
\[
\{Y_{\ell-t+k+i}(1),Y_{i}(2),Y_{\ell-t+i}(1)\}_{i\in[t]}\eqd\{(Y_{i+k},\tilde{Y}_{i},Y_{i})\}_{i\in[t]},
\]
$(c)$ follows since for any pair or r.v.'s $(U,V)$ it holds for
any $r\in\mathbb{R}$ that
\[
\P[U\geq V]\leq\P\left[\{U\geq r\}\cup\{V\leq r\}\right]
\]
and by choosing $r=0$, $(d)$ follows from the union bound, $(e)$
follows since $\{(Y_{k+i},\tilde{Y}_{i})\}_{i\in[t-k]}\eqd\{(Y_{i},\overline{Y}_{i})\}_{i\in[t-k]}$,
$(f)$ follows from the Chernoff-bound analysis in Lemma \ref{lem: Chernoff analysis of crossing events},
for some strictly positive constants $E_{0}\equiv E_{-}(0,P_{XY})>0$
and $E_{+}\equiv E_{+}(P_{XY})>0$, and from Lemma \ref{lem: crossing probability correlated log-likelihoods},
$(g)$ follows since $t-k\ge t_{{\scriptscriptstyle \text{MDO}}}=\frac{\log n_{\ell}}{\log\lambda_{\text{max}}}$
and $t>t_{\epsilon}^{*}\geq t_{{\scriptscriptstyle \text{MDO}}}$
both hold, and $(h)$ holds for some constant $\eta(P_{XY},\beta)>0$. 

Third, and similarly to the previous case, for an error to $\hat{T}=t+k$
for some $k\in\{1,\ldots,\ell-t\}$, it holds that 
\begin{align}
 & \P\left[\hat{T}=t+k\mid T=t\right]\nonumber \\
 & \trre[=,a]\P\left[\log\frac{\prod_{i=1}^{t+k}\lambda(Y_{\ell-t-k+i}(1),Y_{i}(2))}{\prod_{i=1}^{t}\lambda(Y_{\ell-t+i}(1),Y_{i}(2))}\geq0\,\middle\vert\,T=t\right]\\
 & \trre[=,b]\P\left[\log\frac{\prod_{i=1}^{t+k}\lambda(Y_{i-k},\tilde{Y}_{i})}{\prod_{i=1}^{t}\lambda(Y_{i},\tilde{Y}_{i})}\geq0\right]\\
 & =\P\left[\sum_{i=1}^{t+k}\log\lambda(Y_{i-k},\tilde{Y}_{i})\geq\sum_{i=1}^{t}\log\lambda(Y_{i},\tilde{Y}_{i})\right]\\
 & \leq\P\left[\sum_{i=1}^{t+k}\log\lambda(Y_{i-k},\tilde{Y}_{i})\geq0\right]+\P\left[\sum_{i=1}^{t}\log\lambda(Y_{i},\tilde{Y}_{i})<0\right]\\
 & =\P\left[\sum_{i=1}^{t+k}\log\lambda(Y_{i},\overline{Y}_{i})\geq0\right]+\P\left[\sum_{i=1}^{t}\log\lambda(Y_{i},\tilde{Y}_{i})<0\right]\\
 & \leq\exp\left[-\frac{(t+k-1)}{2}\cdot E_{+}\right]+\exp\left[-t\cdot E_{-}\right]\\
 & \leq\exp\left[-\frac{E_{+}}{2\log\lambda_{\text{max}}}\log n_{\ell}\right]+\exp\left[-\frac{E_{-}}{\log\lambda_{\text{max}}}\log n_{\ell}\right]\\
 & \leq\frac{1}{n^{\eta(P_{XY},\beta)}},\label{eq: error conditioned on t overestimate}
\end{align}
using similar reasoning as in the previous case. 

Thus, $p_{\text{type-II}}(\hat{T};t)\leq1$ if $t\leq t_{\epsilon}^{*}$
and the decomposition (\ref{eq: error conditioned on positive t decomposition})
and the bounds (\ref{eq: error conditioned on t to zero}) (\ref{eq: error conditioned on t underestimate})
and (\ref{eq: error conditioned on t overestimate}) result in that
for $t\geq t_{\epsilon}^{*}$
\[
p_{\text{type-II}}(\hat{T};t)\leq\frac{1}{n^{\epsilon/2}}+\frac{\beta\log n}{n^{\eta(P_{XY},\beta)}}\leq\frac{1}{n^{\epsilon/4}}
\]
for all $n\geq n_{0}(P_{XY},\beta,\epsilon)$. This completes the
analysis of $p_{\text{type-II}}(\hat{T};t)$.\textbf{ }

To complete the proof, we bound the Bayesian error probability as
\begin{align}
P_{\text{error}}^{*} & =\sum_{t=0}^{\ell}P_{T}(t)\cdot\P\left[\hat{T}\neq T\mid T=t\right]\\
 & =P_{T}(0)\cdot\P\left[\hat{T}>0\mid T=0\right]+\sum_{t=1}^{t_{\epsilon}^{*}}P_{T}(t)\cdot\P\left[\hat{T}\neq T\mid T=t\right]+\sum_{t=t_{\epsilon}^{*}+1}^{\ell}P_{T}(t)\cdot\P\left[\hat{T}\neq T\mid T=t\right]\\
 & \leq\P\left[\hat{T}>0\mid T=0\right]+\frac{t_{\epsilon}^{*}}{n}+\frac{\ell-t_{\epsilon}^{*}}{n}\max_{t\in\{t_{\epsilon}^{*}+1,\ldots,\ell\}}\P\left[\hat{T}\neq T\mid T=t\right]\\
 & \leq a(P_{XY},\beta)\frac{\sqrt{\log n_{\ell}}}{n}+\theta_{\epsilon}^{*}\frac{\log n_{\ell}}{n}+\frac{\beta\log n_{\ell}}{n^{1+\epsilon/4}}
\end{align}
for all $n\geq n_{0}(P_{XY},\beta,\epsilon)$. The result is then
completed since $\epsilon>0$ is arbitrary, and since $\alpha\to D_{\alpha}(P_{Y\tilde{Y}}\mid\mid P_{Y}\otimes P_{Y})$
is monotonic non-decreasing \cite[Summary table]{van2014renyi}, it
holds that
\[
\theta_{\epsilon}^{*}\to\frac{1}{D(P_{Y\tilde{Y}}\mid\mid P_{Y}\otimes P_{Y})}=\frac{1}{I(Y;\tilde{Y})}
\]
as $\epsilon\downarrow0$. 
\end{IEEEproof}

\subsection{Proof of Theorem \ref{thm:lower bound on Bayesian error}}

The proof of Theorem \ref{thm:lower bound on Bayesian error} follows
similar lines to the proof of Theorem \ref{thm: exponent lower bound noiseless},
and so we just highlight the differences. 
\begin{IEEEproof}[Proof of Theorem \ref{thm:lower bound on Bayesian error}]
 We use the same argument as in the proof of Theorem \ref{thm: exponent lower bound noiseless},
and focus on the case that $\frac{1}{I(Y;\tilde{Y})}\le\beta$ (the
opposite case is proved similarly, and even simpler). Using the notation
therein, we obtain 
\begin{align}
P_{\text{error}}(\hat{T}) & \trre[\geq,a]\min_{t\in[t_{1}]\backslash[t_{0}]}\pi_{t}\cdot\left(\P\left[\hat{D}_{t}=0\mid T=t\right]+\frac{\pi_{0}}{\pi_{t}}\cdot\P\left[\hat{D}_{t}=t\mid T=0\right]\right)\\
 & \trre[\geq,b]\min_{t\in[t_{1}]\backslash[t_{0}]}\pi_{t}\cdot\left(1-\P\left[\sum_{i=1}^{t}\log\lambda(Y_{i},\tilde{Y}_{i})>\log\frac{\pi_{0}}{\pi_{t}}\right]\right)\\
 & \trre[\sim,c]\pi_{t},
\end{align}
where $(a)$ follows using the arguments that lead to (\ref{eq: lower bound on error probability via binary HT})
(proof of Theorem \ref{thm: exponent lower bound noiseless}), $(b)$
follows by setting $\eta=\frac{\pi_{0}}{\pi_{t}}$ in Lemma \ref{lem: strong converse HT}
(or \cite[Theorem 14.9]{polyanskiy2023information}), and as the hypothesis
depends on $Y_{1}^{\ell}(1)$ and $Y_{1}^{\ell}(2)$ only through
their possibly overlapping part. Under the null hypothesis $T=t$,
and each distribution of joint letters is $P_{Y\tilde{Y}}$. Under
the alternative the joint distribution of every pair of letters is
$P_{Y}\otimes P_{Y}$. Finally, $(c)$ holds for any $t\in[t_{1}]\backslash[t_{0}]$
due to the following derivation: 
\begin{align}
 & \P\left[\sum_{i=1}^{t}\log\lambda(Y_{i},\tilde{Y}_{i})>\log\frac{\pi_{0}}{\pi_{t}}\right]\nonumber \\
 & \trre[\leq,i]\P\left[\sum_{i=1}^{t}\log\lambda(Y_{i},\tilde{Y}_{i})>\left(1-\zeta(n)\right)\log n\right]\\
 & \leq\P\left[\sum_{i=1}^{t}\log\lambda(Y_{i},\tilde{Y}_{i})-\E\left[\sum_{i=1}^{t}\log\lambda(Y_{i},\tilde{Y}_{i})\right]>\left(1-\zeta(n)\right)\log n-\E\left[\sum_{i=1}^{t}\log\lambda(Y_{i},\tilde{Y}_{i})\right]\right]\\
 & \trre[\leq,ii]\P\left[\sum_{i=1}^{t}\log\lambda(Y_{i},\tilde{Y}_{i})-\E\left[\sum_{i=1}^{t}\log\lambda(Y_{i},\tilde{Y}_{i})\right]>\frac{\theta_{0}}{2}\log n\right]\\
 & =\P\left[\frac{1}{t}\sum_{i=1}^{t}\log\lambda(Y_{i},\tilde{Y}_{i})-\E\left[\frac{1}{t}\sum_{i=1}^{t}\log\lambda(Y_{i},\tilde{Y}_{i})\right]>\frac{\theta_{0}}{2\theta_{1}}\right]\\
 & \trre[\leq,iii]\frac{4\theta_{1}^{2}\V[\log\lambda(Y_{i},\tilde{Y}_{i})]}{\theta_{0}^{2}}\cdot\frac{1}{t}\\
 & =o_{n}(1),
\end{align}
where here $(i)$ follows by setting, exactly as in (\ref{eq: log prior ratio for lower bound proof}),
\[
\log\frac{\pi_{0}}{\pi_{t}}=\left(1-\zeta(n)\right)\log n,
\]
$(ii)$ follows by choosing
\[
t_{1}=\frac{(1-\theta_{0})}{I(Y;\tilde{Y})}\log n
\]
so that for any $t\in[t_{0}]$
\[
\E\left[\sum_{i=1}^{t}\log\lambda(Y_{i},\tilde{Y}_{i})\right]=t\cdot I(Y;\tilde{Y})\leq(1-\theta_{0})\log n,
\]
and $\theta_{0}-\zeta(n)\geq\theta_{0}/2$, $(iii)$ follows from
Chebyshev\textquoteright s inequality, as $\V[\log\lambda(Y_{i},\tilde{Y}_{i})]<\infty$
(where the latter holds since $0<\lambda_{\text{min}}\leq\lambda(Y,\tilde{Y})\leq\lambda_{\text{max}}<\infty$
with probability $1$). Overall, we obtain that 
\begin{align}
P_{\text{error}}(\hat{T}) & \geq[1-o_{n}(1)]\pi_{t}\\
 & \sim\frac{\left(1-\theta_{0}/2\right)}{I(Y;\tilde{Y})}\frac{\log n}{n},
\end{align}
and the claim of the proof follows by taking $\theta_{0}\downarrow0$.
\end{IEEEproof}
\bibliographystyle{plain}
\bibliography{DNA_alignment}

\end{document}